\documentclass[journal,onecolumn,draftcls,12pt]{IEEEtran}
\usepackage[T1]{fontenc}
\pdfoutput=1

\makeatletter
\def\endthebibliography{%
	\def\@noitemerr{\@latex@warning{Empty `thebibliography' environment}}%
	\endlist
}
\makeatother

\IEEEoverridecommandlockouts
\usepackage{cite}
\usepackage{amsmath,amssymb,amsfonts}
\usepackage{algorithm}
\usepackage{algorithmic}
\usepackage{etoolbox}  
\usepackage{bbm}
\usepackage{amsmath}
\usepackage{tabularx}
\usepackage{amssymb}
\usepackage{threeparttable}
\usepackage{booktabs}
\usepackage{amsthm}
\usepackage{mathtools}
\usepackage{bm}
\makeatletter
\patchcmd{\algorithmic}{\addtolength{\ALC@tlm}{\leftmargin} }{\addtolength{\ALC@tlm}{\leftmargin}}{}{}
\makeatother

\makeatletter
\newcommand\fs@betterruled{%
	\def\@fs@cfont{\bfseries}\let\@fs@capt\floatc@ruled
	\def\@fs@pre{\vspace*{5pt}\hrule height.8pt depth0pt \kern2pt}%
	\def\@fs@post{\kern2pt\hrule\relax}%
	\def\@fs@mid{\kern2pt\hrule\kern2pt}%
	\let\@fs@iftopcapt\iftrue}
\floatstyle{betterruled}
\restylefloat{algorithm}
\makeatother

\usepackage{tikz}
    \usetikzlibrary{shapes.arrows}

\usepackage{pgfplots}
\usepackage{adjustbox}

\usepackage{gincltex}
\usepgfplotslibrary{fillbetween}
\usetikzlibrary{plotmarks}
\usetikzlibrary{patterns}
\usepackage[aboveskip=-5pt]{subcaption}
\usepackage{graphicx}
\usepackage{textcomp}
\usepackage{xcolor}
\usepackage{authblk}
\def\BibTeX{{\rm B\kern-.05em{\sc i\kern-.025em b}\kern-.08em
		T\kern-.1667em\lower.7ex\hbox{E}\kern-.125emX}}
		
\pgfplotsset{compat=1.15}
		
\usepackage{glossaries}

\newacronym{urllc}{URLLC}{Ultra-Reliable Low-Latency Communications}
\newacronym{opf}{OPF}{Oldest Packet First}
\newacronym{isl}{ISL}{Inter-Satellite Link}
\newacronym{jfi}{JFI}{Jain Fairness Index}
\newacronym{aoi}{AoI}{Age of Information}
\newacronym{paoi}{PAoI}{Peak Age of Information}
\newacronym{pdf}{PDF}{Probability Density Function}
\newacronym{pmf}{PMF}{Probability Mass Function}
\newacronym{cdf}{CDF}{Cumulative Density Function}
\newacronym{fcfs}{FCFS}{First Come First Serve}
\newacronym{ar}{AR}{Augmented Reality}
\newacronym{vr}{VR}{Virtual Reality}
\newacronym{qoe}{QoE}{Quality of Experience}
\newacronym{lcfs}{LCFS}{Last Come First Serve}
\newacronym{iot}{IoT}{Internet of Things}
\newacronym{pec}{PEC}{Packet Erasure Channel}
\newacronym[plural=MDPs,firstplural=Markov Decision Processes (MDPs)]{mdp}{MDP}{Markov Decision Process}
\newacronym{mdc}{MDC}{Multiple Description Coding}
\newacronym{fhw}{FHW}{Flatto-Hahn-Wright}
\newacronym[plural=RATs,firstplural=Radio Access Technologies (RATs)]{rat}{RAT}{Radio Access Technology}
\newacronym{mptcp}{MPTCP}{Multipath TCP}
\newacronym{sctp}{SCTP}{Stream Control Transmission Protocol}
\newacronym{leap}{LEAP}{Latency-controlled End-to-End Aggregation Protocol}
\newacronym{rtp}{RTP}{Real-time Transport Protocol}
\newacronym{mpmtp}{MPMTP}{Multipath Multimedia Transport Protocol}
\newacronym{qos}{QoS}{Quality of Service}

\definecolor{red}{HTML}{FFB14E}
\definecolor{blue}{HTML}{FA8775}
\definecolor{violet}{HTML}{EA5F94}
\definecolor{green_D}{HTML}{CD34B5}
\definecolor{cyan}{HTML}{9D02D7}
\definecolor{orange_D}{HTML}{0000FF}

\def \fwidth{0.6\columnwidth}
\def \fheight {0.3\columnwidth}

\def \sfwidth{0.96\linewidth}
\def \sfheight {0.64\linewidth}
\def \ssfheight {0.5\linewidth}

\begin{document}

\title{On the Role of Preemption for Timing Metrics in Coded Multipath Communication}

\author{Federico Chiariotti, Beatriz Soret, Petar Popovski\thanks{Federico Chiariotti (corresponding author, chiariot@dei.unipd.it) is with the Department of Information Engineering, University of Padova, Italy. Petar Popovski (petarp@es.aau.dk) is with the Department of Electronic Systems, Aalborg University, Denmark. Beatriz Soret (bsoret@ic.uma.es) is with the Telecommunication Research Institute (TELMA), Universidad de M\'{a}laga, Spain. Federico Chiariotti and Beatriz Soret are also with the Department of Electronic Systems, Aalborg University, Denmark. This work was partly funded by the IntellIoT project under the H2020 framework grant ID 957218, by the Villum Investigator Grant ``WATER'' from the Velux Foundation, Denmark, and by the Italian Ministry of University and Research as part of the PNRR Seal of Excellence Young Researchers project ``REDIAL.''}
} 

\maketitle

\begin{abstract}
Recent trends in communication networks have focused on \gls{qos} requirements expressed through timing metrics such as latency or \gls{aoi}. A possible way to achieve this is coded multipath communication: redundancy is added to a block of information through a robust packet-level code, transmitting across multiple independent channels to reduce the impact of blockages or rate fluctuation. The number of these links can grow significantly over traditional two-path schemes: in these scenarios, the optimization of the timing metrics is non-trivial, and latency and \gls{aoi} might require different settings. In particular, packet preemption is often the optimal solution to optimize \gls{aoi} in uncoded communication, but can significantly reduce the reliability of individual blocks. In this work, we model the multipath communication as a fork-join $D/M/(K, N )/L$ queue, where $K$ blocks of information are encoded into $N\geq K$ redundant blocks. We derive the latency and \gls{paoi} distributions for different values of the queue size $L$. Our results show that preemption is not always the optimal choice, as dropping a late packet on one path might affect the reliability of the whole block, and that minimizing the \gls{paoi} leads to poor latency performance.
\end{abstract}

\IEEEpeerreviewmaketitle
\glsresetall

\section{Introduction}

The emergence of multiple communication interfaces and the use of a variety of frequency bands has led to an increased interest in multipath communications. Path diversity in communication systems is expressed through the use of independent wired or wireless channels that are not affected by the same blockers and/or fading characteristics, thus leading to increased reliability. In this context, packet-level coding, which allows the encoding of blocks of $K$ information packets into $N$ redundant ones, so that the delivery of any subset of $K$ packets allows the receiver to decode the whole block, is a key enabling technology. Applications with stringent \gls{qos} requirements, defined in terms of timing metrics such as latency or \gls{aoi}~\cite{kaul2012real}, can benefit from the additional reliability provided by path diversity. The use of redundant paths has already been proposed, both for a single link~\cite{suer2019multi} and end-to-end connections~\cite{chiariotti2019analysis,chiariotti2021hop}, to reduce latency and provide reliability guarantees, and path diversity has also been used to provide \gls{urllc} service~\cite{nielsen2018ultra}.

However, the optimization of a multipath coded system presents significant challenges: while packet preemption, i.e., dropping older packets to avoid queuing, minimizes \gls{aoi} in many Markovian systems~\cite{bedewy2019minimizing}, different service distributions have more complex trade-offs, as dropping packets already in service might lead to lower reliability and actually increase the average \gls{aoi}~\cite{wang2018skip}. In a multipath system, the decision over whether and when to preempt is even more complex: straggler packets might be necessary to decode a block due to packet losses, and decisions on one path affect the system as a whole in non-obvious ways. Modeling the effect of these choices is necessary to design effective and reliable systems, whether they aim at minimizing latency or \gls{aoi}.

In this work, we use fork-join queuing theory~\cite{kim1989fork} to model multipath communication: in this model, packets from an individual source are split between different parallel queues, with synchronized arrivals at each queue, and then gathered at the receiver.
While there are some works in the literature studying the latency of coded fork-join queues, the trade-offs between latency and \gls{aoi} are still largely unexplored. Additionally, most works only consider the average latency, while we are interested in the full distribution of the latency and \gls{paoi} (i.e., the maximum value of the \gls{aoi} for each received packet), as studying worst-case performance is crucial for reliability guarantees. We consider a system with deterministic block arrivals, dividing each block into $K$ information packets, which are then encoded and sent over $N$ parallel queuing systems with exponentially distributed service. Each queue has a length $L$, and older packets are discarded when the queue overflows: $L=1$ corresponds to full packet preemption. With a slight abuse of the standard Kendall notation to describe queuing systems, we will refer to this model as $D/M/(K,N)/L$. To the best of our knowledge, this is the first work to study the \gls{paoi} for such a system. The main contributions in this paper are as follows:
\begin{itemize}
 \item We derive the \gls{pdf} of the latency and the block erasure probability for systems with arbitrary rates and for an arbitrary maximum queue length $L$, considering the effect of dropped packets on each individual queuing system on the probability of decoding the whole block;
 \item We extend the analytical model to derive the \gls{pdf} of the \gls{paoi} for the cases with $L=1$, $L=2$, and $L=\infty$;
 \item We give complexity bounds for the computation of these \glspl{pdf} in various cases, discussing the practical applicability of the analytical calculation;
 \item We verify the correctness of our analysis by simulation, drawing important insights on the design of latency- and \gls{paoi}-oriented multipath queuing systems.
\end{itemize}
A preliminary version of this paper, which only considers the cases with $L=1$ and $L=\infty$ in the analysis and presents a limited set of results, was presented in the conference version~\cite{chiariotti2022latency}. This version significantly expands the analysis by considering finite queue sizes, as well as presenting results on the optimization of the queue size and coding scheme, which contribute important design insights for this kind of system.

The rest of the paper is organized as follows: first, the related work on fork-join queuing systems, \gls{aoi}, and multipath communication is presented in Sec.~\ref{sec:related}. We then describe the system model and notations used in Sec.~\ref{sec:system}. The analysis for the cases with finite and infinite queue length is presented in Sec.~\ref{sec:an_L} and Sec.~\ref{sec:an_inf}, respectively, and the simulation settings and results are presented in Sec.~\ref{sec:results}. Finally, Sec.~\ref{sec:conc} concludes the paper and presents some possible avenues of future work.

\section{Related Work}\label{sec:related}

Fork-join queues~\cite{kim1989fork}, in which incoming tasks or blocks of data are divided among multiple queuing systems and served in parallel~\cite{baccelli1989fork}, are a well-studied model in computer science and communication. In particular, their parallel nature makes them well-suited to model parallel computing systems and multipath communication networks~\cite{khudabukhsh2017optimizing}. It is also possible to build complex fork-join networks with series of parallel queues and even loops~\cite{dallery1997properties}.

The main trade-off in redundant fork-join models is between latency and resource efficiency: adding more redundancy, i.e., increasing the number of queues $N$, can reduce the overall latency of each block, at the cost of additional computation or communication resources~\cite{joshi2017cloud}. The use of $N=K$, with no redundancy, can even be the optimal choice if the system is highly loaded and there error-free~\cite{shah2016redundant}. It is also possible to extend the model to more complex systems, with centralized and distributed computational and communication resources over which to schedule blocks of data and tasks~\cite{Sun2016OnDS}. Different queuing policies from the standard \gls{fcfs} can also be considered, taking the network graph and service rates into account~\cite{ozkan2019control}.

There are also some works dealing with worst-case performance in fork-join queues, mostly deriving bounds on the tail of the latency distribution~\cite{rizk2016stochastic} with Markovian arrival processes~\cite{fidler2016non}. An interesting recent work by Raaijmakers \emph{et al.} derives bounds for the latency with heavy-tailed service times~\cite{raaijmakers2021fork}. Another work by Ko and Serfozo~\cite{ko2008sojourn} approximates the complete distribution of latency in $G/M/(N,N)$ fork-join queues, but does not deal with redundant systems.

\gls{aoi} was first introduced in 2012 by Kaul \emph{et al.}~\cite{kaul2012real}, and it has since become a standard metric in real-time applications such as video streaming and \gls{iot} monitoring~\cite{abdelmagid2019ontheroleofaoi}. It is relevant when the receiver application is not just interested in receiving timely updates on a remote process, but also in maintaining fresh knowledge of it at all times. Consequently, the optimization of \gls{aoi} requires a balance between the transmission latency and the packet generation process: transmitting too often can lead to congestion, but transmitting too rarely can lead to high interarrival times, which also have a negative impact on \gls{aoi}. 

Most works in the \gls{aoi} literature are focused on theoretical queuing systems~\cite{kosta2017age}, although some adopt more realistic medium access models such as ALOHA~\cite{chen2021rach,munari2021irsa}, or multi-hop wireless systems~\cite{akar2020finding}. Furthermore, the focus of most of the literature has been on deriving the average \gls{aoi} for different systems, even though some recent works~\cite{chiariotti2020peak,inoue2019general} focus on reliability, deriving the full distribution of the \gls{paoi} or guarantees on its maximal values.

Preemption and the use of shorter queues to improve \gls{aoi} have been the subject of intense study over the past few years. Preemption (i.e., setting the queue length $L=1$) is the optimal solution to minimize the average age in any tandem of $M/M/1$~\cite{bedewy2019multihop} or $M/M/K$~\cite{bedewy2019minimizing} queues with a single source, due to the memorylessness property of Markovian queues. The decision over whether to preempt becomes more complex in $G/G/1$ systems~\cite{wang2018skip}, depending on the service and arrival distributions. $M/M/1/2$ systems, which can keep one packet in the queue and one in service, have been analyzed in~\cite{kosta2019queue,kosta2019age} in combination with the \gls{lcfs} policy.

However, \gls{aoi} is still largely unexplored in fork-join systems, and there is a limited number of works considering it in this setting. The most general of these~\cite{buyukates2020timely} deals with the average \gls{aoi} in what we denote as $M/M/(K,N)$ systems, applying it in the context of distributed computing. The complete distribution of the \gls{paoi} of a fork-join model was first derived in our own previous work~\cite{chiariotti2021latency}, which was limited to the simple $D/M/(K,2)$ and $M/M/(K,2)$ cases.
Another work by Talak \emph{et al.}~\cite{talak2021age} addresses the trade-off between \gls{aoi} and latency, considering the possibility of choosing one or more paths with an intelligent scheduler. The paper shows that age-oriented systems will increase the latency for packets that do not contribute to information freshness (i.e., packets that arrive out of order), increasing both the average and the variance of the latency significantly.

\section{System Model}\label{sec:system}
We first introduce some notation. In the following, we use $p_X(x)$ to indicate the \gls{pdf} of random variable $X$, and $P_X(x)$ to indicate its \gls{cdf}. Random variables are denoted with capital letters, while values are lowecase. Vectors are in bold, e.g., $\mathbf{v}$, and matrices are bold and capitalized, e.g., $\mathbf{M}$.

We consider a parallel queuing system with $N$ parallel queues (referred to throughout the paper as \emph{queuing systems} or simpy \emph{systems}) and synchronized arrivals. A block of $K$ packets is generated every $\tau$ seconds, and encoded using a packet erasure code into $N$ packets. The $N$ packets are then simultaneously queued over the $N$ queues. The block is decoded as soon as any set of $K$ packets is correctly received. 

Each individual system $j$ has an exponentially distributed service time with rate $\mu_j$, and a queue of (potentially infinite) size $L$: at any moment, there can only be up to $L$ packets in each system. In this work, we consider a preemptive \gls{fcfs} queuing policy, so that the packet currently in service is dropped if a new packet is generated and finds $L$ packets ahead of it. Packet dropping is performed independently on each individual queuing system. Therefore, if some packets are dropped on one system, the blocks they belong to might still be decoded correctly if a sufficient number of packets is correctly received on the other systems. Additionally, each system has a packet erasure probability $\varepsilon_j$, so that every transmitted packet might be undecodable at the receiver due to channel impairments.

\begin{figure}
    \centering
    \ifdefined\pdffig
        \includegraphics[width=0.75\linewidth]{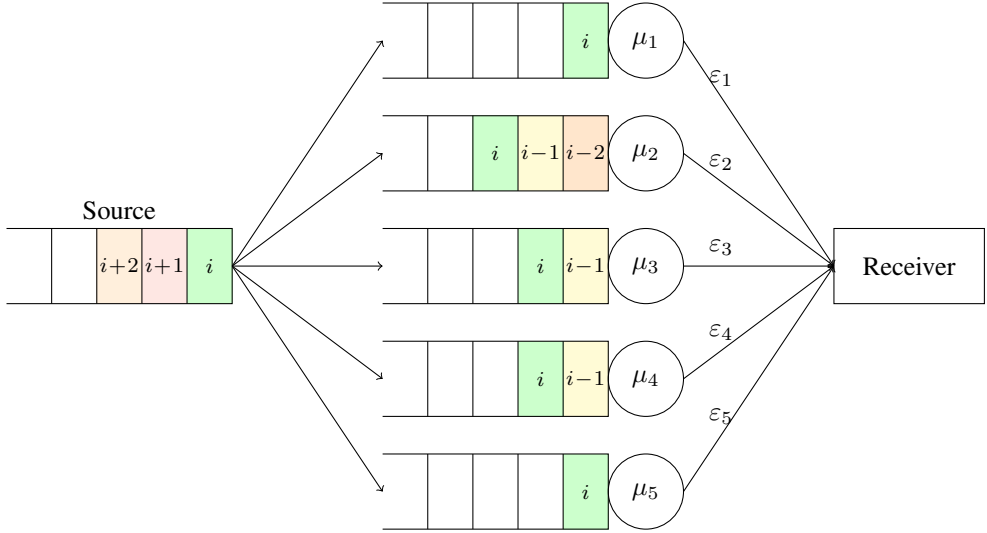}
    \else
        \begin{tikzpicture}

\draw [fill=white!80!green,draw=none](-2.6,1) rectangle (-2,2); 
\draw [fill=white!80!blue,draw=none] (-3.2,1) rectangle (-2.6,2); 
\draw [fill=white!80!red,draw=none] (-3.8,1) rectangle (-3.2,2); 

\node[minimum height=0.7cm] at (-2.3,1.5) {\scriptsize $i$};
\node[minimum height=0.7cm] at (-2.9,1.5) {\scriptsize $i\!+\!1$};
\node[minimum height=0.7cm] at (-3.5,1.5) {\scriptsize $i\!+\!2$};

\draw [fill=white!80!orange,draw=none] (2.4,2.5) rectangle (3,3.5); 
\draw [fill=white!80!yellow,draw=none] (2.4,2.5) rectangle (1.8,3.5); 
\draw [fill=white!80!yellow,draw=none] (2.4,1) rectangle (3,2); 
\draw [fill=white!80!yellow,draw=none] (2.4,-0.5) rectangle (3,0.5); 
\draw [fill=white!80!green,draw=none] (2.4,-2) rectangle (3,-1); 
\draw [fill=white!80!green,draw=none] (2.4,-0.5) rectangle (1.8,0.5); 
\draw [fill=white!80!green,draw=none] (2.4,1) rectangle (1.8,2); 
\draw [fill=white!80!green,draw=none] (1.2,2.5) rectangle (1.8,3.5); 
\draw [fill=white!80!green,draw=none] (2.4,4) rectangle (3,5); 

\node[minimum height=0.7cm] at (2.7,3) {\scriptsize $i\!-\!2$};
\node[minimum height=0.7cm] at (2.1,3) {\scriptsize $i\!-\!1$};
\node[minimum height=0.7cm] at (2.7,1.5) {\scriptsize $i\!-\!1$};
\node[minimum height=0.7cm] at (2.7,0) {\scriptsize $i\!-\!1$};

\node[minimum height=0.7cm] at (2.7,-1.5) {\scriptsize $i$};
\node[minimum height=0.7cm] at (2.1,0) {\scriptsize $i$};
\node[minimum height=0.7cm] at (2.1,1.5) {\scriptsize $i$};
\node[minimum height=0.7cm] at (1.5,3) {\scriptsize $i$};
\node[minimum height=0.7cm] at (2.7,4.5) {\scriptsize $i$};

\draw (-5,2) -- ++(3cm,0) -- ++(0,-1cm) -- ++(-3cm,0);
\foreach \i in {1,...,4}
  \draw (-2cm-\i*0.6cm,2) -- +(0,-1cm);

\foreach \j in {0,...,4}
{
    \draw (0,-1cm+\j*1.5cm) -- ++(3cm,0) -- ++(0,-1cm) -- ++(-3cm,0);
    \foreach \i in {1,...,4}
        \draw (3cm-\i*0.6cm,-1cm+\j*1.5cm) -- +(0,-1cm);
    \draw (3.5,(-1.5cm+\j*1.5cm) circle [radius=0.5cm];
    \draw[->] (-2,1.5) -- (0,-1.5cm+\j*1.5cm);
}

\draw[->] (4,-1.5cm) -- node[near start,above]{\small$\varepsilon_5$} (6,1.5);
\draw[->] (4,0) -- node[near start,above]{\small$\varepsilon_4$} (6,1.5);
\draw[->] (4,1.5) -- node[near start,above]{\small$\varepsilon_3$} (6,1.5);
\draw[->] (4,3) -- node[near start,above]{\small$\varepsilon_2$} (6,1.5);
\draw[->] (4,4.5) -- node[near start,above]{\small$\varepsilon_1$} (6,1.5);

\node[minimum height=1cm,minimum width=2cm,draw] at (7,1.5) {\small Receiver};
\node at (-3.5,2.25) {\small Source};

\node at (3.5,-1.5) {\small $\mu_5$};
\node at (3.5,0) {\small $\mu_4$};
\node at (3.5,1.5) {\small $\mu_3$};
\node at (3.5,3) {\small $\mu_2$};
\node at (3.5,4.5) {\small $\mu_1$};

\end{tikzpicture}
\vspace{0.3cm}
    \fi
    \caption{Schematic of the system model with $N=5$.}
    \label{fig:system}\vspace{-0.6cm}
\end{figure}

If block $i$ is generated at time $g_i$ and its packet is received on system $j$ at time $r_{i,j}$, with $r_{i,j}=\infty$ if the packet is dropped or erased, we can compute the delivery latency $D_i$:
\begin{equation}
 D_i=\inf\left(t\in\mathbb{R}:\sum_{j=1}^N\mathbbm{1}(t-r_{i,j})\geq K\right),
\end{equation}
where $\mathbbm{1}(x)$ is the step function, which is equal to 1 if $x\geq0$ and 0 otherwise. Naturally, some blocks might not be delivered at all, as there might be more than $N-K$ erased or dropped packets. As such, the \gls{cdf} of the latency does not reach 1, except for the error-free case with $L=\infty$, in which no block is lost. The reception instant of a block is denoted by $r_i=g_i+D_i$. We can also define the \gls{aoi} $\theta(t)$ as the time elapsed since the generation of the last correctly received block:
\begin{equation}
 \theta(t)=t-\sup\left(g_i\in\mathbb{R}:g_i+D_i\leq t\right).
\end{equation}
We can then define the \gls{paoi} $\Delta_i$, which is the \gls{aoi} measured at the instant right before decoding for the $i$-th packet:
\begin{equation}
 \Delta_i=\theta(r_i-\epsilon)+\epsilon,
\end{equation}
where $\epsilon$ is an arbitrarily small positive quantity.

\section{Analysis: The $D/M/(K,N)/L$ System with Preemption}\label{sec:an_L}

We can first consider each system to apply the \gls{fcfs} queuing policy over a finite queue of length $L$. If a new packet finds a full queue ahead of it, the oldest packet is dropped and the new one is added at the end of the queue. The case with $L=1$ is the preemptive system that has been shown to be optimal with respect to the average age in the $K=N=1$ case, in which each new set of packets goes into service immediately, regardless of previous packets.

\subsection{Dropping Probability and Latency for a Single Path}

We first define the Poisson \gls{pmf}, which is denoted as $\mathcal{P}_{\mu}(n,t)$:
\begin{equation}
\mathcal{P}_{\mu}(n,t)=\frac{(\mu t)^n e^{-\mu t}}{n!}.
\end{equation}
We can then look at the state of the system $S_{i,j}$, which represents the number of packets currently in system $j$, as seen by the arriving packet $i$  as a Markov chain with timestep $\tau$, whose transition probability is given by:
\begin{equation}
  P^{(L)}_{S_{i+1,j}|S_{i,j}}(s_{i+1,j}|s_{i,j})=\begin{cases}
                          1-\sum_{n=0}^{\min(1+s_{i,j},L)-1}\mathcal{P}_{\mu_j}(n,\tau),&\text{if }s_{i+1,j}=0;\\
                          \mathcal{P}_{\mu_j}(\min(1+s_{i,j},L)-s_{i+1,j},t), &\text{if }0<s_{i+1,j}\leq s_{i,j}+1;\\
                          0, &\text{if }s_{i+1,j}>s_{i,j}+1.
                        \end{cases}
\end{equation}
If we consider the transition matrix $\mathbf{M}^{(L)}_j$ for system $j$, we can derive the steady-state distribution $\bm{\pi}_j$ as its normalized left-eigenvector with eigenvalue 1:
\begin{align}
  (\mathbf{M}^{(L)}_j-\mathbf{I})\bm{\pi}_j=&0;\\
  \sum_{s=0}^L\pi_j(s)=&1,
\end{align}
where $\mathbf{I}$ is the $(L+1)\times(L+1)$ identity matrix. If packet $i$ arrives and finds the system in state $S_j=L$, i.e., with a full queue, the oldest packet is dropped. The variable $Q_{i,j}$, indicating the number of packets ahead of packet $i$ in the queue as it arrives, is then given by:
\begin{equation}
  Q_{i,j}=\min(L-1,S_{i,j}).
\end{equation}

We now consider what happens in system $j$ to a packet arriving and finding $q<L$ packets ahead of it in the queue.
First, we need to compute the dropping probability, as the arrival of new packets might fill up the queue and force the system to drop the packet. In the following, we consider system $j$ individually, as the state of its queue can be separated from the others. We define the dropping probability as $p_{\text{drop},j}^{(L)}(q)$. We can define the dropping probability recursively, knowing that the queue increases after each packet arrival:
\begin{equation}
  p_{\text{drop},j}^{(L)}(q)=\begin{cases}
                       \sum_{d=0}^q \mathcal{P}_{\mu_j}(d,\tau)p_{\text{drop},j}^{(L-1)}(q-d),  &\text{if } 0\leq q<L-1;\\
                       \sum_{d=0}^q \mathcal{P}_{\mu_j}(d,\tau)p_{\text{drop},j}^{(L-1)}(q-\max(d,1)),  &\text{if } q=L-1;\\
                       1, &\text{if }L=0.
                       \end{cases}
\end{equation}

The index $d$ in the sum indicates how many packets depart the queue before the arrival of the next packet, whose distribution is a truncated Poisson~\cite{pack1977output}. After the arrival of the next packet, there is one more occupied spot in the queue after the packet we are considering, and the queue in front of it has decreased by $d$.
We can now look at the delivery latency \gls{pdf} for system $j$, following the same method:
\begin{equation}
\begin{aligned}
  &p_{D_{i,j}|Q_{i,j}}^{(L)}(t|q_{i,j})=\\
  &\begin{cases}
                                         (1-\varepsilon_j)\mu_j\mathcal{P}_{\mu_j}(q_{i,j},t) &\text{if }0\leq t\tau;\\
                                         \sum\limits_{d=0}^{q_{i,j}}(1-\varepsilon_j)\mathcal{P}_{\mu_j}(d,\tau) p_{D_{i,j}|Q_{i,j}}^{(L-1)}(t-\tau|q_{i,j}-d),  &\text{if }t\in (\tau,L\tau],\,{q_{i,j}}<L-1;\\                                         
                                         \sum\limits_{d=0}^{q_{i,j}} (1-\varepsilon_j)\mathcal{P}_{\mu_j}(d,\tau) p_{D_{i,j}|Q_{i,j}}^{(L-1)}(t-\tau|q_{i,j}-\max(d,1)),&\text{if }t\in (\tau,L\tau],\,{q_{i,j}}=L-1;\\
                                         0, &\text{if }t>L\tau.
                                       \end{cases}
\end{aligned}
\end{equation}
Note that the \gls{pdf} does not sum to 1, as we consider the case in which the packet is dropped or erased to have infinite latency.
It is relatively simple to derive the \gls{cdf}, using the known results on the Erlang distribution:
\begin{equation}
\begin{aligned}
  &P_{D_{i,j}|Q_{i,j}}^{(L)}(t|q_{i,j})=\\
  &\begin{cases}
    (1-\varepsilon_j)\frac{\gamma(q+1,\mu_j t)}{q_{i,j}!}, &\text{if }0\leq t\leq\tau;\\
    P_{D_{i,j}|Q_{i,j}}^{(L)}(\tau|q_{i,j})+\sum\limits_{\mathclap{d=0}}^{q_{i,j}} \mathcal{P}_{\mu_j}(d,\tau) P_{D_{i,j}|Q_{i,j}}^{(L-1)}(t-\tau|q_{i,j}-d),  &\text{if }t\in (\tau,L\tau],\,q_{i,j}<L-1;\\                                         
    P_{D_{i,j}|Q_{i,j}}^{(L)}(\tau|q_{i,j})+\sum\limits_{\mathclap{d=0}}^{q_{i,j}} \mathcal{P}_{\mu_j}(d,\tau) P_{D_{i,j}|Q_{i,j}}^{(L-1)}(t-\tau|q_{i,j}-\max(d,1)),  &\text{if }t\in (\tau,L\tau],\,q_{i,j}=L-1;\\
    1, &\text{if } t>L\tau,
  \end{cases}
\end{aligned}
\end{equation}
where $\gamma(s,x)$ is the lower incomplete Gamma function, defined as:
\begin{equation}
  \gamma(s,x)=\int_0^x t^{s-1} e^{-t} dt.
\end{equation}
The conditional probabilities with $S_{i,j}$ as a condition instead of $Q_{i,j}$ can be trivially obtained.
The unconditional dropping probability and latency \gls{pdf} and \gls{cdf} can be computed by applying the law of total probability:
\begin{align}
  p_{\text{drop},j}^{(L)}&=\sum_{s=0}^L \pi_j(s) p_{\text{drop},j}^{(L)}(\min(s,L-1))\\
  p_{D_{i,j}}^{(L)}(t)&=\sum_{s=0}^L \pi_j(s) p_{D_{i,j}|S_{i,j}}^{(L)}(t|\min(s,L-1)).
\end{align}

\subsection{Computing the Latency Distribution}

We can now move from a single queuing system to the overall multipath connection. We first denote the set of numbers from 1 to $N$ as $\mathcal{N}=\{1,\ldots,N\}$. We can now define $\mathcal{S}(K,N)$ as the set of possible unordered sets of non-repeating indices of length $K$:
\begin{equation}
  \mathcal{S}(K,N)=\left\{\mathcal{L}\in\mathcal{N}^K:i\neq j\, \forall i,j\in\mathcal{L}\right\}.
\end{equation}
We can then express the conditioned \gls{pdf} of the overall latency for the $i$-th block in a given state $\mathbf{Q}_i$, which we denote as $p^{(L)}_{D_i}(t)$, as the product of the probability of having decoded the $K-1$ packets in set $\mathcal{L}$ beforehand and getting the $K$-th at time $t$ on path $\ell$, while the packets on the remaining paths in set $\mathcal{N}\setminus\mathcal{L}\setminus\{\ell\}$ have not been received:
\begin{equation}
  p^{(L)}_{D_i|\mathbf{Q}_i}(t|\mathbf{q}_i)=\sum_{\mathclap{\mathcal{L}\in\mathcal{S}(K-1,N)}}\quad\ \prod_{j\in\mathcal{L}}P_{D_{i,j}|Q_{i,j}}(t|q_{i,j})\sum_{\mathclap{\ell\in\mathcal{N}\setminus\mathcal{L}}}(1-\varepsilon_{\ell})p_{D_{i,\ell}|Q_{i,\ell}}(t|q_{i,\ell})\prod_{\mathclap{m\in\mathcal{N}\setminus\mathcal{L}\setminus\{\ell\}}}(1-P_{D_{i,m}|Q_{i,m}}(t|q_{i,m})).
\end{equation}
As the latency distribution is only defined recursively, the latency \gls{cdf} is complex. However, it is relatively simple to get the unconditional \gls{pdf}:
\begin{equation}
  p^{(L)}_{D_i|\mathbf{Q}_i}(t|\mathbf{q}_i)=\sum_{\mathbf{q}\in\{0,\ldots,L-1\}^N}p^{(L)}_{D_i|\mathbf{Q}_i}(t|\mathbf{q})\prod_{j=1}^N \pi_j(q_j).
\end{equation}
As above, it is trivial to obtain $p^{(L)}_{D_i|\mathbf{S}_i}(t|\mathbf{s}_i)$ from $p^{(L)}_{D_i|\mathbf{Q}_i}(t|\mathbf{q}_i)$.

The complexity of computing the delivery latency \gls{pdf} is very high: since the computation involves several nested sums over subsets, its computational complexity is $O(L^N L! N^{K+2})$, as it depends on iterating over all possible states of the queues and subsets of the possible paths. This makes the formulation valid for all values of $L$, but only practical for short queues and a relatively small number of channels.

\subsection{The $L=1$ case}

The equation is much simpler in the $L=1$ case, as we have a single state, and the \gls{pdf} of the delivery latency is given by:
\begin{equation}
    p^{(1)}_{D_i}(t)=\sum_{\mathcal{L}\in\mathcal{S}(K-1,N)}\prod_{j\in\mathcal{L}} (1-\varepsilon_j)\left(1-e^{-\mu_jt}\right)\sum_{\mathclap{\ell\in\mathcal{N}\setminus\mathcal{L}}}(1-\varepsilon_{\ell})\mu_\ell e^{-\mu_\ell t}\prod_{\mathclap{m\in\mathcal{N}\setminus\mathcal{L}\setminus\{\ell\}}}(\varepsilon_m+(1-\varepsilon_m)e^{-\mu_mt}).
\end{equation}
We can now compute the corresponding \gls{cdf}. In the following, we sum over the possible configurations, where, as above, set $\mathcal{L}$ contains the $K-1$ packets delivered before decoding, while path $\ell$ is the one over which the $K$-th packet is delivered. Set $\mathcal{G}$ contains the paths over which packets were delivered before the $K$-th packet, but were erased, and we need to iterate over the possible subsets $\mathcal{M}$ of $\mathcal{L}$:
\begin{equation}
\begin{aligned}
  P^{(1)}_{D_i}(t)=&\int_0^t p_{D_i}(x) dx\\
  =&\sum_{\mathcal{L}\in\mathcal{S}(K-1,N)}\sum_{{\ell\in\mathcal{N}\setminus\mathcal{L}}}\sum_{\mathcal{G}\subseteq\mathcal{N}\setminus\mathcal{L}\setminus\{\ell\}}\sum_{\mathcal{M}\subseteq\mathcal{L}}\int_0^t\left(\prod_{m\in\mathcal{N}\setminus\mathcal{L}\setminus\mathcal{G}\setminus\{\ell\}} \varepsilon_m\right)\mu_\ell (-1)^{|\mathcal{M}|}\\
  &\times\left(\prod_{{j\in\mathcal{L}\cup\mathcal{G}\cup\{\ell\}}}(1-\varepsilon_j)\right)e^{-x\left(\mu_\ell+\sum_{j\in\mathcal{M}\cup\mathcal{G}}\mu_j\right)} dx\\
  =&\sum_{\mathcal{L}\in\mathcal{S}(K-1,N)}\sum_{{\ell\in\mathcal{N}\setminus\mathcal{L}}}\sum_{\mathcal{G}\subseteq\mathcal{N}\setminus\mathcal{L}\setminus\{\ell\}}\sum_{\mathcal{M}\subseteq\mathcal{L}}\left(\prod_{m\in\mathcal{N}\setminus\mathcal{L}\setminus\mathcal{G}\setminus\{\ell\}} \varepsilon_m\right)\mu_\ell (-1)^{|\mathcal{M}|}\\
  &\times\left(\prod_{{j\in\mathcal{L}\cup\mathcal{G}\cup\{\ell\}}}(1-\varepsilon_j)\right)\frac{1-e^{-t\left(\mu_\ell+\sum_{j\in\mathcal{M}\cup\mathcal{G}}\mu_j\right)}}{\mu_\ell+\sum_{j\in\mathcal{M}\cup\mathcal{G}}\mu_j}.
\end{aligned}
\end{equation}
In this case, the success probability for decoding a block is simply given by $p_s^{(1)}=P^{(1)}_{D_i}(\tau)$, as the packet is always in the first position in the queue and the length of the queue is $L=1$. We can simply get the \gls{pdf} of the \gls{paoi} for such a system as:
\begin{equation}
  p_{\Delta_i}^{(1)}(\omega)=(1-p_s^{(1)})^{\max\left(\left\lfloor\frac{\omega}{\tau}\right\rfloor-1,0\right)} p_{D_{i}}^{(1)}(\text{mod}(\omega,\tau)).\label{eq:L1agepdf}
\end{equation}
We can now easily derive the \gls{paoi} \gls{cdf}:
\begin{equation}
  P_{\Delta_i}^{(1)}(\omega)= (p_s^{(1)})^{\left\lfloor\frac{\omega}{\tau}\right\rfloor-1} + (1-p_s^{(1)})^{\left\lfloor\frac{\omega}{\tau}\right\rfloor-1} P_{D_{i}}^{(1)}(\text{mod}(\omega,\tau)).
\end{equation}

The complexity of computing the delivery latency \gls{pdf} and \gls{cdf} is lower than for the $L$-sized queue, but still considerable, as there are several nested sums over subsets. In particular, the complexity of the \gls{pdf} computation is $O(N^{K+1})$, as it depends on iterating over the subsets of size $K-1$ of the set of packets, and the complexity for each cycle is $O(NK)$. The \gls{cdf} computation is even more complex, as it iterates over all possible subsets, requiring $O(N^{K+1}2^N)$ operations.

\subsection{The $L=2$ case}

In order to compute the \gls{paoi}, we need to consider the possibility that the previous block of data was unsuccessful, due to either erasures or dropped packets. This makes the calculation very complicated, so we consider the simple case with $L=2$; cases with longer queues can be analyzed in the same way, but the derivation of the \gls{paoi} is very cumbersome. 

We can first look at $p_{\text{drop},j}^{(2)}(q)$:
\begin{equation}
  p_{\text{drop},j}^{(2)}(q)= e^{-2\mu_j\tau}(1+\mu_j\tau\delta(1-q)),
\end{equation}
where $\delta(x)$ is 1 if $x=0$ and 0 otherwise. We can now give the expression for $p_{D_{i,j}|Q_{i,j}}^{(2)}(t|q_{i,j})$:
\begin{equation}
  p_{D_{i,j}|Q_{i,j}}^{(2)}(t|q_{i,j})=\begin{cases}
                                       (1-\varepsilon_j)\mu_j e^{-\mu_j t},  &\text{if } 0<t\leq2\tau, q_{i,j}=0;\\
                                       (1-\varepsilon_j)\mu_j^2 t e^{-\mu_j t},  &\text{if } 0<t\leq\tau, q_{i,j}=1;\\
                                       (1-\varepsilon_j)(1+\mu_j\tau)\mu_j e^{-\mu_j t},  &\text{if } \tau<t\leq2\tau, q_{i,j}=1;\\
                                       0,  &\text{otherwise.}
                                       \end{cases}
\end{equation}
The \gls{cdf} is also easy to obtain:
\begin{equation}
  P_{D_{i,j}|Q_{i,j}}^{(2)}(t|q_{i,j})=\begin{cases}
                                       (1-\varepsilon_j)(1-e^{-\mu_j t}),  &\text{if } 0<t\leq2\tau, q_{i,j}=0;\\
                                       (1-\varepsilon_j)(1-(1+\mu_j t)\mu_je^{-\mu_j t}),  &\text{if } 0<t\leq\tau, q_{i,j}=1;\\
                                       (1-\varepsilon_j)(1-(1+\mu_j \tau)\mu_je^{-\mu_j \tau}(1-e^{-\mu_jt})),  &\text{if } \tau<t\leq2\tau, q_{i,j}=1;\\
                                       (1-\varepsilon_j)(1-p_{\text{drop},j}^{(2)}(q_{i,j})),  &\text{if }t>2\tau.
                                       \end{cases}
\end{equation}
We now examine the \gls{paoi} for the $L=2$ system. We begin by recalling the probability of the queue emptying out before block $i+1$ is generated:
\begin{equation}
  p_{Q_{i+1,j}|Q_{i,j}}(0|q_{i,j})=1-e^{-\mu_j\tau}(1+\mu\tau q_{i,j}).
\end{equation}
We can then define a sequence of queue states for the packets from $i+1$ to $i+\ell$, defined as $\bm{\Xi}_{j,\ell}\in\{0,1\}^{\ell}$.
The probability of having sequence $\bm{\xi}_{j,\ell}$ is given by:
\begin{equation}
  p_{\bm{\Xi}_{j,\ell}}(\bm{\xi}_{j,\ell})=
  \begin{cases}
    \frac{(1-e^{-\mu_j\tau})(1+\mu_j\tau\pi_j(1))\prod_{m=1}^{\ell-1}p_{Q_{i+1,j}|Q_{i,j}}(\xi_{m+1}|\xi_m)}{\pi_j(0)(1-p_{\text{drop},j}(0)+\pi_j(1)(1-p_{\text{drop},j}(1))},&\text{if }\xi_1=0;\\
    \frac{(e^{\mu_j\tau}-1)(1+\mu_j\tau\pi_j(1))\prod_{m=1}^{\ell-1}p_{Q_{i+1,j}|Q_{i,j}}(\xi_{m+1}|\xi_m)}{\pi_j(0)(1-p_{\text{drop},j}(0)+\pi_j(1)(1-p_{\text{drop},j}(1))},&\text{if }\xi_1=1.                                              
  \end{cases}
\end{equation}
We can now look at the \gls{pdf} of the delivery time of packet $i+m$, with $m<\ell$, for a given sequence $\bm{\xi}_{j,\ell}$:
\begin{equation}
  p_{D_{i+m,j}|\bm{\Xi}_{j,\ell}}(t|\bm{\xi}_{j,\ell})=\begin{cases}
                            \frac{(1-\varepsilon_j)\mu_j^{\xi_m+1}t^{\xi_m}e^{-\mu_jt}}{1-e^{-\mu_j\tau}(1+\mu_j\tau \xi_m)}, &\text{if }t\leq\tau,\xi_{m+1}=0;\\
                            (1-\varepsilon_j)\mu_je^{\mu_j(t-\tau)}, &\text{if } \tau<t\leq2\tau,\xi_{m+1}=1;\\
                            0, &\text{otherwise.}
                           \end{cases}
\end{equation}
We can easily derive the \gls{cdf}:
\begin{equation}
  P_{D_{i+m,j}|\bm{\Xi}_{j,\ell}}(t|\bm{\xi}_{j,\ell})=\begin{cases}
                            \frac{(1-\varepsilon_j)\left(1-e^{-\mu_jt}(1+\mu_j\tau \xi_m)\right)}{1-e^{-\mu_j\tau}(1+\mu_j\tau \xi_m)}, &\text{if }t\leq\tau,\xi_{m+1}=0;\\
                            1-\varepsilon_j, &\text{if }t>\tau,\xi_{m+1}=0;\\
                            0, &\text{if }t\leq\tau,\xi_{m+1}=1;\\
                            (1-\varepsilon_j)(1-e^{\mu_j(t-\tau)}), &\text{if } \tau<t\leq2\tau,\xi_{m+1}=1;\\
                            (1-\varepsilon_j)(1-e^{-\mu_j\tau}),  &\text{if } \tau<t\leq2\tau,\xi_{m+1}=1.
                           \end{cases}
\end{equation}
If we look at the system as a whole, we need at least $K$ packets to be received correctly to decode the block and reset the \gls{paoi} clock. In order to simplify the calculations, we assume that $N<2K$, which means that blocks are always either decoded in order or lost. We then have:
\begin{equation}
\begin{aligned}
  p_{\Delta_{i,j}}(\ell\tau+\omega)=&\sum_{\bm{\xi}_{\ell+1}\in\{0,1\}^{N(\ell+1)}}\prod_{j=1}^N p_{\bm{\Xi}_{j,\ell}}(\bm{\xi}_{j,\ell})\sum_{k=0}^{K-1}\sum_{\mathcal{M}\in\mathcal{S}(k,N)}\sum_{n=1}^{\ell-1}\prod_{m\in\mathcal{M}}P_{D_{i+n,m}|\bm{\Xi}_{\ell+1}}(\tau|\bm{\xi}_{\ell+1})\\
  &\prod_{\mathclap{m\in\mathcal{N}\setminus\mathcal{M}}}(1-P_{D_{i+n,m}|\bm{\Xi}_{\ell+1}}(\tau|\bm{\xi}_{\ell+1}))
  \sum_{\mathcal{L}\in\mathcal{S}(K-1,N)}\prod_{j\in\mathcal{L}} P_{D_{i+\ell,j}|\bm{\Xi}_{j,\ell+1}}(\omega|\bm{\xi}_{j,\ell+1})\\
  &\sum_{\mathclap{m\in\mathcal{N}\setminus\mathcal{L}}}p_{D_{i+\ell}|\bm{\Xi}_{\ell+1}}(\omega|\bm{\xi}_{\ell+1})\prod_{\mathclap{m\in\mathcal{N}\setminus\mathcal{L}\setminus\{\ell\}}}(1-P_{D_{i+\ell}|\bm{\Xi}_{\ell+1}}(\omega|\bm{\xi}_{\ell+1})).
\end{aligned}
\end{equation}
As the \gls{pdf} depends on multiple iterations, it is computationally very heavy, as its complexity is $O(\ell N^{K+1}2^{\ell+2N})$. The solution is then very impractical to compute even for small systems and values of the \gls{paoi}.

\section{Analysis: The $D/M/(K,N)/\infty$ System}\label{sec:an_inf}

Finally, we analyze a classical queuing system with infinite buffers and \gls{fcfs} queuing. The steady-state probability right before a new arrival was derived in~\cite{pinotsi2005synchronized} using Palm probability theory~\cite{baccelli2012palm}, and is given by:
\begin{equation}
 p_{Q_{i,j}}^{(\infty)}(q_{i,j})=(1-\sigma_j)\sigma_j^{q_{i,j}},\label{eq:p_md1_alt}
\end{equation} 
where the parameter $\sigma_j$ is the solution in $(0,1)$ to the following equation:
\begin{equation}
 x=e^{2\mu_j \tau(x-1)}.\label{eq:sigma}
\end{equation}
We know that the delivery time for packet $i$, which finds $q_{i,j}$ packets queued ahead of it in system $j$, follows an Erlang distribution~\cite{erlang1917losning}, weighted by the success probability on the link:
\begin{equation}
  p_{D_{i,j}|Q_{i,j}}^{(\infty)}(t|q_{i,j})=(1-\varepsilon_j)\frac{(\mu_jt)^{q_{i,j}}e^{-\mu_jt}}{q_{i,j}!}.\label{eq:cond_single}
\end{equation}
We then compute the corresponding \gls{cdf}, $P_{D_{i,j}|Q_{i,j}}^{(\infty)}(t|q_{i.j})$: 
\begin{equation}
  P_{D_{i,j}|Q_{i,j}}^{(\infty)}(t|q_{i.j})=(1-\varepsilon_j)\mu_j\left(1-\sum_{n=0}^{q_{i,j}}\mathcal{P}_{\mu_j}(n,t)\right).\label{eq:cond_del}
\end{equation}
Note that the \gls{cdf} sums to $(1-\varepsilon_j)$, as erased packets have infinite latency. 
In order for the transmission of the coded block to be successful, the packets from the links contained in one of the vectors in $\mathcal{S}(K,N)$ must be delivered correctly. We can then express the probability of decoding the block after latency $t$ for a given state $\mathbf{Q}_i$ as the product of the probability of having decoded $K-1$ packets beforehand and getting the $K$-th at time $t$:
\begin{equation}
  p_{D_i|\mathbf{Q}_i}^{(\infty)}(t|\mathbf{q}_i)=\sum_{\mathcal{L}\in\mathcal{S}(K-1,N)}\prod_{j\in\mathcal{L}} P_{D_{i,j}|Q_{i,j}}^{(\infty)}(t|q_{i,j})\sum_{\mathclap{\ell\in\mathcal{N}\setminus\mathcal{L}}}p_{D_{i,\ell}|Q_{i,\ell}}^{(\infty)}(t|q_{i,\ell})\prod_{\mathclap{m\in\mathcal{N}\setminus\mathcal{L}\setminus\{\ell\}}}(1-P_{D_{i,m}|Q_{i,m}}^{(\infty)}(t|q_{i,m})).\label{eq:cond_del_full}
\end{equation}
As the systems are separable, the steady-state probability of being in a given state for the $j$-th system is independent from all others, and follows~\eqref{eq:p_md1_alt}. We can then remove the condition on~\eqref{eq:cond_del} for each individual system:
\begin{equation}
  \begin{aligned}
    P_{D_{i,j}}^{(\infty)}(t)=&\sum_{q_{i,j}=0}^\infty \pi_j(q_{i,j})P_{D_{i,j}|Q_{i,j}}^{(\infty)}(t|q_{i.j})\\
    =&\sum_{q_{i,j}=0}^\infty(1-\sigma_j)(1-\varepsilon_j)\sigma_j^{q_{i,j}}\left(1-\sum_{n=0}^{q_{i,j}}\mathcal{P}_{\mu_j}(n,t)\right)\\
    =&(1-\varepsilon_j)\left(1-e^{-\mu_jt}\sum_{q_{i,j}=0}^\infty\sum_{n=0}^{q_{i,j}}\frac{(1-\sigma_j)\sigma_j^{q_{i,j}}(\mu_jt)^n}{n!}\right)\\
    =&(1-\varepsilon_j)\left(1-e^{-\mu_jt}\sum_{n=0}^\infty\frac{(\mu_j\sigma_jt)^n}{n!}\right)\\
    =&(1-\varepsilon_j)\left(1-e^{-\mu_j(1-\sigma_j)t}\right).\label{eq:del_uncond}
  \end{aligned}
\end{equation}
We can now substitute~\eqref{eq:del_uncond} into~\eqref{eq:cond_del_full} to get the latency \gls{pdf} for the block:
\begin{equation}
\begin{aligned}
    p_{D_i}^{(\infty)}(t)=&\sum_{\mathcal{L}\in\mathcal{S}(K-1,N)}\prod_{j\in\mathcal{L}} (1-\varepsilon_j)\left(1-e^{-\mu_j(1-\sigma_j)t}\right)\sum_{\mathclap{\ell\in\mathcal{N}\setminus\mathcal{L}}}(1-\varepsilon_{\ell})(1-\sigma_{\ell})\mu_{\ell}e^{-\mu_{\ell}(1-\sigma_{\ell})t}\\
    &\times\prod_{\mathclap{m\in\mathcal{N}\setminus\mathcal{L}\setminus\{\ell\}}}\left(\varepsilon_m+(1-\varepsilon_m)e^{-\mu_j(1-\sigma_j)t}\right).\label{eq:del_full}
\end{aligned}
\end{equation}
We can easily compute the corresponding \gls{cdf}:
\begin{equation}
\begin{aligned}
  P^{(\infty)}_{D_i}(t)=&\int_0^t p_{D_i}^{(\infty)}(x) dx\\
   =&\sum_{M=K}^N\sum_{\mathcal{L}\in\mathcal{S}(M,\mathcal{N})}\prod_{j\in\mathcal{L}}(1-\varepsilon_j)\left(1-e^{-\mu_j(1-\sigma_j) t}\right)\prod_{\mathclap{m\in\mathcal{N}\setminus\mathcal{L}}} \left(\varepsilon_m+(1-\varepsilon_m)e^{-\mu_m(1-\sigma_m) t}\right).
\end{aligned}
\end{equation}
As no packets are dropped from the queue, the success probability $p_s$ for a block is given by:
\begin{equation}
  p_s^{(\infty)}=\sum_{M=K}^{N}\sum_{\mathcal{L}\in\mathcal{S}(m,N)}\prod_{j\in\mathcal{L}}\prod_{m\in\mathcal{N}\setminus\mathcal{L}}(1-\varepsilon_j)\varepsilon_{m}.
\end{equation}
In order to get the latency distribution for successful blocks, i.e., conditioning on the block's success, it is sufficient to divide $p_{D_i}^{(\infty)}(t)$ by $p_s^{(\infty)}$. We can then compute the \gls{paoi} \gls{pdf}, considering that failures are independent:
\begin{equation}
p_{\Delta_i}^{(\infty)}(\omega)=\sum_{e=0}^{\left\lfloor\frac{\omega}{\tau}\right\rfloor}(1-p_s^{(\infty)})^e p_{D_{i}}^{(\infty)}(\omega-e\tau).
\end{equation}
The \gls{paoi} \gls{cdf} calculation is also straightforward:
\begin{equation}
P_{\Delta_i}^{(\infty)}(\omega)=\sum_{e=0}^{\left\lfloor\frac{\omega}{\tau}\right\rfloor}(1-p_s^{(\infty)})^e P_{D_{i}}^{(\infty)}(\omega-e\tau).
\end{equation}

\section{Simulation Settings and Results}\label{sec:results}

In the following, we verify our analytical calculations by comparing them with the results of Monte Carlo simulations on the system. The simulations were run for $N_p=10^6$ packets in each case, and the empirical and analytical \glspl{cdf} match perfectly.

\begin{figure}
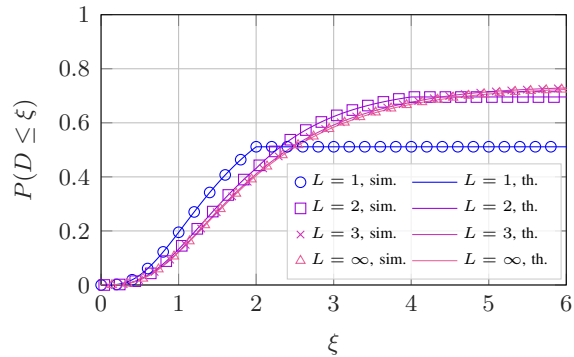
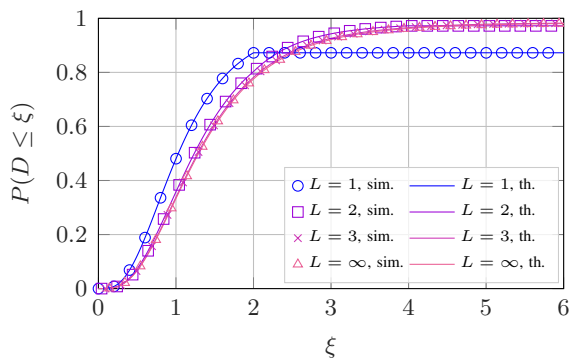
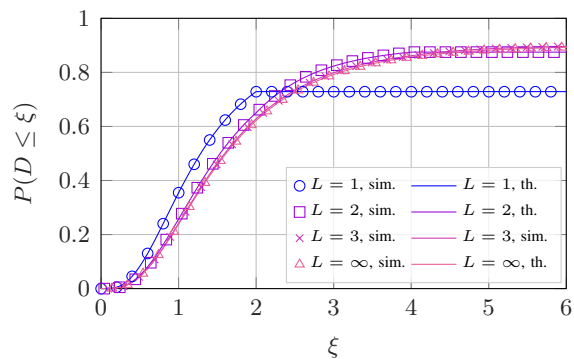
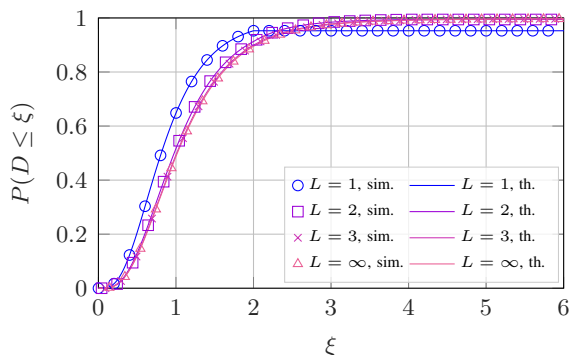
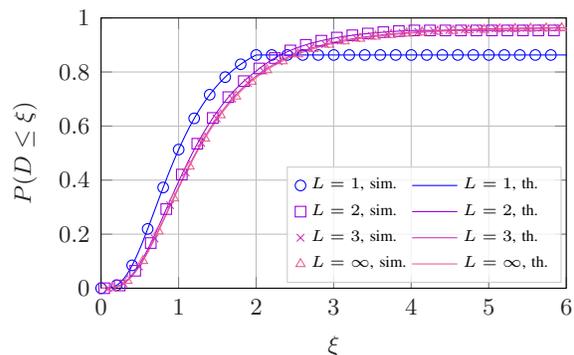

    \centering
	\begin{subfigure}[b]{.49\linewidth}
	    \centering    
	    \ifdefined\pdffig
            \includegraphics[width=\linewidth]{tikz/paper-figure1}
        \else
            \input{fig/lat_45_rho_0.5_e_0.1.tex}
        \fi
        \caption{Latency \gls{cdf} for the $(4,5)$ system with $\varepsilon=0.1$.}
        \label{fig:45_01_rho05}
    \end{subfigure}	
	\begin{subfigure}[b]{.49\linewidth}
	    \centering
        \ifdefined\pdffig
            \includegraphics[width=\linewidth]{tikz/paper-figure2}
        \else
            \input{fig/lat_45_rho_0.5_e_0.2.tex}
        \fi
        \caption{Latency \gls{cdf} for the $(4,5)$ system with $\varepsilon=0.2$.}
        \label{fig:45_02_rho05}
    \end{subfigure}	
	\begin{subfigure}[b]{.49\linewidth}
	    \centering
        \ifdefined\pdffig
            \includegraphics[width=\linewidth]{tikz/paper-figure3}
        \else
            \input{fig/lat_46_rho_0.5_e_0.1.tex}
        \fi
        \caption{Latency \gls{cdf} for the $(4,6)$ system with $\varepsilon=0.1$.}
        \label{fig:46_01_rho05}
    \end{subfigure}	
    \begin{subfigure}[b]{.49\linewidth}
	    \centering
        \ifdefined\pdffig
            \includegraphics[width=\linewidth]{tikz/paper-figure4}
        \else
            \input{fig/lat_46_rho_0.5_e_0.2.tex}
        \fi
        \caption{Latency \gls{cdf} for the $(4,6)$ system with $\varepsilon=0.2$.}
        \label{fig:46_02_rho05}
    \end{subfigure}
    \begin{subfigure}[b]{.49\linewidth}
	    \centering
        \ifdefined\pdffig
            \includegraphics[width=\linewidth]{tikz/paper-figure5}
        \else
            \input{fig/lat_47_rho_0.5_e_0.1.tex}
        \fi        
        \caption{Latency \gls{cdf} for the $(4,7)$ system with $\varepsilon=0.1$.}
        \label{fig:47_01_rho05}
    \end{subfigure}
    \begin{subfigure}[b]{.49\linewidth}
	    \centering
        \ifdefined\pdffig
            \includegraphics[width=\linewidth]{tikz/paper-figure6}
        \else
            \input{fig/lat_47_rho_0.5_e_0.2.tex}
        \fi        
        \caption{Latency \gls{cdf} for the $(4,7)$ system with $\varepsilon=0.2$.}
        \label{fig:47_02_rho05}
    \end{subfigure}
     \caption{Latency \gls{cdf} for different queue sizes and codes with $\mu=1$ and $\tau=2$. Markers are used for the simulations and solid lines for the theoretical analysis.}\vspace{-0.6cm}
 \label{fig:lat_CDF}
\end{figure}

In the case for $N\geq2K$, which is not pictured in our plots, the analytical \gls{pdf} of the \gls{paoi} for $L>1$ is actually an upper bound, as we discussed in the previous section, but we could not find a setting for which the difference was significant except for the $(1,2)$ case, discussed in our previous work~\cite{chiariotti2021latency}. This can be easily explained: in order for block $i+1$ to be decoded before block $i$ in a $(K,N)$ system, a series of circumstances must arise, the combination of which is extremely rare (aside for the $(1,2)$ case):
\begin{itemize}
 \item $E_i\geq K$ packets from the $i$-th block are erased;
 \item At least $K$ packets from the $i+1$-th block are delivered successfully;
 \item At least $K$ packets from the $i$-th block are delivered successfully, but the $K$-th packet from block $i$ arrives after the $K$-th packet from block $i+1$.
\end{itemize}
Finally, we consider theoretical curves for the delivery latency for the blocks for the cases in which $L=2$ and $L=3$, but the significant computational complexity of the \gls{paoi} calculation led us to only consider the Monte Carlo simulation results for that metric.

\subsection{Balanced Scenario}

We first consider a balanced scenario, in which $\mu_j=1\,\forall j$. In this case, no path is faster or slower than any of the others. Fig.~\ref{fig:lat_CDF} shows the \gls{cdf} of the latency for this scenario in the case where $\tau=2$. In all cases, the system with $L=3$ is almost indistinguishable from the one with $L=\infty$, as there are almost never more than 3 packets in the queue, and the effect of packet dropping is then unnoticeable. The system with $L=2$, on the other hand, has a slightly faster delivery, but lower reliability: the cases in which the $L=2$ system drops packets are the ones that have a higher latency in $L>2$ systems. 

\begin{figure}
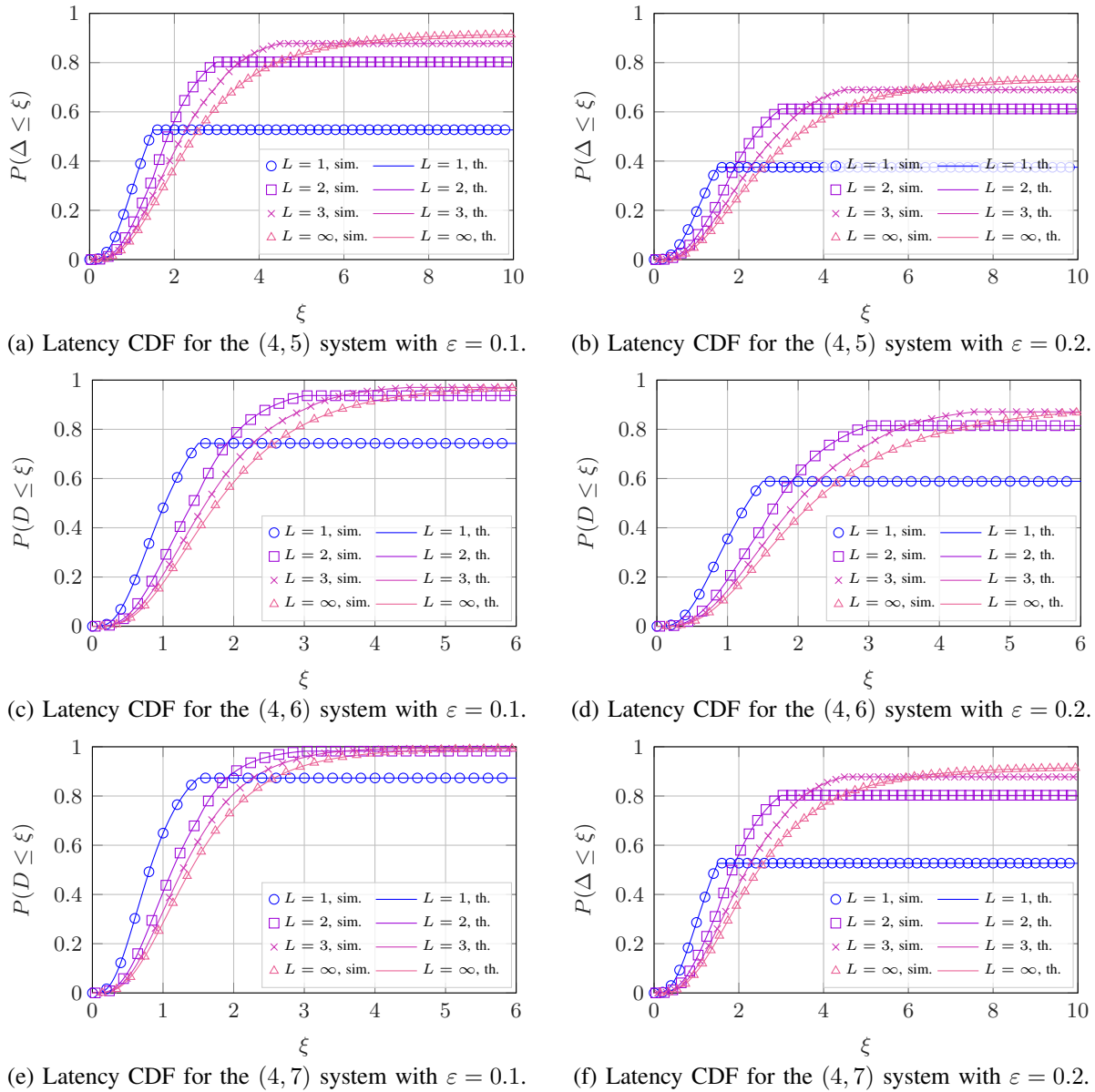

    \centering
	\begin{subfigure}[b]{.49\linewidth}
	    \centering
        \ifdefined\pdffig
            \includegraphics[width=\linewidth]{tikz/paper-figure7}
        \else
            \input{fig/lat_45_rho_0.67_e_0.1.tex}
        \fi
        \caption{Latency \gls{cdf} for the $(4,5)$ system with $\varepsilon=0.1$.}
        \label{fig:45_01_rho066}
    \end{subfigure}	
	\begin{subfigure}[b]{.49\linewidth}
	    \centering
        \ifdefined\pdffig
            \includegraphics[width=\linewidth]{tikz/paper-figure8}
        \else
            \input{fig/lat_45_rho_0.67_e_0.2.tex}
        \fi        
        \caption{Latency \gls{cdf} for the $(4,5)$ system with $\varepsilon=0.2$.}
        \label{fig:45_02_rho066}
    \end{subfigure}	
	\begin{subfigure}[b]{.49\linewidth}
	    \centering
        \ifdefined\pdffig
            \includegraphics[width=\linewidth]{tikz/paper-figure9}
        \else
            \input{fig/lat_46_rho_0.67_e_0.1.tex}
        \fi
        \caption{Latency \gls{cdf} for the $(4,6)$ system with $\varepsilon=0.1$.}
        \label{fig:46_01_rho066}
    \end{subfigure}	
    \begin{subfigure}[b]{.49\linewidth}
	    \centering
        \ifdefined\pdffig
            \includegraphics[width=\linewidth]{tikz/paper-figure10}
        \else
            \input{fig/lat_46_rho_0.67_e_0.2.tex}
        \fi
        \caption{Latency \gls{cdf} for the $(4,6)$ system with $\varepsilon=0.2$.}
        \label{fig:46_02_rho066}
    \end{subfigure}
    \begin{subfigure}[b]{.49\linewidth}
	    \centering
        \ifdefined\pdffig
            \includegraphics[width=\linewidth]{tikz/paper-figure11}
        \else
            \input{fig/lat_47_rho_0.67_e_0.1.tex}
        \fi
        \caption{Latency \gls{cdf} for the $(4,7)$ system with $\varepsilon=0.1$.}
        \label{fig:47_01_rho066}
    \end{subfigure}
    \begin{subfigure}[b]{.49\linewidth}
	    \centering
        \ifdefined\pdffig
            \includegraphics[width=\linewidth]{tikz/paper-figure12}
        \else
            \input{fig/lat_45_rho_0.67_e_0.1.tex}
        \fi
        \caption{Latency \gls{cdf} for the $(4,7)$ system with $\varepsilon=0.2$.}
        \label{fig:47_02_rho066}
    \end{subfigure}
     \caption{Latency \gls{cdf} for different queue sizes and codes with $\mu=1$ and $\tau=1.5$.}\vspace{-0.6cm}
 \label{fig:lat_CDF_highload}
\end{figure}

The preemptive system with $L=1$ is a clearer example of this: since the arrival of a new packet makes each system drop the one in service if it has not been delivered, the \gls{cdf} of the latency becomes flat after $\tau$. This can speed up the delivery of the blocks that are decoded, but reduces the success probability, particularly for systems with lower redundancy: in this case, dropped packets can have a significant impact on the overall decoding probability of the block. In fact, selecting the correct queue length and avoiding dropping either too few or too many packets becomes even more critical when the system load increas, as shown in Fig.~\ref{fig:lat_CDF_highload}: in this case, the scenario is the same, but the load is increased from 0.5 to 0.67 by setting $\tau=1.5$. As the figure shows, the difference between the systems becomes clearer: setting a lower $L$ can lead to delivering blocks significantly faster, but correspondingly losing significantly more, particularly when the redundancy is limited. The optimal queue length then depends on the priorities of the system, i.e., whether latency for the delivered blocks or reliability is more important.

\begin{figure}
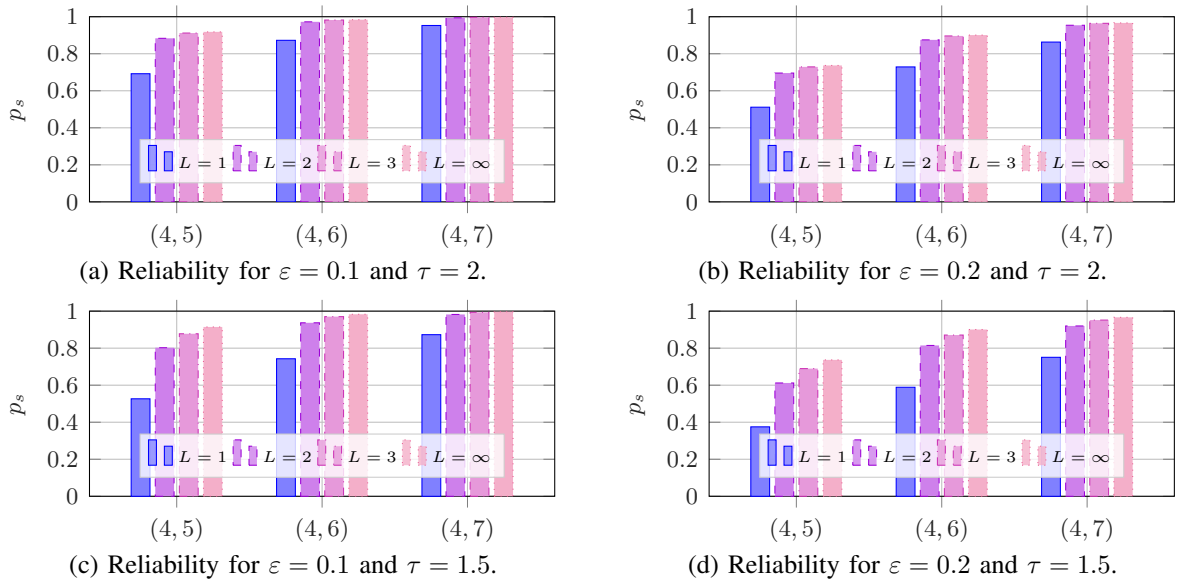

    \centering
	\begin{subfigure}[b]{.49\linewidth}
	    \centering
        \ifdefined\pdffig
            \includegraphics[width=\linewidth]{tikz/paper-figure13}
        \else
            \begin{tikzpicture} 
\begin{axis}
[ybar, 
enlarge x limits=0.3, 
bar width=0.25cm,
axis background/.style={fill=white},
xmajorgrids,
ymajorgrids,
ymin=0, 
width=\sfwidth, height=\ssfheight, 
ymax=1, 
legend style={font=\tiny, at={(0.5,0.1)}, anchor=south,fill opacity=0.8, legend columns=4,draw opacity=1, text opacity=1, draw=white!80!black},
ylabel={$p_s$}, 
symbolic x coords={{$(4,5)$}, {$(4,6)$}, {$(4,7)$}}, 
xtick=data,  
ylabel near ticks,
ylabel style={font=\footnotesize\color{white!15!black}},
xticklabel style={font=\footnotesize\color{white!15!black}},
yticklabel style={font=\footnotesize\color{white!15!black}}
] 

\addplot[color=orange_D,fill={white!50!orange_D}] coordinates {({$(4,5)$},0.6921) ({$(4,6)$},0.8725) ({$(4,7)$},0.95257)};
\addlegendentry{$L=1$}
\addplot[color=cyan,fill={white!50!cyan},dashed] coordinates {({$(4,5)$},0.88326) ({$(4,6)$},0.97233) ({$(4,7)$},0.994179)};
\addlegendentry{$L=2$}
\addplot[color=green_D,fill={white!50!green_D},dashdotted] coordinates {({$(4,5)$},0.91187) ({$(4,6)$},0.9821) ({$(4,7)$},0.99678)};
\addlegendentry{$L=3$}
\addplot[color=violet,fill={white!50!violet},dotted] coordinates {({$(4,5)$},0.91808) ({$(4,6)$},0.9840) ({$(4,7)$},0.99723)};
\addlegendentry{$L=\infty$}
 

\end{axis}
\end{tikzpicture}
        \fi
        \caption{Reliability for $\varepsilon=0.1$ and $\tau=2$.}
        \label{fig:rel_01}
    \end{subfigure}	
	\begin{subfigure}[b]{.49\linewidth}
	    \centering
        \ifdefined\pdffig
            \includegraphics[width=\linewidth]{tikz/paper-figure14}
        \else
            \begin{tikzpicture} 
\begin{axis}
[ybar, 
enlarge x limits=0.3, 
bar width=0.25cm,
axis background/.style={fill=white},
xmajorgrids,
ymajorgrids,
ymin=0, 
width=\sfwidth, height=\ssfheight, 
ymax=1, 
legend style={font=\tiny, at={(0.5,0.1)}, anchor=south,fill opacity=0.8, legend columns=4,draw opacity=1, text opacity=1, draw=white!80!black},
ylabel={$p_s$}, 
symbolic x coords={{$(4,5)$}, {$(4,6)$}, {$(4,7)$}}, 
xtick=data,  
ylabel near ticks,
ylabel style={font=\footnotesize\color{white!15!black}},
xticklabel style={font=\footnotesize\color{white!15!black}},
yticklabel style={font=\footnotesize\color{white!15!black}}
] 

\addplot[color=orange_D,fill={white!50!orange_D}] coordinates {({$(4,5)$},0.51127) ({$(4,6)$},0.7288) ({$(4,7)$},0.86299)};
\addlegendentry{$L=1$}
\addplot[color=cyan,fill={white!50!cyan},dashed] coordinates {({$(4,5)$},0.695696) ({$(4,6)$},0.8749) ({$(4,7)$},0.95378)};
\addlegendentry{$L=2$}
\addplot[color=green_D,fill={white!50!green_D},dashdotted] coordinates {({$(4,5)$},0.72906) ({$(4,6)$},0.896138) ({$(4,7)$},0.964307)};
\addlegendentry{$L=3$}
\addplot[color=violet,fill={white!50!violet},dotted] coordinates {({$(4,5)$},0.736708) ({$(4,6)$},0.90077) ({$(4,7)$},0.96649)};
\addlegendentry{$L=\infty$}
 

\end{axis}
\end{tikzpicture}
        \fi
        \caption{Reliability for $\varepsilon=0.2$ and $\tau=2$.}
        \label{fig:rel_02}
    \end{subfigure}	
    \begin{subfigure}[b]{.49\linewidth}
	    \centering
        \ifdefined\pdffig
            \includegraphics[width=\linewidth]{tikz/paper-figure15}
        \else
            \begin{tikzpicture} 
\begin{axis}
[ybar, 
enlarge x limits=0.3, 
bar width=0.25cm,
axis background/.style={fill=white},
xmajorgrids,
ymajorgrids,
ymin=0, 
width=\sfwidth, height=\ssfheight, 
ymax=1, 
legend style={font=\tiny, at={(0.5,0.1)}, anchor=south,fill opacity=0.8, legend columns=4,draw opacity=1, text opacity=1, draw=white!80!black},
ylabel={$p_s$}, 
symbolic x coords={{$(4,5)$}, {$(4,6)$}, {$(4,7)$}}, 
xtick=data,  
ylabel near ticks,
ylabel style={font=\footnotesize\color{white!15!black}},
xticklabel style={font=\footnotesize\color{white!15!black}},
yticklabel style={font=\footnotesize\color{white!15!black}}
] 

\addplot[color=orange_D,fill={white!50!orange_D}] coordinates {({$(4,5)$},0.52654) ({$(4,6)$},0.74279) ({$(4,7)$},0.87290)};
\addlegendentry{$L=1$}
\addplot[color=cyan,fill={white!50!cyan},dashed] coordinates {({$(4,5)$},0.80244) ({$(4,6)$},0.93701) ({$(4,7)$},0.982057)};
\addlegendentry{$L=2$}
\addplot[color=green_D,fill={white!50!green_D},dashdotted] coordinates {({$(4,5)$},0.87782) ({$(4,6)$},0.97031) ({$(4,7)$},0.993595)};
\addlegendentry{$L=3$}
\addplot[color=violet,fill={white!50!violet},dotted] coordinates {({$(4,5)$},0.91462) ({$(4,6)$},0.98296) ({$(4,7)$},0.996990)};
\addlegendentry{$L=\infty$}
 

\end{axis}
\end{tikzpicture}
        \fi
        \caption{Reliability for $\varepsilon=0.1$ and $\tau=1.5$.}
        \label{fig:rel_01_hl}
    \end{subfigure}	
	\begin{subfigure}[b]{.49\linewidth}
	    \centering
        \ifdefined\pdffig
            \includegraphics[width=\linewidth]{tikz/paper-figure16}
        \else
            \begin{tikzpicture} 
\begin{axis}
[ybar, 
enlarge x limits=0.3, 
bar width=0.25cm,
axis background/.style={fill=white},
xmajorgrids,
ymajorgrids,
ymin=0, 
width=\sfwidth, height=\ssfheight, 
ymax=1, 
legend style={font=\tiny, at={(0.5,0.1)}, anchor=south,fill opacity=0.8, legend columns=4,draw opacity=1, text opacity=1, draw=white!80!black},
ylabel={$p_s$}, 
symbolic x coords={{$(4,5)$}, {$(4,6)$}, {$(4,7)$}}, 
xtick=data,  
ylabel near ticks,
ylabel style={font=\footnotesize\color{white!15!black}},
xticklabel style={font=\footnotesize\color{white!15!black}},
yticklabel style={font=\footnotesize\color{white!15!black}}
] 

\addplot[color=orange_D,fill={white!50!orange_D}] coordinates {({$(4,5)$},0.375078) ({$(4,6)$},0.588822) ({$(4,7)$},0.7506278)};
\addlegendentry{$L=1$}
\addplot[color=cyan,fill={white!50!cyan},dashed] coordinates {({$(4,5)$},0.612027) ({$(4,6)$},0.814644) ({$(4,7)$},0.919942)};
\addlegendentry{$L=2$}
\addplot[color=green_D,fill={white!50!green_D},dashdotted] coordinates {({$(4,5)$},0.68964) ({$(4,6)$},0.8708821) ({$(4,7)$},0.9517124)};
\addlegendentry{$L=3$}
\addplot[color=violet,fill={white!50!violet},dotted] coordinates {({$(4,5)$},0.736708) ({$(4,6)$},0.90077) ({$(4,7)$},0.96649)};
\addlegendentry{$L=\infty$}
 

\end{axis}
\end{tikzpicture}
        \fi
        \caption{Reliability for $\varepsilon=0.2$ and $\tau=1.5$.}
        \label{fig:rel_02_hl}
    \end{subfigure}	
     \caption{Reliability for different amounts of redundancy and queue length configurations with $K=4$.}\vspace{-0.6cm}
 \label{fig:lat_rel}
\end{figure}

This phenomenon is clearly visible from Fig.~\ref{fig:lat_rel}, which shows the block decoding success probability for each system. The upper part of the figure, which corresponds to the $\tau=2$ scenario, shows an almost imperceptible difference between $L=3$ and $L=\infty$, and a small one between $L=2$ and $L=3$. The difference between $L=1$ and $L=2$, on the other hand, is stark, as packets are often dropped by the former, leading to a higher vulnerability to errors. In the lower part of the figure, we can see that the difference between the systems is accentuated for $\tau=1.5$, although the main trends remain the same. 

\begin{figure}
    \centering
	\begin{subfigure}[b]{.49\linewidth}
	    \centering
        \ifdefined\pdffig
            \includegraphics[width=\linewidth]{tikz/paper-figure17}
        \else
            \input{fig/age_45_rho_0.5_e_0.1.tex}
        \fi
        \caption{\gls{paoi} \gls{cdf} for the $(4,5)$ system with $\varepsilon=0.1$.}
        \label{fig:age_45_01_rho05}
    \end{subfigure}	
	\begin{subfigure}[b]{.49\linewidth}
	    \centering
        \ifdefined\pdffig
            \includegraphics[width=\linewidth]{tikz/paper-figure18}
        \else
            \input{fig/age_45_rho_0.5_e_0.2.tex}
        \fi        
        \caption{\gls{paoi} \gls{cdf} for the $(4,5)$ system with $\varepsilon=0.2$.}
        \label{fig:age_45_02_rho05}
    \end{subfigure}	
	\begin{subfigure}[b]{.49\linewidth}
	    \centering
        \ifdefined\pdffig
            \includegraphics[width=\linewidth]{tikz/paper-figure19}
        \else
            \input{fig/age_46_rho_0.5_e_0.1.tex}
        \fi        
        \caption{\gls{paoi} \gls{cdf} for the $(4,6)$ system with $\varepsilon=0.1$.}
        \label{fig:age_46_01_rho05}
    \end{subfigure}	
    \begin{subfigure}[b]{.49\linewidth}
	    \centering
        \ifdefined\pdffig
            \includegraphics[width=\linewidth]{tikz/paper-figure20}
        \else
            \input{fig/age_46_rho_0.5_e_0.2.tex}
        \fi        
        \caption{\gls{paoi} \gls{cdf} for the $(4,6)$ system with $\varepsilon=0.2$.}
        \label{fig:age_46_02_rho05}
    \end{subfigure}
    \begin{subfigure}[b]{.49\linewidth}
	    \centering
        \ifdefined\pdffig
            \includegraphics[width=\linewidth]{tikz/paper-figure21}
        \else
            \input{fig/age_47_rho_0.5_e_0.1.tex}
        \fi        
        \caption{\gls{paoi} \gls{cdf} for the $(4,7)$ system with $\varepsilon=0.1$.}
        \label{fig:age_47_01_rho05}
    \end{subfigure}
    \begin{subfigure}[b]{.49\linewidth}
	    \centering
        \ifdefined\pdffig
            \includegraphics[width=\linewidth]{tikz/paper-figure22}
        \else
            \input{fig/age_47_rho_0.5_e_0.2.tex}
        \fi        
        \caption{\gls{paoi} \gls{cdf} for the $(4,7)$ system with $\varepsilon=0.2$.}
        \label{fig:age_47_02_rho05}
    \end{subfigure}
     \caption{\gls{paoi} \gls{cdf} for different queue sizes and codes with $\mu=1$ and $\tau=2$.}\vspace{-0.6cm}
 \label{fig:age_CDF}
\end{figure}
In general, systems with lower redundancy or a higher erasure probability suffer more from the occasional dropped packet: having enough redundant paths, or a lower erasure probability on the paths, allows the use of a shorter queue, taking advantage of the redundancy to drop stragglers and limit latency. This problem becomes even more significant if the load on the system increases, as even a small delay on one packet can cause potentially long queues, leading to more frequent dropping and more lost blocks.

We can now examine the \gls{paoi} in the same conditions, in the case with $\tau=2$. The \gls{cdf} of the peak age is shown in Fig.~\ref{fig:age_CDF}, and in this case, the loss of some blocks can be compensated by the lower overall latency. In fact, the $L=1$ system becomes more effective than ones with longer queues for the $(4,7)$ system, as the few lost blocks are compensated for by the significantly lower latency. It is interesting to note that the \gls{cdf} for the $L=1$ system is not smooth: this indicates that the \gls{pdf} of the \gls{paoi} has jump discontinuities in those points, which correspond to all multiples of $\tau$. This corresponds to the discontinuities in~\eqref{eq:L1agepdf}.

\begin{figure}
    \centering
	\begin{subfigure}[b]{.49\linewidth}
	    \centering
        \ifdefined\pdffig
            \includegraphics[width=\linewidth]{tikz/paper-figure23}
        \else
            \input{fig/age_45_rho_0.67_e_0.1.tex}
        \fi        
        \caption{\gls{paoi} \gls{cdf} for the $(4,5)$ system with $\varepsilon=0.1$.}
        \label{fig:age_45_01_rho06}
    \end{subfigure}	
	\begin{subfigure}[b]{.49\linewidth}
	    \centering
        \ifdefined\pdffig
            \includegraphics[width=\linewidth]{tikz/paper-figure24}
        \else
            \input{fig/age_45_rho_0.67_e_0.2.tex}
        \fi        
        \caption{\gls{paoi} \gls{cdf} for the $(4,5)$ system with $\varepsilon=0.2$.}
        \label{fig:age_45_02_rho06}
    \end{subfigure}	
	\begin{subfigure}[b]{.49\linewidth}
	    \centering
        \ifdefined\pdffig
            \includegraphics[width=\linewidth]{tikz/paper-figure25}
        \else
            \input{fig/age_46_rho_0.67_e_0.1.tex}
        \fi        
        \caption{\gls{paoi} \gls{cdf} for the $(4,6)$ system with $\varepsilon=0.1$.}
        \label{fig:age_46_01_rho06}
    \end{subfigure}	
    \begin{subfigure}[b]{.49\linewidth}
	    \centering
        \ifdefined\pdffig
            \includegraphics[width=\linewidth]{tikz/paper-figure26}
        \else
            \input{fig/age_46_rho_0.67_e_0.2.tex}
        \fi        
        \caption{\gls{paoi} \gls{cdf} for the $(4,6)$ system with $\varepsilon=0.2$.}
        \label{fig:age_46_02_rho06}
    \end{subfigure}
    \begin{subfigure}[b]{.49\linewidth}
	    \centering
        \ifdefined\pdffig
            \includegraphics[width=\linewidth]{tikz/paper-figure27}
        \else
            \input{fig/age_47_rho_0.67_e_0.1.tex}
        \fi        
        \caption{\gls{paoi} \gls{cdf} for the $(4,7)$ system with $\varepsilon=0.1$.}
        \label{fig:age_47_01_rho06}
    \end{subfigure}
    \begin{subfigure}[b]{.49\linewidth}
	    \centering
        \ifdefined\pdffig
            \includegraphics[width=\linewidth]{tikz/paper-figure28}
        \else
            \input{fig/age_47_rho_0.67_e_0.2.tex}
        \fi        
        \caption{\gls{paoi} \gls{cdf} for the $(4,7)$ system with $\varepsilon=0.2$.}
        \label{fig:age_47_02_rho06}
    \end{subfigure}
     \caption{\gls{paoi} \gls{cdf} for different queue sizes and codes with $\mu=1$ and $\tau=1.5$.}\vspace{-0.6cm}
 \label{fig:age_CDF_hl}
\end{figure}

\begin{figure}
    \centering
	\begin{subfigure}[b]{.49\linewidth}
	    \centering
        \ifdefined\pdffig
            \includegraphics[width=\linewidth]{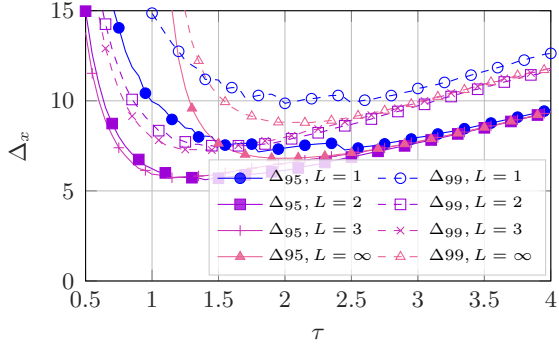}
        \else
            \begin{tikzpicture}
\begin{axis}[%
width=\sfwidth,
height=\sfheight,
xmin=0.5,
xmax=4,
ymin=0,
ymax=15,
axis background/.style={fill=white},
xlabel style={font=\footnotesize\color{white!15!black}},
xlabel={$\tau$},
ylabel near ticks,
ylabel style={font=\footnotesize\color{white!15!black}},
xticklabel style={font=\footnotesize\color{white!15!black}},
yticklabel style={font=\footnotesize\color{white!15!black}},
ylabel={$\Delta_x$},
axis background/.style={fill=white},
xmajorgrids,
ymajorgrids,
legend style={font=\tiny, at={(0.99,0.02)}, anchor=south east, legend columns=2,legend cell align=left, align=left,fill opacity=0.8, draw opacity=1, text opacity=1, draw=white!80!black}
]
\addplot [color=orange_D, mark=*,,mark options={solid}, mark repeat = 4]
  table[row sep=crcr]{%
0.5	16\\
0.55	16\\
0.6	16\\
0.65	16\\
0.7	15.349\\
0.75	14.05\\
0.8	12.751\\
0.85	11.868\\
0.9	11.381\\
0.95	10.422\\
1	9.956\\
1.05	9.77\\
1.1	9.433\\
1.15	8.966\\
1.2	8.86\\
1.25	8.494\\
1.3	8.443\\
1.35	7.995\\
1.4	8.041\\
1.45	8.01\\
1.5	7.496\\
1.55	7.543\\
1.6	7.59\\
1.65	7.632\\
1.7	7.639\\
1.75	7.596\\
1.8	7.166\\
1.85	7.217\\
1.9	7.276\\
1.95	7.318\\
2	7.381\\
2.05	7.432\\
2.1	7.478\\
2.15	7.532\\
2.2	7.576\\
2.25	7.606\\
2.3	7.63\\
2.35	7.639\\
2.4	7.546\\
2.45	7.271\\
2.5	7.319\\
2.55	7.375\\
2.6	7.452\\
2.65	7.49\\
2.7	7.565\\
2.75	7.607\\
2.8	7.687\\
2.85	7.744\\
2.9	7.811\\
2.95	7.878\\
3	7.967\\
3.05	8.019\\
3.1	8.082\\
3.15	8.156\\
3.2	8.234\\
3.25	8.295\\
3.3	8.389\\
3.35	8.451\\
3.4	8.524\\
3.45	8.608\\
3.5	8.695\\
3.55	8.765\\
3.6	8.844\\
3.65	8.927\\
3.7	9.011\\
3.75	9.091\\
3.8	9.175\\
3.85	9.257\\
3.9	9.349\\
3.95	9.426\\
4	9.509\\
};
\addlegendentry{$\Delta_{95}, L=1$}

\addplot [color=orange_D,dashed,mark=o,,mark options={solid}, mark repeat=4]
  table[row sep=crcr]{%
0.5	16\\
0.55	16\\
0.6	16\\
0.65	16\\
0.7	16\\
0.75	16\\
0.8	16\\
0.85	16\\
0.9	16\\
0.95	15.789\\
1	14.864\\
1.05	14.29\\
1.1	13.729\\
1.15	13.189\\
1.2	12.746\\
1.25	12.237\\
1.3	11.977\\
1.35	11.74\\
1.4	11.189\\
1.45	11.193\\
1.5	11.15\\
1.55	10.671\\
1.6	10.731\\
1.65	10.724\\
1.7	10.174\\
1.75	10.221\\
1.8	10.294\\
1.85	10.348\\
1.9	10.365\\
1.95	10.182\\
2	9.86\\
2.05	9.942\\
2.1	9.996\\
2.15	10.078\\
2.2	10.116\\
2.25	10.16\\
2.3	10.248\\
2.35	10.294\\
2.4	10.295\\
2.45	10.162\\
2.5	9.885\\
2.55	9.971\\
2.6	10.032\\
2.65	10.067\\
2.7	10.165\\
2.75	10.246\\
2.8	10.29\\
2.85	10.363\\
2.9	10.463\\
2.95	10.536\\
3	10.684\\
3.05	10.754\\
3.1	10.827\\
3.15	10.915\\
3.2	11.019\\
3.25	11.104\\
3.3	11.231\\
3.35	11.308\\
3.4	11.424\\
3.45	11.526\\
3.5	11.632\\
3.55	11.73\\
3.6	11.84\\
3.65	11.94\\
3.7	12.043\\
3.75	12.175\\
3.8	12.256\\
3.85	12.38\\
3.9	12.464\\
3.95	12.578\\
4	12.631\\
};
\addlegendentry{$\Delta_{99}, L=1$}

\addplot [color=cyan,mark=square*,mark options={solid}, mark repeat=4,mark phase=1]
  table[row sep=crcr]{%
0.5	14.973\\
0.55	12.59\\
0.6	10.935\\
0.65	9.619\\
0.7	8.737\\
0.75	8.016\\
0.8	7.468\\
0.85	7.052\\
0.9	6.743\\
0.95	6.412\\
1	6.23\\
1.05	6.053\\
1.1	6\\
1.15	5.744\\
1.2	5.759\\
1.25	5.747\\
1.3	5.732\\
1.35	5.711\\
1.4	5.619\\
1.45	5.674\\
1.5	5.701\\
1.55	5.746\\
1.6	5.795\\
1.65	5.834\\
1.7	5.883\\
1.75	5.922\\
1.8	5.956\\
1.85	6.002\\
1.9	6.054\\
1.95	6.103\\
2	6.147\\
2.05	6.2\\
2.1	6.27\\
2.15	6.351\\
2.2	6.412\\
2.25	6.493\\
2.3	6.577\\
2.35	6.657\\
2.4	6.728\\
2.45	6.806\\
2.5	6.877\\
2.55	6.956\\
2.6	7.034\\
2.65	7.119\\
2.7	7.196\\
2.75	7.266\\
2.8	7.351\\
2.85	7.425\\
2.9	7.499\\
2.95	7.581\\
3	7.668\\
3.05	7.742\\
3.1	7.826\\
3.15	7.907\\
3.2	7.987\\
3.25	8.075\\
3.3	8.163\\
3.35	8.247\\
3.4	8.328\\
3.45	8.412\\
3.5	8.501\\
3.55	8.584\\
3.6	8.676\\
3.65	8.757\\
3.7	8.853\\
3.75	8.937\\
3.8	9.027\\
3.85	9.107\\
3.9	9.211\\
3.95	9.298\\
4	9.394\\
};
\addlegendentry{$\Delta_{95}, L=2$}

\addplot [color=cyan,dashed,mark=square,,mark options={solid}, mark repeat=4,mark phase=1]
  table[row sep=crcr]{%
0.5	16\\
0.55	16\\
0.6	16\\
0.65	14.253\\
0.7	12.854\\
0.75	11.782\\
0.8	10.816\\
0.85	10.072\\
0.9	9.573\\
0.95	9.103\\
1	8.718\\
1.05	8.343\\
1.1	8.247\\
1.15	7.922\\
1.2	7.86\\
1.25	7.719\\
1.3	7.585\\
1.35	7.588\\
1.4	7.502\\
1.45	7.508\\
1.5	7.406\\
1.55	7.434\\
1.6	7.502\\
1.65	7.505\\
1.7	7.513\\
1.75	7.562\\
1.8	7.558\\
1.85	7.605\\
1.9	7.634\\
1.95	7.72\\
2	7.802\\
2.05	7.892\\
2.1	7.976\\
2.15	8.06\\
2.2	8.145\\
2.25	8.226\\
2.3	8.315\\
2.35	8.434\\
2.4	8.523\\
2.45	8.613\\
2.5	8.691\\
2.55	8.788\\
2.6	8.901\\
2.65	8.988\\
2.7	9.108\\
2.75	9.225\\
2.8	9.313\\
2.85	9.407\\
2.9	9.516\\
2.95	9.618\\
3	9.723\\
3.05	9.805\\
3.1	9.925\\
3.15	10.049\\
3.2	10.168\\
3.25	10.238\\
3.3	10.356\\
3.35	10.466\\
3.4	10.582\\
3.45	10.642\\
3.5	10.797\\
3.55	10.831\\
3.6	10.95\\
3.65	11.043\\
3.7	11.132\\
3.75	11.205\\
3.8	11.373\\
3.85	11.438\\
3.9	11.468\\
3.95	11.633\\
4	11.712\\
};
\addlegendentry{$\Delta_{99}, L=2$}

\addplot [color=green_D,mark=+,mark options={solid}, mark repeat=4,mark phase=2]
  table[row sep=crcr]{%
0.5	13.777\\
0.55	11.525\\
0.6	10.003\\
0.65	8.881\\
0.7	8.08\\
0.75	7.395\\
0.8	6.948\\
0.85	6.562\\
0.9	6.244\\
0.95	6.137\\
1	5.925\\
1.05	5.887\\
1.1	5.827\\
1.15	5.733\\
1.2	5.757\\
1.25	5.79\\
1.3	5.805\\
1.35	5.823\\
1.4	5.847\\
1.45	5.879\\
1.5	5.924\\
1.55	5.998\\
1.6	6.072\\
1.65	6.134\\
1.7	6.189\\
1.75	6.259\\
1.8	6.306\\
1.85	6.374\\
1.9	6.421\\
1.95	6.486\\
2	6.535\\
2.05	6.59\\
2.1	6.638\\
2.15	6.701\\
2.2	6.761\\
2.25	6.806\\
2.3	6.85\\
2.35	6.911\\
2.4	6.961\\
2.45	7.014\\
2.5	7.069\\
2.55	7.135\\
2.6	7.191\\
2.65	7.256\\
2.7	7.314\\
2.75	7.383\\
2.8	7.455\\
2.85	7.524\\
2.9	7.595\\
2.95	7.665\\
3	7.728\\
3.05	7.802\\
3.1	7.879\\
3.15	7.966\\
3.2	8.041\\
3.25	8.113\\
3.3	8.193\\
3.35	8.267\\
3.4	8.363\\
3.45	8.449\\
3.5	8.524\\
3.55	8.607\\
3.6	8.685\\
3.65	8.782\\
3.7	8.866\\
3.75	8.946\\
3.8	9.039\\
3.85	9.127\\
3.9	9.215\\
3.95	9.3\\
4	9.391\\
};
\addlegendentry{$\Delta_{95}, L=3$}

\addplot [color=green_D,mark=x,dashed,mark options={solid}, mark repeat=4,mark phase=2]  
table[row sep=crcr]{%
0.5	16\\
0.55	16\\
0.6	14.72\\
0.65	12.906\\
0.7	11.673\\
0.75	10.617\\
0.8	9.799\\
0.85	9.15\\
0.9	8.664\\
0.95	8.301\\
1	7.947\\
1.05	7.813\\
1.1	7.556\\
1.15	7.479\\
1.2	7.348\\
1.25	7.3\\
1.3	7.292\\
1.35	7.288\\
1.4	7.268\\
1.45	7.266\\
1.5	7.32\\
1.55	7.39\\
1.6	7.47\\
1.65	7.535\\
1.7	7.57\\
1.75	7.649\\
1.8	7.702\\
1.85	7.777\\
1.9	7.807\\
1.95	7.88\\
2	7.972\\
2.05	8.067\\
2.1	8.159\\
2.15	8.293\\
2.2	8.376\\
2.25	8.435\\
2.3	8.539\\
2.35	8.636\\
2.4	8.736\\
2.45	8.83\\
2.5	8.92\\
2.55	9.007\\
2.6	9.098\\
2.65	9.205\\
2.7	9.294\\
2.75	9.383\\
2.8	9.475\\
2.85	9.587\\
2.9	9.684\\
2.95	9.798\\
3	9.872\\
3.05	9.986\\
3.1	10.086\\
3.15	10.196\\
3.2	10.286\\
3.25	10.426\\
3.3	10.501\\
3.35	10.562\\
3.4	10.73\\
3.45	10.799\\
3.5	10.931\\
3.55	11.024\\
3.6	11.061\\
3.65	11.245\\
3.7	11.315\\
3.75	11.319\\
3.8	11.499\\
3.85	11.545\\
3.9	11.613\\
3.95	11.738\\
4	11.801\\
};
\addlegendentry{$\Delta_{99}, L=3$}

\addplot [color=violet,mark=triangle*,mark options={solid}, mark repeat=4,mark phase=3]
  table[row sep=crcr]{%
0.5	16\\
0.55	16\\
0.6	16\\
0.65	16\\
0.7	16\\
0.75	16\\
0.8	16\\
0.85	16\\
0.9	16\\
0.95	16\\
1	16\\
1.05	16\\
1.1	16\\
1.15	15.045\\
1.2	12.318\\
1.25	10.72\\
1.3	9.58\\
1.35	8.89\\
1.4	8.32\\
1.45	7.938\\
1.5	7.632\\
1.55	7.421\\
1.6	7.24\\
1.65	7.116\\
1.7	7.017\\
1.75	6.942\\
1.8	6.88\\
1.85	6.85\\
1.9	6.818\\
1.95	6.811\\
2	6.814\\
2.05	6.813\\
2.1	6.843\\
2.15	6.845\\
2.2	6.87\\
2.25	6.909\\
2.3	6.929\\
2.35	6.968\\
2.4	7.022\\
2.45	7.053\\
2.5	7.107\\
2.55	7.167\\
2.6	7.218\\
2.65	7.271\\
2.7	7.341\\
2.75	7.399\\
2.8	7.469\\
2.85	7.527\\
2.9	7.601\\
2.95	7.667\\
3	7.74\\
3.05	7.815\\
3.1	7.884\\
3.15	7.967\\
3.2	8.044\\
3.25	8.116\\
3.3	8.192\\
3.35	8.273\\
3.4	8.362\\
3.45	8.44\\
3.5	8.518\\
3.55	8.611\\
3.6	8.691\\
3.65	8.784\\
3.7	8.861\\
3.75	8.952\\
3.8	9.039\\
3.85	9.133\\
3.9	9.219\\
3.95	9.301\\
4	9.397\\
};
\addlegendentry{$\Delta{95}, L = \infty$}

\addplot [color=violet,dashed,mark=triangle,mark options={solid}, mark repeat=4,mark phase=3]
  table[row sep=crcr]{%
0.5	16\\
0.55	16\\
0.6	16\\
0.65	16\\
0.7	16\\
0.75	16\\
0.8	16\\
0.85	16\\
0.9	16\\
0.95	16\\
1	16\\
1.05	16\\
1.1	16\\
1.15	16\\
1.2	16\\
1.25	14.935\\
1.3	13.264\\
1.35	12.218\\
1.4	11.335\\
1.45	10.724\\
1.5	10.278\\
1.55	9.936\\
1.6	9.644\\
1.65	9.43\\
1.7	9.256\\
1.75	9.141\\
1.8	9.011\\
1.85	8.921\\
1.9	8.862\\
1.95	8.811\\
2	8.813\\
2.05	8.797\\
2.1	8.804\\
2.15	8.789\\
2.2	8.794\\
2.25	8.853\\
2.3	8.872\\
2.35	8.904\\
2.4	8.97\\
2.45	9.029\\
2.5	9.074\\
2.55	9.176\\
2.6	9.223\\
2.65	9.303\\
2.7	9.387\\
2.75	9.457\\
2.8	9.563\\
2.85	9.657\\
2.9	9.752\\
2.95	9.825\\
3	9.946\\
3.05	10.039\\
3.1	10.123\\
3.15	10.229\\
3.2	10.339\\
3.25	10.435\\
3.3	10.509\\
3.35	10.611\\
3.4	10.711\\
3.45	10.803\\
3.5	10.928\\
3.55	11.037\\
3.6	11.077\\
3.65	11.215\\
3.7	11.299\\
3.75	11.417\\
3.8	11.462\\
3.85	11.592\\
3.9	11.633\\
3.95	11.714\\
4	11.795\\
};
\addlegendentry{$\Delta{99}, L = \infty$}

\end{axis}
\end{tikzpicture}%
        \fi        
        \caption{\gls{paoi} percentiles for the $(4,5)$ system with $\varepsilon=0.1$.}
        \label{fig:age_45_01_perc}
    \end{subfigure}	
	\begin{subfigure}[b]{.49\linewidth}
	    \centering
        \ifdefined\pdffig
            \includegraphics[width=\linewidth]{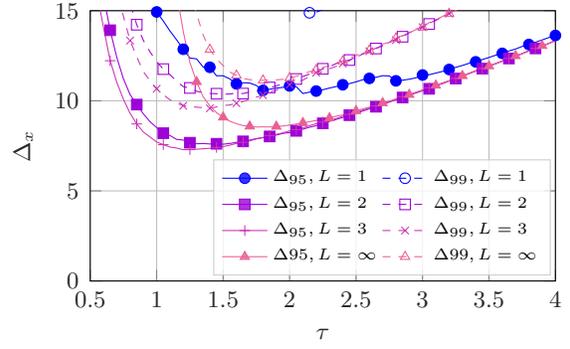}
        \else
            \begin{tikzpicture}
\begin{axis}[%
width=\sfwidth,
height=\sfheight,
xmin=0.5,
xmax=4,
ymin=0,
ymax=15,
axis background/.style={fill=white},
xlabel style={font=\footnotesize\color{white!15!black}},
xlabel={$\tau$},
ylabel near ticks,
ylabel style={font=\footnotesize\color{white!15!black}},
xticklabel style={font=\footnotesize\color{white!15!black}},
yticklabel style={font=\footnotesize\color{white!15!black}},
ylabel={$\Delta_x$},
axis background/.style={fill=white},
xmajorgrids,
ymajorgrids,
legend style={font=\tiny, at={(0.99,0.02)}, anchor=south east, legend columns=2,legend cell align=left, align=left,fill opacity=0.8, draw opacity=1, text opacity=1, draw=white!80!black}
]
\addplot [color=orange_D, mark=*,mark options={solid}, mark repeat = 4]
  table[row sep=crcr]{%
0.5	16\\
0.55	16\\
0.6	16\\
0.65	16\\
0.7	16\\
0.75	16\\
0.8	16\\
0.85	16\\
0.9	16\\
0.95	15.876\\
1	14.931\\
1.05	14.427\\
1.1	13.874\\
1.15	13.383\\
1.2	12.861\\
1.25	12.371\\
1.3	12.327\\
1.35	11.866\\
1.4	11.86\\
1.45	11.39\\
1.5	11.39\\
1.55	11.214\\
1.6	10.94\\
1.65	11.001\\
1.7	10.986\\
1.75	10.85\\
1.8	10.579\\
1.85	10.639\\
1.9	10.704\\
1.95	10.746\\
2	10.825\\
2.05	10.786\\
2.1	10.403\\
2.15	10.481\\
2.2	10.54\\
2.25	10.614\\
2.3	10.706\\
2.35	10.791\\
2.4	10.89\\
2.45	10.954\\
2.5	11.051\\
2.55	11.168\\
2.6	11.244\\
2.65	11.304\\
2.7	11.393\\
2.75	11.379\\
2.8	11.108\\
2.85	11.141\\
2.9	11.199\\
2.95	11.273\\
3	11.431\\
3.05	11.448\\
3.1	11.537\\
3.15	11.631\\
3.2	11.738\\
3.25	11.858\\
3.3	11.961\\
3.35	12.081\\
3.4	12.162\\
3.45	12.29\\
3.5	12.405\\
3.55	12.53\\
3.6	12.632\\
3.65	12.765\\
3.7	12.873\\
3.75	13.019\\
3.8	13.125\\
3.85	13.252\\
3.9	13.378\\
3.95	13.508\\
4	13.627\\
};
\addlegendentry{$\Delta_{95}, L=1$}

\addplot [color=orange_D,dashed,mark=o,,mark options={solid}, mark repeat=4]
  table[row sep=crcr]{%
0.5	16\\
0.55	16\\
0.6	16\\
0.65	16\\
0.7	16\\
0.75	16\\
0.8	16\\
0.85	16\\
0.9	16\\
0.95	16\\
1	16\\
1.05	16\\
1.1	16\\
1.15	16\\
1.2	16\\
1.25	16\\
1.3	16\\
1.35	16\\
1.4	16\\
1.45	16\\
1.5	16\\
1.55	16\\
1.6	15.784\\
1.65	15.865\\
1.7	15.594\\
1.75	15.399\\
1.8	15.442\\
1.85	15.51\\
1.9	15.05\\
1.95	15.09\\
2	15.244\\
2.05	15.266\\
2.1	15.293\\
2.15	14.884\\
2.2	14.918\\
2.25	15.069\\
2.3	15.164\\
2.35	15.289\\
2.4	15.388\\
2.45	15.415\\
2.5	15.434\\
2.55	15.208\\
2.6	15.205\\
2.65	15.315\\
2.7	15.466\\
2.75	15.558\\
2.8	15.698\\
2.85	15.791\\
2.9	15.966\\
2.95	16\\
3	16\\
3.05	16\\
3.1	16\\
3.15	16\\
3.2	16\\
3.25	16\\
3.3	16\\
3.35	16\\
3.4	16\\
3.45	16\\
3.5	16\\
3.55	16\\
3.6	16\\
3.65	16\\
3.7	16\\
3.75	16\\
3.8	16\\
3.85	16\\
3.9	16\\
3.95	16\\
4	16\\
};
\addlegendentry{$\Delta_{99}, L=1$}

\addplot [color=cyan,mark=square*,mark options={solid}, mark repeat=4,mark phase=1]
  table[row sep=crcr]{%
0.5	16\\
0.55	16\\
0.6	15.805\\
0.65	13.916\\
0.7	12.41\\
0.75	11.218\\
0.8	10.345\\
0.85	9.79\\
0.9	9.276\\
0.95	8.838\\
1	8.515\\
1.05	8.202\\
1.1	8.075\\
1.15	7.858\\
1.2	7.822\\
1.25	7.673\\
1.3	7.615\\
1.35	7.628\\
1.4	7.605\\
1.45	7.616\\
1.5	7.591\\
1.55	7.631\\
1.6	7.694\\
1.65	7.748\\
1.7	7.816\\
1.75	7.906\\
1.8	7.953\\
1.85	8.021\\
1.9	8.098\\
1.95	8.193\\
2	8.235\\
2.05	8.331\\
2.1	8.426\\
2.15	8.532\\
2.2	8.621\\
2.25	8.748\\
2.3	8.859\\
2.35	8.965\\
2.4	9.076\\
2.45	9.203\\
2.5	9.304\\
2.55	9.431\\
2.6	9.554\\
2.65	9.67\\
2.7	9.799\\
2.75	9.931\\
2.8	10.05\\
2.85	10.181\\
2.9	10.303\\
2.95	10.433\\
3	10.558\\
3.05	10.664\\
3.1	10.838\\
3.15	10.952\\
3.2	11.096\\
3.25	11.235\\
3.3	11.375\\
3.35	11.51\\
3.4	11.645\\
3.45	11.785\\
3.5	11.935\\
3.55	12.053\\
3.6	12.211\\
3.65	12.349\\
3.7	12.471\\
3.75	12.614\\
3.8	12.767\\
3.85	12.907\\
3.9	13.049\\
3.95	13.178\\
4	13.341\\
};
\addlegendentry{$\Delta_{95}, L=2$}

\addplot [color=cyan,dashed,mark=square,,mark options={solid}, mark repeat=4,mark phase=1]
  table[row sep=crcr]{%
0.5	16\\
0.55	16\\
0.6	16\\
0.65	16\\
0.7	16\\
0.75	16\\
0.8	15.417\\
0.85	14.223\\
0.9	13.348\\
0.95	12.718\\
1	12.167\\
1.05	11.758\\
1.1	11.449\\
1.15	11.099\\
1.2	10.785\\
1.25	10.714\\
1.3	10.525\\
1.35	10.464\\
1.4	10.409\\
1.45	10.392\\
1.5	10.367\\
1.55	10.375\\
1.6	10.439\\
1.65	10.401\\
1.7	10.493\\
1.75	10.53\\
1.8	10.602\\
1.85	10.739\\
1.9	10.823\\
1.95	10.951\\
2	11.038\\
2.05	11.211\\
2.1	11.348\\
2.15	11.485\\
2.2	11.57\\
2.25	11.775\\
2.3	11.892\\
2.35	12.066\\
2.4	12.175\\
2.45	12.262\\
2.5	12.434\\
2.55	12.609\\
2.6	12.73\\
2.65	12.902\\
2.7	13.057\\
2.75	13.275\\
2.8	13.407\\
2.85	13.556\\
2.9	13.784\\
2.95	13.941\\
3	14.1\\
3.05	14.263\\
3.1	14.463\\
3.15	14.669\\
3.2	14.876\\
3.25	15.048\\
3.3	15.211\\
3.35	15.388\\
3.4	15.59\\
3.45	15.783\\
3.5	15.938\\
3.55	16\\
3.6	16\\
3.65	16\\
3.7	16\\
3.75	16\\
3.8	16\\
3.85	16\\
3.9	16\\
3.95	16\\
4	16\\
};
\addlegendentry{$\Delta_{99}, L=2$}

\addplot [color=green_D,mark=+,mark options={solid}, mark repeat=4,mark phase=2]
  table[row sep=crcr]{%
0.5	16\\
0.55	16\\
0.6	14.073\\
0.65	12.224\\
0.7	11.021\\
0.75	10.082\\
0.8	9.307\\
0.85	8.716\\
0.9	8.268\\
0.95	7.988\\
1	7.714\\
1.05	7.568\\
1.1	7.432\\
1.15	7.377\\
1.2	7.288\\
1.25	7.292\\
1.3	7.312\\
1.35	7.342\\
1.4	7.363\\
1.45	7.412\\
1.5	7.458\\
1.55	7.536\\
1.6	7.633\\
1.65	7.698\\
1.7	7.779\\
1.75	7.882\\
1.8	7.965\\
1.85	8.072\\
1.9	8.14\\
1.95	8.223\\
2	8.333\\
2.05	8.403\\
2.1	8.488\\
2.15	8.62\\
2.2	8.729\\
2.25	8.819\\
2.3	8.912\\
2.35	9.017\\
2.4	9.144\\
2.45	9.247\\
2.5	9.363\\
2.55	9.484\\
2.6	9.596\\
2.65	9.714\\
2.7	9.84\\
2.75	9.971\\
2.8	10.065\\
2.85	10.229\\
2.9	10.341\\
2.95	10.467\\
3	10.585\\
3.05	10.716\\
3.1	10.846\\
3.15	10.977\\
3.2	11.105\\
3.25	11.239\\
3.3	11.4\\
3.35	11.514\\
3.4	11.665\\
3.45	11.795\\
3.5	11.925\\
3.55	12.063\\
3.6	12.192\\
3.65	12.354\\
3.7	12.484\\
3.75	12.623\\
3.8	12.783\\
3.85	12.905\\
3.9	13.038\\
3.95	13.193\\
4	13.351\\
};
\addlegendentry{$\Delta_{95}, L=3$}

\addplot [color=green_D,mark=x,dashed,mark options={solid}, mark repeat=4,mark phase=2]
  table[row sep=crcr]{%
0.5	16\\
0.55	16\\
0.6	16\\
0.65	16\\
0.7	16\\
0.75	14.686\\
0.8	13.342\\
0.85	12.441\\
0.9	11.617\\
0.95	11.108\\
1	10.67\\
1.05	10.337\\
1.1	10.056\\
1.15	9.883\\
1.2	9.724\\
1.25	9.689\\
1.3	9.644\\
1.35	9.621\\
1.4	9.581\\
1.45	9.66\\
1.5	9.731\\
1.55	9.771\\
1.6	9.866\\
1.65	9.961\\
1.7	10.068\\
1.75	10.204\\
1.8	10.312\\
1.85	10.483\\
1.9	10.592\\
1.95	10.731\\
2	10.891\\
2.05	11.014\\
2.1	11.15\\
2.15	11.352\\
2.2	11.561\\
2.25	11.61\\
2.3	11.776\\
2.35	11.913\\
2.4	12.072\\
2.45	12.222\\
2.5	12.42\\
2.55	12.587\\
2.6	12.748\\
2.65	12.923\\
2.7	13.061\\
2.75	13.243\\
2.8	13.369\\
2.85	13.608\\
2.9	13.759\\
2.95	13.973\\
3	14.083\\
3.05	14.282\\
3.1	14.499\\
3.15	14.647\\
3.2	14.835\\
3.25	15.007\\
3.3	15.23\\
3.35	15.389\\
3.4	15.551\\
3.45	15.766\\
3.5	15.922\\
3.55	16\\
3.6	16\\
3.65	16\\
3.7	16\\
3.75	16\\
3.8	16\\
3.85	16\\
3.9	16\\
3.95	16\\
4	16\\
};
\addlegendentry{$\Delta_{99}, L=3$}

\addplot [color=violet,mark=triangle*,mark options={solid}, mark repeat=4,mark phase=3]
  table[row sep=crcr]{%
0.5	16\\
0.55	16\\
0.6	16\\
0.65	16\\
0.7	16\\
0.75	16\\
0.8	16\\
0.85	16\\
0.9	16\\
0.95	16\\
1	16\\
1.05	16\\
1.1	16\\
1.15	16\\
1.2	13.848\\
1.25	12.125\\
1.3	11.054\\
1.35	10.226\\
1.4	9.716\\
1.45	9.341\\
1.5	9.086\\
1.55	8.87\\
1.6	8.727\\
1.65	8.626\\
1.7	8.584\\
1.75	8.545\\
1.8	8.554\\
1.85	8.547\\
1.9	8.575\\
1.95	8.6\\
2	8.656\\
2.05	8.709\\
2.1	8.778\\
2.15	8.842\\
2.2	8.908\\
2.25	8.987\\
2.3	9.045\\
2.35	9.149\\
2.4	9.215\\
2.45	9.33\\
2.5	9.437\\
2.55	9.55\\
2.6	9.651\\
2.65	9.767\\
2.7	9.881\\
2.75	9.981\\
2.8	10.111\\
2.85	10.227\\
2.9	10.343\\
2.95	10.476\\
3	10.591\\
3.05	10.746\\
3.1	10.855\\
3.15	10.994\\
3.2	11.113\\
3.25	11.243\\
3.3	11.384\\
3.35	11.535\\
3.4	11.664\\
3.45	11.791\\
3.5	11.928\\
3.55	12.078\\
3.6	12.207\\
3.65	12.341\\
3.7	12.489\\
3.75	12.618\\
3.8	12.77\\
3.85	12.925\\
3.9	13.055\\
3.95	13.18\\
4	13.345\\
};
\addlegendentry{$\Delta{95}, L = \infty$}

\addplot [color=violet,dashed,mark=triangle,mark options={solid}, mark repeat=4,mark phase=3]
  table[row sep=crcr]{%
0.5	16\\
0.55	16\\
0.6	16\\
0.65	16\\
0.7	16\\
0.75	16\\
0.8	16\\
0.85	16\\
0.9	16\\
0.95	16\\
1	16\\
1.05	16\\
1.1	16\\
1.15	16\\
1.2	16\\
1.25	16\\
1.3	14.831\\
1.35	13.641\\
1.4	12.847\\
1.45	12.326\\
1.5	11.951\\
1.55	11.602\\
1.6	11.412\\
1.65	11.238\\
1.7	11.179\\
1.75	11.11\\
1.8	11.141\\
1.85	11.149\\
1.9	11.161\\
1.95	11.193\\
2	11.267\\
2.05	11.364\\
2.1	11.482\\
2.15	11.617\\
2.2	11.703\\
2.25	11.85\\
2.3	11.949\\
2.35	12.102\\
2.4	12.197\\
2.45	12.384\\
2.5	12.517\\
2.55	12.675\\
2.6	12.804\\
2.65	12.986\\
2.7	13.148\\
2.75	13.246\\
2.8	13.468\\
2.85	13.621\\
2.9	13.765\\
2.95	13.951\\
3	14.074\\
3.05	14.342\\
3.1	14.482\\
3.15	14.689\\
3.2	14.839\\
3.25	15.022\\
3.3	15.193\\
3.35	15.437\\
3.4	15.579\\
3.45	15.768\\
3.5	15.957\\
3.55	16\\
3.6	16\\
3.65	16\\
3.7	16\\
3.75	16\\
3.8	16\\
3.85	16\\
3.9	16\\
3.95	16\\
4	16\\
};
\addlegendentry{$\Delta{99}, L = \infty$}

\end{axis}
\end{tikzpicture}%
        \fi        
        \caption{\gls{paoi} percentiles for the $(4,5)$ system with $\varepsilon=0.2$.}
        \label{fig:age_45_02_perc}
    \end{subfigure}	
	\begin{subfigure}[b]{.49\linewidth}
	    \centering
        \ifdefined\pdffig
            \includegraphics[width=\linewidth]{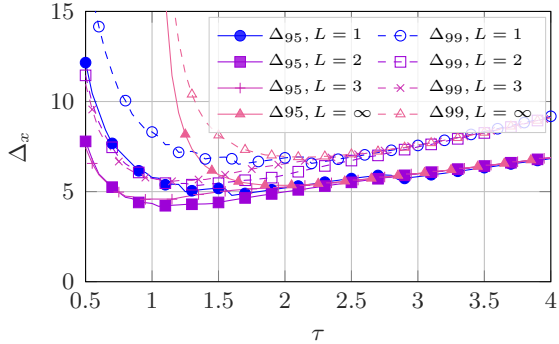}
        \else
            \begin{tikzpicture}
\begin{axis}[%
width=\sfwidth,
height=\sfheight,
xmin=0.5,
xmax=4,
ymin=0,
ymax=15,
axis background/.style={fill=white},
xlabel style={font=\footnotesize\color{white!15!black}},
xlabel={$\tau$},
ylabel near ticks,
ylabel style={font=\footnotesize\color{white!15!black}},
xticklabel style={font=\footnotesize\color{white!15!black}},
yticklabel style={font=\footnotesize\color{white!15!black}},
ylabel={$\Delta_x$},
axis background/.style={fill=white},
xmajorgrids,
ymajorgrids,
legend style={font=\tiny, at={(0.99,0.98)}, anchor=north east, legend columns=2,legend cell align=left, align=left,fill opacity=0.8, draw opacity=1, text opacity=1, draw=white!80!black}
]
\addplot [color=orange_D, mark=*,mark options={solid}, mark repeat = 4]
  table[row sep=crcr]{%
0.5	12.159\\
0.55	10.448\\
0.6	9.434\\
0.65	8.423\\
0.7	7.676\\
0.75	7.279\\
0.8	6.901\\
0.85	6.532\\
0.9	6.155\\
0.95	6.063\\
1	5.784\\
1.05	5.743\\
1.1	5.381\\
1.15	5.411\\
1.2	5.401\\
1.25	4.998\\
1.3	5.048\\
1.35	5.093\\
1.4	5.133\\
1.45	5.164\\
1.5	5.176\\
1.55	5.157\\
1.6	4.79\\
1.65	4.84\\
1.7	4.898\\
1.75	4.945\\
1.8	4.995\\
1.85	5.048\\
1.9	5.102\\
1.95	5.152\\
2	5.207\\
2.05	5.257\\
2.1	5.308\\
2.15	5.371\\
2.2	5.415\\
2.25	5.472\\
2.3	5.524\\
2.35	5.58\\
2.4	5.628\\
2.45	5.683\\
2.5	5.723\\
2.55	5.77\\
2.6	5.815\\
2.65	5.855\\
2.7	5.896\\
2.75	5.923\\
2.8	5.906\\
2.85	5.674\\
2.9	5.733\\
2.95	5.787\\
3	5.828\\
3.05	5.875\\
3.1	5.936\\
3.15	5.974\\
3.2	6.02\\
3.25	6.078\\
3.3	6.129\\
3.35	6.175\\
3.4	6.23\\
3.45	6.274\\
3.5	6.329\\
3.55	6.376\\
3.6	6.431\\
3.65	6.469\\
3.7	6.535\\
3.75	6.583\\
3.8	6.63\\
3.85	6.677\\
3.9	6.729\\
3.95	6.775\\
4	6.834\\
};
\addlegendentry{$\Delta_{95}, L=1$}

\addplot [color=orange_D,dashed,mark=o,,mark options={solid}, mark repeat=4]
  table[row sep=crcr]{%
0.5	16\\
0.55	15.796\\
0.6	14.155\\
0.65	12.718\\
0.7	11.601\\
0.75	10.621\\
0.8	10.048\\
0.85	9.306\\
0.9	8.863\\
0.95	8.455\\
1	8.308\\
1.05	8.051\\
1.1	7.605\\
1.15	7.613\\
1.2	7.178\\
1.25	7.214\\
1.3	7.233\\
1.35	7.101\\
1.4	6.814\\
1.45	6.872\\
1.5	6.904\\
1.55	6.916\\
1.6	6.823\\
1.65	6.519\\
1.7	6.57\\
1.75	6.613\\
1.8	6.672\\
1.85	6.716\\
1.9	6.763\\
1.95	6.819\\
2	6.874\\
2.05	6.906\\
2.1	6.924\\
2.15	6.912\\
2.2	6.568\\
2.25	6.622\\
2.3	6.674\\
2.35	6.747\\
2.4	6.797\\
2.45	6.863\\
2.5	6.909\\
2.55	6.973\\
2.6	7.04\\
2.65	7.101\\
2.7	7.167\\
2.75	7.241\\
2.8	7.293\\
2.85	7.358\\
2.9	7.441\\
2.95	7.507\\
3	7.562\\
3.05	7.646\\
3.1	7.72\\
3.15	7.791\\
3.2	7.847\\
3.25	7.938\\
3.3	8.009\\
3.35	8.095\\
3.4	8.173\\
3.45	8.257\\
3.5	8.324\\
3.55	8.412\\
3.6	8.49\\
3.65	8.574\\
3.7	8.664\\
3.75	8.742\\
3.8	8.832\\
3.85	8.909\\
3.9	8.991\\
3.95	9.079\\
4	9.178\\
};
\addlegendentry{$\Delta_{99}, L=1$}

\addplot [color=cyan,mark=square*,mark options={solid}, mark repeat=4,mark phase=1]
  table[row sep=crcr]{%
0.5	7.787\\
0.55	6.796\\
0.6	5.981\\
0.65	5.617\\
0.7	5.247\\
0.75	4.96\\
0.8	4.684\\
0.85	4.611\\
0.9	4.399\\
0.95	4.386\\
1	4.342\\
1.05	4.183\\
1.1	4.216\\
1.15	4.242\\
1.2	4.263\\
1.25	4.289\\
1.3	4.298\\
1.35	4.309\\
1.4	4.316\\
1.45	4.335\\
1.5	4.399\\
1.55	4.465\\
1.6	4.528\\
1.65	4.588\\
1.7	4.649\\
1.75	4.703\\
1.8	4.762\\
1.85	4.818\\
1.9	4.872\\
1.95	4.933\\
2	4.988\\
2.05	5.039\\
2.1	5.099\\
2.15	5.143\\
2.2	5.203\\
2.25	5.253\\
2.3	5.309\\
2.35	5.365\\
2.4	5.416\\
2.45	5.461\\
2.5	5.515\\
2.55	5.567\\
2.6	5.62\\
2.65	5.663\\
2.7	5.714\\
2.75	5.771\\
2.8	5.799\\
2.85	5.85\\
2.9	5.904\\
2.95	5.949\\
3	5.984\\
3.05	6.034\\
3.1	6.079\\
3.15	6.116\\
3.2	6.16\\
3.25	6.202\\
3.3	6.248\\
3.35	6.288\\
3.4	6.326\\
3.45	6.379\\
3.5	6.42\\
3.55	6.481\\
3.6	6.51\\
3.65	6.559\\
3.7	6.602\\
3.75	6.655\\
3.8	6.691\\
3.85	6.745\\
3.9	6.79\\
3.95	6.84\\
4	6.884\\
};
\addlegendentry{$\Delta_{95}, L=2$}

\addplot [color=cyan,dashed,mark=square,,mark options={solid}, mark repeat=4,mark phase=1]
  table[row sep=crcr]{%
0.5	11.452\\
0.55	9.894\\
0.6	8.843\\
0.65	8.074\\
0.7	7.459\\
0.75	6.909\\
0.8	6.546\\
0.85	6.312\\
0.9	6.064\\
0.95	5.869\\
1	5.738\\
1.05	5.668\\
1.1	5.469\\
1.15	5.469\\
1.2	5.452\\
1.25	5.452\\
1.3	5.385\\
1.35	5.346\\
1.4	5.391\\
1.45	5.432\\
1.5	5.467\\
1.55	5.506\\
1.6	5.547\\
1.65	5.578\\
1.7	5.631\\
1.75	5.646\\
1.8	5.697\\
1.85	5.734\\
1.9	5.775\\
1.95	5.847\\
2	5.929\\
2.05	6.013\\
2.1	6.104\\
2.15	6.18\\
2.2	6.263\\
2.25	6.339\\
2.3	6.426\\
2.35	6.506\\
2.4	6.578\\
2.45	6.657\\
2.5	6.724\\
2.55	6.812\\
2.6	6.876\\
2.65	6.952\\
2.7	7.01\\
2.75	7.1\\
2.8	7.167\\
2.85	7.251\\
2.9	7.33\\
2.95	7.399\\
3	7.48\\
3.05	7.564\\
3.1	7.621\\
3.15	7.701\\
3.2	7.778\\
3.25	7.865\\
3.3	7.929\\
3.35	8.014\\
3.4	8.102\\
3.45	8.191\\
3.5	8.256\\
3.55	8.346\\
3.6	8.431\\
3.65	8.506\\
3.7	8.585\\
3.75	8.693\\
3.8	8.76\\
3.85	8.853\\
3.9	8.946\\
3.95	9.025\\
4	9.104\\
};
\addlegendentry{$\Delta_{99}, L=2$}

\addplot [color=green_D,mark=+,mark options={solid}, mark repeat=4,mark phase=2]
  table[row sep=crcr]{%
0.5	7.444\\
0.55	6.563\\
0.6	5.932\\
0.65	5.567\\
0.7	5.236\\
0.75	4.973\\
0.8	4.747\\
0.85	4.718\\
0.9	4.598\\
0.95	4.573\\
1	4.587\\
1.05	4.585\\
1.1	4.584\\
1.15	4.579\\
1.2	4.653\\
1.25	4.723\\
1.3	4.778\\
1.35	4.84\\
1.4	4.884\\
1.45	4.935\\
1.5	4.98\\
1.55	5.02\\
1.6	5.068\\
1.65	5.097\\
1.7	5.135\\
1.75	5.154\\
1.8	5.175\\
1.85	5.191\\
1.9	5.211\\
1.95	5.229\\
2	5.253\\
2.05	5.293\\
2.1	5.318\\
2.15	5.343\\
2.2	5.37\\
2.25	5.41\\
2.3	5.447\\
2.35	5.486\\
2.4	5.522\\
2.45	5.56\\
2.5	5.598\\
2.55	5.643\\
2.6	5.684\\
2.65	5.727\\
2.7	5.77\\
2.75	5.808\\
2.8	5.846\\
2.85	5.892\\
2.9	5.929\\
2.95	5.977\\
3	6.004\\
3.05	6.041\\
3.1	6.093\\
3.15	6.123\\
3.2	6.162\\
3.25	6.205\\
3.3	6.257\\
3.35	6.294\\
3.4	6.344\\
3.45	6.388\\
3.5	6.427\\
3.55	6.469\\
3.6	6.517\\
3.65	6.564\\
3.7	6.618\\
3.75	6.656\\
3.8	6.707\\
3.85	6.749\\
3.9	6.799\\
3.95	6.847\\
4	6.888\\
};
\addlegendentry{$\Delta_{95}, L=3$}

\addplot [color=green_D,mark=x,dashed,mark options={solid}, mark repeat=4,mark phase=2]
  table[row sep=crcr]{%
0.5	10.957\\
0.55	9.558\\
0.6	8.408\\
0.65	7.722\\
0.7	7.197\\
0.75	6.669\\
0.8	6.354\\
0.85	6.135\\
0.9	5.977\\
0.95	5.779\\
1	5.738\\
1.05	5.666\\
1.1	5.582\\
1.15	5.572\\
1.2	5.613\\
1.25	5.638\\
1.3	5.623\\
1.35	5.672\\
1.4	5.672\\
1.45	5.736\\
1.5	5.837\\
1.55	5.933\\
1.6	6.022\\
1.65	6.1\\
1.7	6.183\\
1.75	6.252\\
1.8	6.313\\
1.85	6.384\\
1.9	6.437\\
1.95	6.494\\
2	6.543\\
2.05	6.618\\
2.1	6.659\\
2.15	6.713\\
2.2	6.757\\
2.25	6.804\\
2.3	6.858\\
2.35	6.896\\
2.4	6.937\\
2.45	6.969\\
2.5	7.01\\
2.55	7.073\\
2.6	7.112\\
2.65	7.17\\
2.7	7.217\\
2.75	7.279\\
2.8	7.334\\
2.85	7.406\\
2.9	7.456\\
2.95	7.525\\
3	7.568\\
3.05	7.652\\
3.1	7.707\\
3.15	7.778\\
3.2	7.853\\
3.25	7.924\\
3.3	7.989\\
3.35	8.064\\
3.4	8.136\\
3.45	8.219\\
3.5	8.298\\
3.55	8.373\\
3.6	8.457\\
3.65	8.533\\
3.7	8.608\\
3.75	8.702\\
3.8	8.786\\
3.85	8.863\\
3.9	8.951\\
3.95	9.034\\
4	9.12\\
};
\addlegendentry{$\Delta_{99}, L=3$}

\addplot [color=violet,mark=triangle*,mark options={solid}, mark repeat=4,mark phase=3]
  table[row sep=crcr]{%
0.5	16\\
0.55	16\\
0.6	16\\
0.65	16\\
0.7	16\\
0.75	16\\
0.8	16\\
0.85	16\\
0.9	16\\
0.95	16\\
1	16\\
1.05	16\\
1.1	15.704\\
1.15	11.442\\
1.2	9.373\\
1.25	8.176\\
1.3	7.369\\
1.35	6.862\\
1.4	6.418\\
1.45	6.171\\
1.5	5.938\\
1.55	5.793\\
1.6	5.671\\
1.65	5.558\\
1.7	5.488\\
1.75	5.436\\
1.8	5.395\\
1.85	5.362\\
1.9	5.349\\
1.95	5.352\\
2	5.348\\
2.05	5.359\\
2.1	5.376\\
2.15	5.387\\
2.2	5.406\\
2.25	5.438\\
2.3	5.472\\
2.35	5.493\\
2.4	5.536\\
2.45	5.573\\
2.5	5.621\\
2.55	5.652\\
2.6	5.686\\
2.65	5.73\\
2.7	5.768\\
2.75	5.815\\
2.8	5.858\\
2.85	5.892\\
2.9	5.922\\
2.95	5.97\\
3	6.01\\
3.05	6.038\\
3.1	6.087\\
3.15	6.123\\
3.2	6.169\\
3.25	6.208\\
3.3	6.252\\
3.35	6.296\\
3.4	6.348\\
3.45	6.389\\
3.5	6.421\\
3.55	6.481\\
3.6	6.519\\
3.65	6.559\\
3.7	6.614\\
3.75	6.649\\
3.8	6.696\\
3.85	6.749\\
3.9	6.794\\
3.95	6.832\\
4	6.887\\
};
\addlegendentry{$\Delta{95}, L = \infty$}

\addplot [color=violet,dashed,mark=triangle,mark options={solid}, mark repeat=4,mark phase=3]
  table[row sep=crcr]{%
0.5	16\\
0.55	16\\
0.6	16\\
0.65	16\\
0.7	16\\
0.75	16\\
0.8	16\\
0.85	16\\
0.9	16\\
0.95	16\\
1	16\\
1.05	16\\
1.1	16\\
1.15	16\\
1.2	13.356\\
1.25	11.577\\
1.3	10.397\\
1.35	9.543\\
1.4	8.905\\
1.45	8.501\\
1.5	8.128\\
1.55	7.9\\
1.6	7.69\\
1.65	7.451\\
1.7	7.323\\
1.75	7.225\\
1.8	7.135\\
1.85	7.038\\
1.9	7.015\\
1.95	6.981\\
2	6.95\\
2.05	6.928\\
2.1	6.922\\
2.15	6.912\\
2.2	6.921\\
2.25	6.929\\
2.3	6.957\\
2.35	6.97\\
2.4	7.004\\
2.45	7.041\\
2.5	7.078\\
2.55	7.111\\
2.6	7.142\\
2.65	7.183\\
2.7	7.23\\
2.75	7.278\\
2.8	7.364\\
2.85	7.411\\
2.9	7.447\\
2.95	7.516\\
3	7.598\\
3.05	7.642\\
3.1	7.7\\
3.15	7.765\\
3.2	7.835\\
3.25	7.915\\
3.3	7.983\\
3.35	8.073\\
3.4	8.149\\
3.45	8.214\\
3.5	8.28\\
3.55	8.368\\
3.6	8.449\\
3.65	8.543\\
3.7	8.621\\
3.75	8.68\\
3.8	8.78\\
3.85	8.858\\
3.9	8.94\\
3.95	9.023\\
4	9.111\\
};
\addlegendentry{$\Delta{99}, L = \infty$}

\end{axis}
\end{tikzpicture}%
        \fi        
        \caption{\gls{paoi} percentiles for the $(4,6)$ system with $\varepsilon=0.1$.}
        \label{fig:age_46_01_perc}
    \end{subfigure}	
    \begin{subfigure}[b]{.49\linewidth}
	    \centering
        \ifdefined\pdffig
            \includegraphics[width=\linewidth]{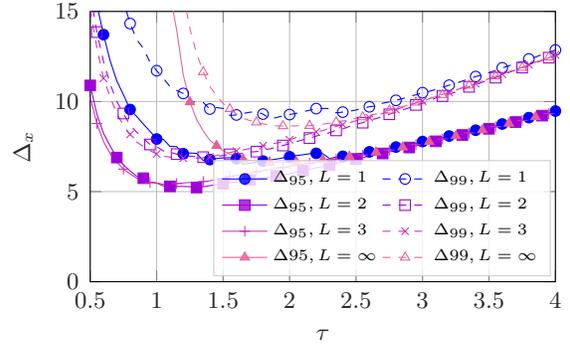}
        \else
            \begin{tikzpicture}
\begin{axis}[%
width=\sfwidth,
height=\sfheight,
xmin=0.5,
xmax=4,
ymin=0,
ymax=15,
axis background/.style={fill=white},
xlabel style={font=\footnotesize\color{white!15!black}},
xlabel={$\tau$},
ylabel near ticks,
ylabel style={font=\footnotesize\color{white!15!black}},
xticklabel style={font=\footnotesize\color{white!15!black}},
yticklabel style={font=\footnotesize\color{white!15!black}},
ylabel={$\Delta_x$},
axis background/.style={fill=white},
xmajorgrids,
ymajorgrids,
legend style={font=\tiny, at={(0.99,0.02)}, anchor=south east, legend columns=2,legend cell align=left, align=left,fill opacity=0.8, draw opacity=1, text opacity=1, draw=white!80!black}
]
\addplot [color=orange_D, mark=*,mark options={solid}, mark repeat = 4]
  table[row sep=crcr]{%
0.5	16\\
0.55	15.38\\
0.6	13.711\\
0.65	12.261\\
0.7	11.147\\
0.75	10.376\\
0.8	9.557\\
0.85	9.167\\
0.9	8.745\\
0.95	8.347\\
1	7.918\\
1.05	7.9\\
1.1	7.508\\
1.15	7.499\\
1.2	7.103\\
1.25	7.142\\
1.3	7.122\\
1.35	6.727\\
1.4	6.778\\
1.45	6.835\\
1.5	6.876\\
1.55	6.88\\
1.6	6.83\\
1.65	6.526\\
1.7	6.578\\
1.75	6.634\\
1.8	6.695\\
1.85	6.75\\
1.9	6.806\\
1.95	6.87\\
2	6.933\\
2.05	7.001\\
2.1	7.047\\
2.15	7.1\\
2.2	7.123\\
2.25	7.104\\
2.3	6.839\\
2.35	6.915\\
2.4	6.959\\
2.45	7.021\\
2.5	7.076\\
2.55	7.133\\
2.6	7.208\\
2.65	7.27\\
2.7	7.353\\
2.75	7.41\\
2.8	7.474\\
2.85	7.552\\
2.9	7.63\\
2.95	7.685\\
3	7.779\\
3.05	7.838\\
3.1	7.924\\
3.15	8.003\\
3.2	8.085\\
3.25	8.162\\
3.3	8.242\\
3.35	8.328\\
3.4	8.408\\
3.45	8.493\\
3.5	8.578\\
3.55	8.657\\
3.6	8.752\\
3.65	8.837\\
3.7	8.937\\
3.75	9.016\\
3.8	9.105\\
3.85	9.192\\
3.9	9.28\\
3.95	9.369\\
4	9.47\\
};
\addlegendentry{$\Delta_{95}, L=1$}

\addplot [color=orange_D,dashed,mark=o,,mark options={solid}, mark repeat=4]
  table[row sep=crcr]{%
0.5	16\\
0.55	16\\
0.6	16\\
0.65	16\\
0.7	16\\
0.75	15.575\\
0.8	14.327\\
0.85	13.437\\
0.9	12.95\\
0.95	12.176\\
1	11.71\\
1.05	11.29\\
1.1	10.836\\
1.15	10.611\\
1.2	10.447\\
1.25	9.959\\
1.3	9.987\\
1.35	9.948\\
1.4	9.582\\
1.45	9.661\\
1.5	9.617\\
1.55	9.23\\
1.6	9.255\\
1.65	9.332\\
1.7	9.407\\
1.75	9.428\\
1.8	9.304\\
1.85	9.045\\
1.9	9.107\\
1.95	9.206\\
2	9.276\\
2.05	9.344\\
2.1	9.435\\
2.15	9.515\\
2.2	9.606\\
2.25	9.616\\
2.3	9.612\\
2.35	9.374\\
2.4	9.412\\
2.45	9.458\\
2.5	9.531\\
2.55	9.635\\
2.6	9.707\\
2.65	9.758\\
2.7	9.905\\
2.75	9.977\\
2.8	10.066\\
2.85	10.16\\
2.9	10.292\\
2.95	10.369\\
3	10.489\\
3.05	10.565\\
3.1	10.671\\
3.15	10.788\\
3.2	10.905\\
3.25	11.004\\
3.3	11.134\\
3.35	11.263\\
3.4	11.385\\
3.45	11.482\\
3.5	11.607\\
3.55	11.724\\
3.6	11.87\\
3.65	11.978\\
3.7	12.128\\
3.75	12.202\\
3.8	12.36\\
3.85	12.467\\
3.9	12.612\\
3.95	12.733\\
4	12.858\\
};
\addlegendentry{$\Delta_{99}, L=1$}

\addplot [color=cyan,mark=square*,,mark options={solid}, mark repeat=4,mark phase=1]
  table[row sep=crcr]{%
0.5	10.892\\
0.55	9.301\\
0.6	8.237\\
0.65	7.534\\
0.7	6.893\\
0.75	6.488\\
0.8	6.145\\
0.85	5.849\\
0.9	5.743\\
0.95	5.527\\
1	5.475\\
1.05	5.316\\
1.1	5.29\\
1.15	5.296\\
1.2	5.284\\
1.25	5.272\\
1.3	5.228\\
1.35	5.275\\
1.4	5.329\\
1.45	5.382\\
1.5	5.437\\
1.55	5.492\\
1.6	5.541\\
1.65	5.586\\
1.7	5.645\\
1.75	5.702\\
1.8	5.753\\
1.85	5.808\\
1.9	5.869\\
1.95	5.916\\
2	5.987\\
2.05	6.071\\
2.1	6.162\\
2.15	6.236\\
2.2	6.324\\
2.25	6.396\\
2.3	6.469\\
2.35	6.565\\
2.4	6.636\\
2.45	6.723\\
2.5	6.811\\
2.55	6.885\\
2.6	6.969\\
2.65	7.033\\
2.7	7.117\\
2.75	7.207\\
2.8	7.281\\
2.85	7.371\\
2.9	7.448\\
2.95	7.542\\
3	7.622\\
3.05	7.705\\
3.1	7.791\\
3.15	7.875\\
3.2	7.961\\
3.25	8.052\\
3.3	8.142\\
3.35	8.233\\
3.4	8.312\\
3.45	8.404\\
3.5	8.486\\
3.55	8.58\\
3.6	8.677\\
3.65	8.765\\
3.7	8.86\\
3.75	8.949\\
3.8	9.032\\
3.85	9.122\\
3.9	9.216\\
3.95	9.322\\
4	9.406\\
};
\addlegendentry{$\Delta_{95}, L=2$}

\addplot [color=cyan,dashed,mark=square,,mark options={solid}, mark repeat=4,mark phase=1]
  table[row sep=crcr]{%
0.5	16\\
0.55	13.853\\
0.6	12.228\\
0.65	10.977\\
0.7	10.01\\
0.75	9.334\\
0.8	8.699\\
0.85	8.293\\
0.9	7.95\\
0.95	7.616\\
1	7.497\\
1.05	7.259\\
1.1	7.193\\
1.15	7.081\\
1.2	6.969\\
1.25	6.967\\
1.3	6.933\\
1.35	6.912\\
1.4	6.905\\
1.45	6.947\\
1.5	7.004\\
1.55	7.068\\
1.6	7.083\\
1.65	7.125\\
1.7	7.193\\
1.75	7.221\\
1.8	7.302\\
1.85	7.37\\
1.9	7.462\\
1.95	7.558\\
2	7.661\\
2.05	7.768\\
2.1	7.876\\
2.15	7.973\\
2.2	8.081\\
2.25	8.184\\
2.3	8.274\\
2.35	8.396\\
2.4	8.482\\
2.45	8.615\\
2.5	8.741\\
2.55	8.825\\
2.6	8.962\\
2.65	9.048\\
2.7	9.183\\
2.75	9.29\\
2.8	9.409\\
2.85	9.538\\
2.9	9.649\\
2.95	9.81\\
3	9.901\\
3.05	10.028\\
3.1	10.201\\
3.15	10.299\\
3.2	10.417\\
3.25	10.525\\
3.3	10.666\\
3.35	10.814\\
3.4	10.925\\
3.45	11.075\\
3.5	11.201\\
3.55	11.322\\
3.6	11.486\\
3.65	11.614\\
3.7	11.748\\
3.75	11.883\\
3.8	12.006\\
3.85	12.125\\
3.9	12.287\\
3.95	12.434\\
4	12.526\\
};
\addlegendentry{$\Delta_{99}, L=2$}

\addplot [color=green_D,mark=+,mark options={solid}, mark repeat=4,mark phase=2]
  table[row sep=crcr]{%
0.5	10.229\\
0.55	8.778\\
0.6	7.779\\
0.65	7.105\\
0.7	6.638\\
0.75	6.247\\
0.8	5.986\\
0.85	5.744\\
0.9	5.618\\
0.95	5.508\\
1	5.476\\
1.05	5.411\\
1.1	5.404\\
1.15	5.443\\
1.2	5.479\\
1.25	5.506\\
1.3	5.529\\
1.35	5.559\\
1.4	5.598\\
1.45	5.68\\
1.5	5.758\\
1.55	5.835\\
1.6	5.906\\
1.65	5.971\\
1.7	6.042\\
1.75	6.104\\
1.8	6.169\\
1.85	6.23\\
1.9	6.295\\
1.95	6.347\\
2	6.402\\
2.05	6.471\\
2.1	6.528\\
2.15	6.587\\
2.2	6.628\\
2.25	6.675\\
2.3	6.742\\
2.35	6.794\\
2.4	6.86\\
2.45	6.915\\
2.5	6.981\\
2.55	7.034\\
2.6	7.104\\
2.65	7.173\\
2.7	7.234\\
2.75	7.308\\
2.8	7.378\\
2.85	7.448\\
2.9	7.53\\
2.95	7.61\\
3	7.672\\
3.05	7.756\\
3.1	7.838\\
3.15	7.92\\
3.2	8.003\\
3.25	8.086\\
3.3	8.17\\
3.35	8.258\\
3.4	8.337\\
3.45	8.426\\
3.5	8.518\\
3.55	8.609\\
3.6	8.694\\
3.65	8.775\\
3.7	8.861\\
3.75	8.955\\
3.8	9.048\\
3.85	9.146\\
3.9	9.23\\
3.95	9.324\\
4	9.41\\
};
\addlegendentry{$\Delta_{95}, L=3$}

\addplot [color=green_D,mark=x,dashed,mark options={solid}, mark repeat=4,mark phase=2]
  table[row sep=crcr]{%
0.5	15.135\\
0.55	12.995\\
0.6	11.29\\
0.65	10.251\\
0.7	9.33\\
0.75	8.671\\
0.8	8.176\\
0.85	7.755\\
0.9	7.5\\
0.95	7.276\\
1	7.089\\
1.05	6.961\\
1.1	6.917\\
1.15	6.832\\
1.2	6.858\\
1.25	6.86\\
1.3	6.86\\
1.35	6.869\\
1.4	6.924\\
1.45	7.006\\
1.5	7.09\\
1.55	7.163\\
1.6	7.251\\
1.65	7.318\\
1.7	7.388\\
1.75	7.457\\
1.8	7.548\\
1.85	7.62\\
1.9	7.695\\
1.95	7.778\\
2	7.879\\
2.05	8.002\\
2.1	8.109\\
2.15	8.206\\
2.2	8.315\\
2.25	8.383\\
2.3	8.514\\
2.35	8.614\\
2.4	8.72\\
2.45	8.825\\
2.5	8.932\\
2.55	9.007\\
2.6	9.107\\
2.65	9.241\\
2.7	9.337\\
2.75	9.425\\
2.8	9.544\\
2.85	9.675\\
2.9	9.791\\
2.95	9.917\\
3	10.017\\
3.05	10.15\\
3.1	10.276\\
3.15	10.387\\
3.2	10.503\\
3.25	10.627\\
3.3	10.771\\
3.35	10.865\\
3.4	10.992\\
3.45	11.122\\
3.5	11.272\\
3.55	11.398\\
3.6	11.517\\
3.65	11.659\\
3.7	11.756\\
3.75	11.901\\
3.8	12.047\\
3.85	12.193\\
3.9	12.314\\
3.95	12.423\\
4	12.589\\
};
\addlegendentry{$\Delta_{99}, L=3$}

\addplot [color=violet,mark=triangle*,mark options={solid}, mark repeat=4,mark phase=3]
  table[row sep=crcr]{%
0.5	16\\
0.55	16\\
0.6	16\\
0.65	16\\
0.7	16\\
0.75	16\\
0.8	16\\
0.85	16\\
0.9	16\\
0.95	16\\
1	16\\
1.05	16\\
1.1	16\\
1.15	13.979\\
1.2	11.481\\
1.25	9.971\\
1.3	9.004\\
1.35	8.368\\
1.4	7.853\\
1.45	7.552\\
1.5	7.274\\
1.55	7.072\\
1.6	6.927\\
1.65	6.803\\
1.7	6.725\\
1.75	6.675\\
1.8	6.651\\
1.85	6.621\\
1.9	6.621\\
1.95	6.625\\
2	6.648\\
2.05	6.654\\
2.1	6.683\\
2.15	6.695\\
2.2	6.73\\
2.25	6.76\\
2.3	6.816\\
2.35	6.848\\
2.4	6.906\\
2.45	6.949\\
2.5	7.01\\
2.55	7.057\\
2.6	7.128\\
2.65	7.182\\
2.7	7.257\\
2.75	7.322\\
2.8	7.403\\
2.85	7.464\\
2.9	7.533\\
2.95	7.616\\
3	7.696\\
3.05	7.766\\
3.1	7.837\\
3.15	7.922\\
3.2	8.005\\
3.25	8.085\\
3.3	8.176\\
3.35	8.26\\
3.4	8.338\\
3.45	8.415\\
3.5	8.5\\
3.55	8.607\\
3.6	8.685\\
3.65	8.784\\
3.7	8.871\\
3.75	8.943\\
3.8	9.049\\
3.85	9.141\\
3.9	9.223\\
3.95	9.321\\
4	9.425\\
};
\addlegendentry{$\Delta{95}, L = \infty$}

\addplot [color=violet,dashed,mark=triangle,mark options={solid}, mark repeat=4,mark phase=3]
  table[row sep=crcr]{%
0.5	16\\
0.55	16\\
0.6	16\\
0.65	16\\
0.7	16\\
0.75	16\\
0.8	16\\
0.85	16\\
0.9	16\\
0.95	16\\
1	16\\
1.05	16\\
1.1	16\\
1.15	16\\
1.2	16\\
1.25	14.045\\
1.3	12.675\\
1.35	11.641\\
1.4	10.843\\
1.45	10.313\\
1.5	9.899\\
1.55	9.586\\
1.6	9.312\\
1.65	9.069\\
1.7	8.946\\
1.75	8.848\\
1.8	8.757\\
1.85	8.705\\
1.9	8.644\\
1.95	8.642\\
2	8.646\\
2.05	8.645\\
2.1	8.682\\
2.15	8.657\\
2.2	8.711\\
2.25	8.743\\
2.3	8.794\\
2.35	8.857\\
2.4	8.926\\
2.45	8.971\\
2.5	9.074\\
2.55	9.134\\
2.6	9.227\\
2.65	9.305\\
2.7	9.426\\
2.75	9.503\\
2.8	9.628\\
2.85	9.72\\
2.9	9.816\\
2.95	9.948\\
3	10.043\\
3.05	10.174\\
3.1	10.269\\
3.15	10.372\\
3.2	10.51\\
3.25	10.629\\
3.3	10.761\\
3.35	10.891\\
3.4	10.989\\
3.45	11.146\\
3.5	11.254\\
3.55	11.404\\
3.6	11.5\\
3.65	11.658\\
3.7	11.792\\
3.75	11.88\\
3.8	12.038\\
3.85	12.177\\
3.9	12.307\\
3.95	12.455\\
4	12.587\\
};
\addlegendentry{$\Delta{99}, L = \infty$}

\end{axis}
\end{tikzpicture}%
        \fi        
        \caption{\gls{paoi} percentiles for the $(4,6)$ system with $\varepsilon=0.2$.}
        \label{fig:age_46_02_perc}
    \end{subfigure}
    \begin{subfigure}[b]{.49\linewidth}
	    \centering
        \ifdefined\pdffig
            \includegraphics[width=\linewidth]{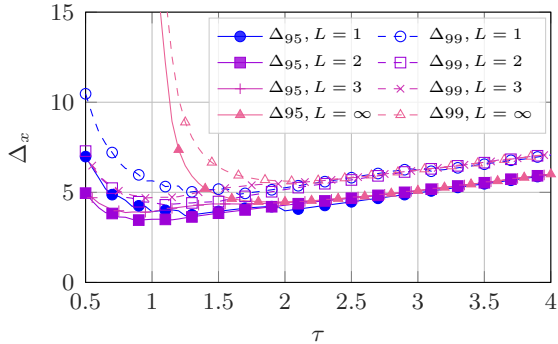}
        \else
            \begin{tikzpicture}
\begin{axis}[%
width=\sfwidth,
height=\sfheight,
xmin=0.5,
xmax=4,
ymin=0,
ymax=15,
axis background/.style={fill=white},
xlabel style={font=\footnotesize\color{white!15!black}},
xlabel={$\tau$},
ylabel near ticks,
ylabel style={font=\footnotesize\color{white!15!black}},
xticklabel style={font=\footnotesize\color{white!15!black}},
yticklabel style={font=\footnotesize\color{white!15!black}},
ylabel={$\Delta_x$},
axis background/.style={fill=white},
xmajorgrids,
ymajorgrids,
legend style={font=\tiny, at={(0.99,0.98)}, anchor=north east, legend columns=2,legend cell align=left, align=left,fill opacity=0.8, draw opacity=1, text opacity=1, draw=white!80!black}
]
\addplot [color=orange_D, mark=*,mark options={solid}, mark repeat = 4]
  table[row sep=crcr]{%
0.5	6.983\\
0.55	6.37\\
0.6	5.789\\
0.65	5.19\\
0.7	4.871\\
0.75	4.769\\
0.8	4.557\\
0.85	4.223\\
0.9	4.252\\
0.95	4.222\\
1	3.925\\
1.05	3.971\\
1.1	4.003\\
1.15	4.024\\
1.2	3.974\\
1.25	3.705\\
1.3	3.753\\
1.35	3.798\\
1.4	3.848\\
1.45	3.897\\
1.5	3.939\\
1.55	3.988\\
1.6	4.033\\
1.65	4.075\\
1.7	4.117\\
1.75	4.152\\
1.8	4.175\\
1.85	4.199\\
1.9	4.205\\
1.95	4.178\\
2	3.972\\
2.05	4.022\\
2.1	4.071\\
2.15	4.12\\
2.2	4.17\\
2.25	4.223\\
2.3	4.272\\
2.35	4.321\\
2.4	4.37\\
2.45	4.42\\
2.5	4.473\\
2.55	4.522\\
2.6	4.576\\
2.65	4.621\\
2.7	4.667\\
2.75	4.721\\
2.8	4.773\\
2.85	4.819\\
2.9	4.872\\
2.95	4.925\\
3	4.967\\
3.05	5.022\\
3.1	5.075\\
3.15	5.124\\
3.2	5.171\\
3.25	5.221\\
3.3	5.272\\
3.35	5.322\\
3.4	5.372\\
3.45	5.424\\
3.5	5.473\\
3.55	5.521\\
3.6	5.567\\
3.65	5.62\\
3.7	5.669\\
3.75	5.721\\
3.8	5.771\\
3.85	5.817\\
3.9	5.875\\
3.95	5.922\\
4	5.975\\
};
\addlegendentry{$\Delta_{95}, L=1$}

\addplot [color=orange_D,dashed,mark=o,,mark options={solid}, mark repeat=4]
  table[row sep=crcr]{%
0.5	10.468\\
0.55	9.319\\
0.6	8.341\\
0.65	7.707\\
0.7	7.208\\
0.75	6.677\\
0.8	6.326\\
0.85	6.236\\
0.9	5.97\\
0.95	5.631\\
1	5.638\\
1.05	5.558\\
1.1	5.319\\
1.15	5.351\\
1.2	5.347\\
1.25	4.978\\
1.3	5.028\\
1.35	5.074\\
1.4	5.116\\
1.45	5.157\\
1.5	5.178\\
1.55	5.192\\
1.6	5.123\\
1.65	4.89\\
1.7	4.941\\
1.75	4.994\\
1.8	5.038\\
1.85	5.091\\
1.9	5.14\\
1.95	5.207\\
2	5.252\\
2.05	5.311\\
2.1	5.366\\
2.15	5.422\\
2.2	5.479\\
2.25	5.546\\
2.3	5.592\\
2.35	5.644\\
2.4	5.705\\
2.45	5.761\\
2.5	5.813\\
2.55	5.878\\
2.6	5.941\\
2.65	5.993\\
2.7	6.044\\
2.75	6.098\\
2.8	6.166\\
2.85	6.21\\
2.9	6.255\\
2.95	6.297\\
3	6.255\\
3.05	6.322\\
3.1	6.17\\
3.15	6.229\\
3.2	6.26\\
3.25	6.302\\
3.3	6.359\\
3.35	6.416\\
3.4	6.458\\
3.45	6.524\\
3.5	6.559\\
3.55	6.622\\
3.6	6.652\\
3.65	6.706\\
3.7	6.778\\
3.75	6.806\\
3.8	6.844\\
3.85	6.908\\
3.9	6.962\\
3.95	7.017\\
4	7.062\\
};
\addlegendentry{$\Delta_{99}, L=1$}

\addplot [color=cyan,mark=square*,mark options={solid}, mark repeat=4,mark phase=1]
  table[row sep=crcr]{%
0.5	4.956\\
0.55	4.519\\
0.6	4.139\\
0.65	3.875\\
0.7	3.825\\
0.75	3.632\\
0.8	3.617\\
0.85	3.503\\
0.9	3.459\\
0.95	3.482\\
1	3.505\\
1.05	3.507\\
1.1	3.498\\
1.15	3.473\\
1.2	3.522\\
1.25	3.585\\
1.3	3.635\\
1.35	3.694\\
1.4	3.743\\
1.45	3.794\\
1.5	3.842\\
1.55	3.891\\
1.6	3.936\\
1.65	3.979\\
1.7	4.024\\
1.75	4.071\\
1.8	4.114\\
1.85	4.155\\
1.9	4.2\\
1.95	4.239\\
2	4.284\\
2.05	4.327\\
2.1	4.368\\
2.15	4.409\\
2.2	4.448\\
2.25	4.491\\
2.3	4.522\\
2.35	4.559\\
2.4	4.597\\
2.45	4.638\\
2.5	4.673\\
2.55	4.715\\
2.6	4.751\\
2.65	4.791\\
2.7	4.831\\
2.75	4.878\\
2.8	4.911\\
2.85	4.953\\
2.9	5\\
2.95	5.046\\
3	5.086\\
3.05	5.128\\
3.1	5.173\\
3.15	5.214\\
3.2	5.26\\
3.25	5.31\\
3.3	5.35\\
3.35	5.401\\
3.4	5.448\\
3.45	5.49\\
3.5	5.531\\
3.55	5.585\\
3.6	5.63\\
3.65	5.679\\
3.7	5.728\\
3.75	5.775\\
3.8	5.821\\
3.85	5.867\\
3.9	5.918\\
3.95	5.964\\
4	6.008\\
};
\addlegendentry{$\Delta_{95}, L=2$}

\addplot [color=cyan,dashed,mark=square,,mark options={solid}, mark repeat=4,mark phase=1]
  table[row sep=crcr]{%
0.5	7.288\\
0.55	6.458\\
0.6	5.88\\
0.65	5.544\\
0.7	5.233\\
0.75	4.984\\
0.8	4.721\\
0.85	4.674\\
0.9	4.477\\
0.95	4.462\\
1	4.45\\
1.05	4.369\\
1.1	4.304\\
1.15	4.344\\
1.2	4.369\\
1.25	4.407\\
1.3	4.412\\
1.35	4.444\\
1.4	4.457\\
1.45	4.473\\
1.5	4.489\\
1.55	4.569\\
1.6	4.643\\
1.65	4.713\\
1.7	4.777\\
1.75	4.848\\
1.8	4.906\\
1.85	4.965\\
1.9	5.033\\
1.95	5.091\\
2	5.155\\
2.05	5.209\\
2.1	5.261\\
2.15	5.321\\
2.2	5.381\\
2.25	5.426\\
2.3	5.478\\
2.35	5.537\\
2.4	5.587\\
2.45	5.645\\
2.5	5.687\\
2.55	5.749\\
2.6	5.802\\
2.65	5.85\\
2.7	5.915\\
2.75	5.956\\
2.8	6.011\\
2.85	6.066\\
2.9	6.13\\
2.95	6.176\\
3	6.218\\
3.05	6.27\\
3.1	6.303\\
3.15	6.338\\
3.2	6.385\\
3.25	6.434\\
3.3	6.469\\
3.35	6.509\\
3.4	6.574\\
3.45	6.613\\
3.5	6.645\\
3.55	6.705\\
3.6	6.759\\
3.65	6.792\\
3.7	6.835\\
3.75	6.898\\
3.8	6.93\\
3.85	6.992\\
3.9	7.029\\
3.95	7.065\\
4	7.105\\
};
\addlegendentry{$\Delta_{99}, L=2$}

\addplot [color=green_D,mark=+,mark options={solid}, mark repeat=4,mark phase=2]
  table[row sep=crcr]{%
0.5	5.053\\
0.55	4.702\\
0.6	4.397\\
0.65	4.212\\
0.7	4.031\\
0.75	3.978\\
0.8	3.877\\
0.85	3.907\\
0.9	3.91\\
0.95	3.899\\
1	3.924\\
1.05	3.993\\
1.1	4.058\\
1.15	4.114\\
1.2	4.163\\
1.25	4.21\\
1.3	4.25\\
1.35	4.29\\
1.4	4.324\\
1.45	4.356\\
1.5	4.354\\
1.55	4.356\\
1.6	4.357\\
1.65	4.353\\
1.7	4.358\\
1.75	4.369\\
1.8	4.375\\
1.85	4.386\\
1.9	4.398\\
1.95	4.413\\
2	4.434\\
2.05	4.448\\
2.1	4.471\\
2.15	4.497\\
2.2	4.522\\
2.25	4.55\\
2.3	4.579\\
2.35	4.608\\
2.4	4.639\\
2.45	4.67\\
2.5	4.702\\
2.55	4.742\\
2.6	4.774\\
2.65	4.813\\
2.7	4.854\\
2.75	4.889\\
2.8	4.931\\
2.85	4.97\\
2.9	5.009\\
2.95	5.051\\
3	5.096\\
3.05	5.143\\
3.1	5.179\\
3.15	5.222\\
3.2	5.271\\
3.25	5.309\\
3.3	5.359\\
3.35	5.407\\
3.4	5.451\\
3.45	5.495\\
3.5	5.541\\
3.55	5.585\\
3.6	5.639\\
3.65	5.685\\
3.7	5.728\\
3.75	5.773\\
3.8	5.822\\
3.85	5.87\\
3.9	5.915\\
3.95	5.965\\
4	6.008\\
};
\addlegendentry{$\Delta_{95}, L=3$}

\addplot [color=green_D,mark=x,dashed,mark options={solid}, mark repeat=4,mark phase=2]
  table[row sep=crcr]{%
0.5	7.251\\
0.55	6.483\\
0.6	5.917\\
0.65	5.607\\
0.7	5.325\\
0.75	5.085\\
0.8	4.955\\
0.85	4.839\\
0.9	4.79\\
0.95	4.686\\
1	4.706\\
1.05	4.722\\
1.1	4.726\\
1.15	4.727\\
1.2	4.76\\
1.25	4.844\\
1.3	4.921\\
1.35	5.004\\
1.4	5.065\\
1.45	5.124\\
1.5	5.174\\
1.55	5.235\\
1.6	5.283\\
1.65	5.316\\
1.7	5.359\\
1.75	5.394\\
1.8	5.432\\
1.85	5.446\\
1.9	5.466\\
1.95	5.47\\
2	5.496\\
2.05	5.511\\
2.1	5.547\\
2.15	5.57\\
2.2	5.602\\
2.25	5.625\\
2.3	5.661\\
2.35	5.691\\
2.4	5.731\\
2.45	5.76\\
2.5	5.802\\
2.55	5.844\\
2.6	5.879\\
2.65	5.929\\
2.7	5.974\\
2.75	6.018\\
2.8	6.067\\
2.85	6.101\\
2.9	6.148\\
2.95	6.184\\
3	6.243\\
3.05	6.289\\
3.1	6.32\\
3.15	6.357\\
3.2	6.404\\
3.25	6.44\\
3.3	6.493\\
3.35	6.526\\
3.4	6.563\\
3.45	6.619\\
3.5	6.668\\
3.55	6.714\\
3.6	6.785\\
3.65	6.816\\
3.7	6.841\\
3.75	6.889\\
3.8	6.946\\
3.85	6.989\\
3.9	7.029\\
3.95	7.065\\
4	7.11\\
};
\addlegendentry{$\Delta_{99}, L=3$}

\addplot [color=violet,mark=triangle*,mark options={solid}, mark repeat=4,mark phase=3]
  table[row sep=crcr]{%
0.5	16\\
0.55	16\\
0.6	16\\
0.65	16\\
0.7	16\\
0.75	16\\
0.8	16\\
0.85	16\\
0.9	16\\
0.95	16\\
1	16\\
1.05	16\\
1.1	12.196\\
1.15	8.943\\
1.2	7.363\\
1.25	6.475\\
1.3	5.906\\
1.35	5.507\\
1.4	5.2\\
1.45	5.006\\
1.5	4.847\\
1.55	4.734\\
1.6	4.645\\
1.65	4.586\\
1.7	4.535\\
1.75	4.509\\
1.8	4.48\\
1.85	4.475\\
1.9	4.465\\
1.95	4.464\\
2	4.471\\
2.05	4.483\\
2.1	4.496\\
2.15	4.508\\
2.2	4.537\\
2.25	4.557\\
2.3	4.588\\
2.35	4.619\\
2.4	4.649\\
2.45	4.679\\
2.5	4.711\\
2.55	4.74\\
2.6	4.775\\
2.65	4.816\\
2.7	4.853\\
2.75	4.894\\
2.8	4.934\\
2.85	4.974\\
2.9	5.014\\
2.95	5.054\\
3	5.093\\
3.05	5.138\\
3.1	5.182\\
3.15	5.228\\
3.2	5.27\\
3.25	5.314\\
3.3	5.36\\
3.35	5.41\\
3.4	5.45\\
3.45	5.492\\
3.5	5.54\\
3.55	5.588\\
3.6	5.631\\
3.65	5.681\\
3.7	5.731\\
3.75	5.773\\
3.8	5.821\\
3.85	5.868\\
3.9	5.922\\
3.95	5.962\\
4	6.009\\
};
\addlegendentry{$\Delta{95}, L = \infty$}

\addplot [color=violet,dashed,mark=triangle,mark options={solid}, mark repeat=4,mark phase=3]
  table[row sep=crcr]{%
0.5	16\\
0.55	16\\
0.6	16\\
0.65	16\\
0.7	16\\
0.75	16\\
0.8	16\\
0.85	16\\
0.9	16\\
0.95	16\\
1	16\\
1.05	16\\
1.1	16\\
1.15	12.536\\
1.2	10.193\\
1.25	8.889\\
1.3	8.051\\
1.35	7.448\\
1.4	6.964\\
1.45	6.659\\
1.5	6.404\\
1.55	6.21\\
1.6	6.05\\
1.65	5.94\\
1.7	5.845\\
1.75	5.783\\
1.8	5.718\\
1.85	5.689\\
1.9	5.66\\
1.95	5.633\\
2	5.625\\
2.05	5.614\\
2.1	5.634\\
2.15	5.62\\
2.2	5.652\\
2.25	5.661\\
2.3	5.691\\
2.35	5.73\\
2.4	5.749\\
2.45	5.787\\
2.5	5.838\\
2.55	5.855\\
2.6	5.887\\
2.65	5.946\\
2.7	5.981\\
2.75	6.033\\
2.8	6.077\\
2.85	6.114\\
2.9	6.17\\
2.95	6.199\\
3	6.236\\
3.05	6.286\\
3.1	6.323\\
3.15	6.366\\
3.2	6.403\\
3.25	6.453\\
3.3	6.507\\
3.35	6.539\\
3.4	6.586\\
3.45	6.611\\
3.5	6.647\\
3.55	6.705\\
3.6	6.741\\
3.65	6.812\\
3.7	6.84\\
3.75	6.876\\
3.8	6.936\\
3.85	6.987\\
3.9	7.048\\
3.95	7.064\\
4	7.111\\
};
\addlegendentry{$\Delta{99}, L = \infty$}

\end{axis}
\end{tikzpicture}%
        \fi        
        \caption{\gls{paoi} percentiles for the $(4,7)$ system with $\varepsilon=0.1$.}
        \label{fig:age_47_01_perc}
    \end{subfigure}
    \begin{subfigure}[b]{.49\linewidth}
	    \centering
        \ifdefined\pdffig
            \includegraphics[width=\linewidth]{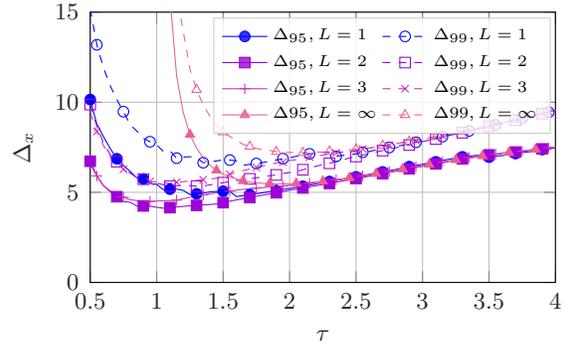}
        \else
            \begin{tikzpicture}
\begin{axis}[%
width=\sfwidth,
height=\sfheight,
xmin=0.5,
xmax=4,
ymin=0,
ymax=15,
axis background/.style={fill=white},
xlabel style={font=\footnotesize\color{white!15!black}},
xlabel={$\tau$},
ylabel near ticks,
ylabel style={font=\footnotesize\color{white!15!black}},
xticklabel style={font=\footnotesize\color{white!15!black}},
yticklabel style={font=\footnotesize\color{white!15!black}},
ylabel={$\Delta_x$},
axis background/.style={fill=white},
xmajorgrids,
ymajorgrids,
legend style={font=\tiny, at={(0.99,0.98)}, anchor=north east, legend columns=2,legend cell align=left, align=left,fill opacity=0.8, draw opacity=1, text opacity=1, draw=white!80!black}
]
\addplot [color=orange_D, mark=*,mark options={solid}, mark repeat = 4]
  table[row sep=crcr]{%
0.5	10.149\\
0.55	8.784\\
0.6	8.089\\
0.65	7.442\\
0.7	6.862\\
0.75	6.492\\
0.8	6.165\\
0.85	5.811\\
0.9	5.744\\
0.95	5.493\\
1	5.444\\
1.05	5.143\\
1.1	5.176\\
1.15	5.178\\
1.2	5.109\\
1.25	4.877\\
1.3	4.916\\
1.35	4.964\\
1.4	5.02\\
1.45	5.045\\
1.5	5.061\\
1.55	5.052\\
1.6	4.764\\
1.65	4.823\\
1.7	4.871\\
1.75	4.927\\
1.8	4.982\\
1.85	5.036\\
1.9	5.098\\
1.95	5.155\\
2	5.211\\
2.05	5.273\\
2.1	5.337\\
2.15	5.402\\
2.2	5.463\\
2.25	5.532\\
2.3	5.598\\
2.35	5.66\\
2.4	5.73\\
2.45	5.791\\
2.5	5.858\\
2.55	5.934\\
2.6	6.001\\
2.65	6.074\\
2.7	6.131\\
2.75	6.203\\
2.8	6.268\\
2.85	6.352\\
2.9	6.424\\
2.95	6.493\\
3	6.555\\
3.05	6.633\\
3.1	6.692\\
3.15	6.775\\
3.2	6.838\\
3.25	6.897\\
3.3	6.969\\
3.35	7\\
3.4	7.049\\
3.45	7.046\\
3.5	6.951\\
3.55	7.012\\
3.6	7.039\\
3.65	7.089\\
3.7	7.153\\
3.75	7.208\\
3.8	7.246\\
3.85	7.295\\
3.9	7.375\\
3.95	7.401\\
4	7.464\\
};
\addlegendentry{$\Delta_{95}, L=1$}

\addplot [color=orange_D,dashed,mark=o,,mark options={solid}, mark repeat=4]
  table[row sep=crcr]{%
0.5	15.191\\
0.55	13.188\\
0.6	11.907\\
0.65	10.925\\
0.7	10.153\\
0.75	9.478\\
0.8	8.768\\
0.85	8.399\\
0.9	7.998\\
0.95	7.807\\
1	7.631\\
1.05	7.273\\
1.1	7.305\\
1.15	6.895\\
1.2	6.957\\
1.25	6.976\\
1.3	6.931\\
1.35	6.644\\
1.4	6.685\\
1.45	6.744\\
1.5	6.793\\
1.55	6.791\\
1.6	6.739\\
1.65	6.503\\
1.7	6.551\\
1.75	6.599\\
1.8	6.656\\
1.85	6.723\\
1.9	6.794\\
1.95	6.866\\
2	6.912\\
2.05	7\\
2.1	7.052\\
2.15	7.135\\
2.2	7.168\\
2.25	7.222\\
2.3	7.223\\
2.35	6.995\\
2.4	7.046\\
2.45	7.086\\
2.5	7.134\\
2.55	7.221\\
2.6	7.263\\
2.65	7.356\\
2.7	7.396\\
2.75	7.488\\
2.8	7.545\\
2.85	7.618\\
2.9	7.714\\
2.95	7.774\\
3	7.841\\
3.05	7.93\\
3.1	7.993\\
3.15	8.093\\
3.2	8.164\\
3.25	8.244\\
3.3	8.332\\
3.35	8.407\\
3.4	8.478\\
3.45	8.576\\
3.5	8.648\\
3.55	8.748\\
3.6	8.825\\
3.65	8.92\\
3.7	9.012\\
3.75	9.103\\
3.8	9.184\\
3.85	9.259\\
3.9	9.364\\
3.95	9.448\\
4	9.551\\
};
\addlegendentry{$\Delta_{99}, L=1$}

\addplot [color=cyan,mark=square*,mark options={solid}, mark repeat=4,mark phase=1]
  table[row sep=crcr]{%
0.5	6.725\\
0.55	5.926\\
0.6	5.362\\
0.65	5.037\\
0.7	4.756\\
0.75	4.486\\
0.8	4.452\\
0.85	4.24\\
0.9	4.246\\
0.95	4.226\\
1	4.171\\
1.05	4.118\\
1.1	4.151\\
1.15	4.191\\
1.2	4.219\\
1.25	4.252\\
1.3	4.271\\
1.35	4.304\\
1.4	4.325\\
1.45	4.356\\
1.5	4.428\\
1.55	4.508\\
1.6	4.578\\
1.65	4.643\\
1.7	4.71\\
1.75	4.782\\
1.8	4.85\\
1.85	4.911\\
1.9	4.978\\
1.95	5.044\\
2	5.115\\
2.05	5.175\\
2.1	5.244\\
2.15	5.306\\
2.2	5.368\\
2.25	5.435\\
2.3	5.493\\
2.35	5.565\\
2.4	5.627\\
2.45	5.701\\
2.5	5.769\\
2.55	5.827\\
2.6	5.898\\
2.65	5.967\\
2.7	6.037\\
2.75	6.112\\
2.8	6.176\\
2.85	6.241\\
2.9	6.305\\
2.95	6.382\\
3	6.443\\
3.05	6.514\\
3.1	6.581\\
3.15	6.634\\
3.2	6.717\\
3.25	6.784\\
3.3	6.857\\
3.35	6.915\\
3.4	6.966\\
3.45	7.003\\
3.5	7.078\\
3.55	7.104\\
3.6	7.165\\
3.65	7.194\\
3.7	7.236\\
3.75	7.292\\
3.8	7.331\\
3.85	7.382\\
3.9	7.409\\
3.95	7.49\\
4	7.515\\
};
\addlegendentry{$\Delta_{95}, L=2$}

\addplot [color=cyan,dashed,mark=square,,mark options={solid}, mark repeat=4,mark phase=1]
  table[row sep=crcr]{%
0.5	9.87\\
0.55	8.66\\
0.6	7.746\\
0.65	7.096\\
0.7	6.702\\
0.75	6.374\\
0.8	6.066\\
0.85	5.829\\
0.9	5.729\\
0.95	5.547\\
1	5.514\\
1.05	5.42\\
1.1	5.344\\
1.15	5.367\\
1.2	5.38\\
1.25	5.375\\
1.3	5.34\\
1.35	5.354\\
1.4	5.413\\
1.45	5.485\\
1.5	5.536\\
1.55	5.596\\
1.6	5.652\\
1.65	5.691\\
1.7	5.746\\
1.75	5.81\\
1.8	5.845\\
1.85	5.921\\
1.9	5.966\\
1.95	6.041\\
2	6.1\\
2.05	6.154\\
2.1	6.253\\
2.15	6.345\\
2.2	6.45\\
2.25	6.534\\
2.3	6.612\\
2.35	6.708\\
2.4	6.799\\
2.45	6.906\\
2.5	6.987\\
2.55	7.062\\
2.6	7.143\\
2.65	7.245\\
2.7	7.316\\
2.75	7.4\\
2.8	7.479\\
2.85	7.557\\
2.9	7.627\\
2.95	7.732\\
3	7.81\\
3.05	7.892\\
3.1	7.961\\
3.15	8.05\\
3.2	8.145\\
3.25	8.235\\
3.3	8.309\\
3.35	8.417\\
3.4	8.493\\
3.45	8.567\\
3.5	8.663\\
3.55	8.752\\
3.6	8.842\\
3.65	8.931\\
3.7	9\\
3.75	9.106\\
3.8	9.183\\
3.85	9.282\\
3.9	9.362\\
3.95	9.466\\
4	9.543\\
};
\addlegendentry{$\Delta_{99}, L=2$}

\addplot [color=green_D,mark=+,mark options={solid}, mark repeat=4,mark phase=2]
  table[row sep=crcr]{%
0.5	6.528\\
0.55	5.909\\
0.6	5.365\\
0.65	5.085\\
0.7	4.818\\
0.75	4.729\\
0.8	4.588\\
0.85	4.544\\
0.9	4.459\\
0.95	4.494\\
1	4.513\\
1.05	4.533\\
1.1	4.541\\
1.15	4.567\\
1.2	4.646\\
1.25	4.721\\
1.3	4.791\\
1.35	4.858\\
1.4	4.909\\
1.45	4.97\\
1.5	5.021\\
1.55	5.065\\
1.6	5.118\\
1.65	5.163\\
1.7	5.203\\
1.75	5.234\\
1.8	5.262\\
1.85	5.297\\
1.9	5.319\\
1.95	5.351\\
2	5.395\\
2.05	5.418\\
2.1	5.46\\
2.15	5.501\\
2.2	5.547\\
2.25	5.588\\
2.3	5.642\\
2.35	5.693\\
2.4	5.737\\
2.45	5.788\\
2.5	5.852\\
2.55	5.915\\
2.6	5.963\\
2.65	6.031\\
2.7	6.087\\
2.75	6.15\\
2.8	6.22\\
2.85	6.275\\
2.9	6.338\\
2.95	6.415\\
3	6.475\\
3.05	6.557\\
3.1	6.608\\
3.15	6.685\\
3.2	6.744\\
3.25	6.803\\
3.3	6.854\\
3.35	6.928\\
3.4	6.993\\
3.45	7.038\\
3.5	7.061\\
3.55	7.122\\
3.6	7.191\\
3.65	7.188\\
3.7	7.254\\
3.75	7.281\\
3.8	7.346\\
3.85	7.389\\
3.9	7.44\\
3.95	7.471\\
4	7.504\\
};
\addlegendentry{$\Delta_{95}, L=3$}

\addplot [color=green_D,mark=x,dashed,mark options={solid}, mark repeat=4,mark phase=2]
  table[row sep=crcr]{%
0.5	9.442\\
0.55	8.377\\
0.6	7.593\\
0.65	6.98\\
0.7	6.61\\
0.75	6.278\\
0.8	6.03\\
0.85	5.813\\
0.9	5.733\\
0.95	5.596\\
1	5.576\\
1.05	5.554\\
1.1	5.494\\
1.15	5.564\\
1.2	5.593\\
1.25	5.629\\
1.3	5.664\\
1.35	5.696\\
1.4	5.727\\
1.45	5.785\\
1.5	5.899\\
1.55	6.005\\
1.6	6.113\\
1.65	6.208\\
1.7	6.292\\
1.75	6.387\\
1.8	6.463\\
1.85	6.552\\
1.9	6.628\\
1.95	6.682\\
2	6.761\\
2.05	6.811\\
2.1	6.886\\
2.15	6.957\\
2.2	7.014\\
2.25	7.064\\
2.3	7.13\\
2.35	7.186\\
2.4	7.238\\
2.45	7.294\\
2.5	7.337\\
2.55	7.405\\
2.6	7.438\\
2.65	7.515\\
2.7	7.551\\
2.75	7.616\\
2.8	7.687\\
2.85	7.755\\
2.9	7.797\\
2.95	7.878\\
3	7.948\\
3.05	8.032\\
3.1	8.097\\
3.15	8.164\\
3.2	8.242\\
3.25	8.304\\
3.3	8.385\\
3.35	8.47\\
3.4	8.545\\
3.45	8.624\\
3.5	8.691\\
3.55	8.798\\
3.6	8.887\\
3.65	8.961\\
3.7	9.042\\
3.75	9.116\\
3.8	9.21\\
3.85	9.301\\
3.9	9.386\\
3.95	9.476\\
4	9.562\\
};
\addlegendentry{$\Delta_{99}, L=3$}

\addplot [color=violet,mark=triangle*,mark options={solid}, mark repeat=4,mark phase=3]
  table[row sep=crcr]{%
0.5	16\\
0.55	16\\
0.6	16\\
0.65	16\\
0.7	16\\
0.75	16\\
0.8	16\\
0.85	16\\
0.9	16\\
0.95	16\\
1	16\\
1.05	16\\
1.1	15.816\\
1.15	11.465\\
1.2	9.425\\
1.25	8.209\\
1.3	7.443\\
1.35	6.916\\
1.4	6.507\\
1.45	6.223\\
1.5	6.009\\
1.55	5.864\\
1.6	5.743\\
1.65	5.638\\
1.7	5.575\\
1.75	5.535\\
1.8	5.495\\
1.85	5.484\\
1.9	5.47\\
1.95	5.463\\
2	5.482\\
2.05	5.495\\
2.1	5.514\\
2.15	5.542\\
2.2	5.583\\
2.25	5.613\\
2.3	5.665\\
2.35	5.709\\
2.4	5.758\\
2.45	5.807\\
2.5	5.857\\
2.55	5.914\\
2.6	5.97\\
2.65	6.031\\
2.7	6.097\\
2.75	6.152\\
2.8	6.22\\
2.85	6.292\\
2.9	6.349\\
2.95	6.417\\
3	6.477\\
3.05	6.547\\
3.1	6.61\\
3.15	6.675\\
3.2	6.739\\
3.25	6.801\\
3.3	6.868\\
3.35	6.938\\
3.4	6.979\\
3.45	7.02\\
3.5	7.082\\
3.55	7.118\\
3.6	7.156\\
3.65	7.209\\
3.7	7.253\\
3.75	7.267\\
3.8	7.348\\
3.85	7.387\\
3.9	7.409\\
3.95	7.484\\
4	7.501\\
};
\addlegendentry{$\Delta{95}, L = \infty$}

\addplot [color=violet,dashed,mark=triangle,mark options={solid}, mark repeat=4,mark phase=3]
  table[row sep=crcr]{%
0.5	16\\
0.55	16\\
0.6	16\\
0.65	16\\
0.7	16\\
0.75	16\\
0.8	16\\
0.85	16\\
0.9	16\\
0.95	16\\
1	16\\
1.05	16\\
1.1	16\\
1.15	16\\
1.2	13.919\\
1.25	11.849\\
1.3	10.697\\
1.35	9.861\\
1.4	9.228\\
1.45	8.701\\
1.5	8.346\\
1.55	8.102\\
1.6	7.882\\
1.65	7.659\\
1.7	7.545\\
1.75	7.441\\
1.8	7.359\\
1.85	7.327\\
1.9	7.255\\
1.95	7.21\\
2	7.211\\
2.05	7.227\\
2.1	7.197\\
2.15	7.195\\
2.2	7.231\\
2.25	7.239\\
2.3	7.274\\
2.35	7.284\\
2.4	7.328\\
2.45	7.361\\
2.5	7.392\\
2.55	7.445\\
2.6	7.473\\
2.65	7.544\\
2.7	7.594\\
2.75	7.634\\
2.8	7.701\\
2.85	7.773\\
2.9	7.824\\
2.95	7.899\\
3	7.95\\
3.05	8.016\\
3.1	8.078\\
3.15	8.172\\
3.2	8.235\\
3.25	8.312\\
3.3	8.384\\
3.35	8.469\\
3.4	8.541\\
3.45	8.617\\
3.5	8.705\\
3.55	8.777\\
3.6	8.861\\
3.65	8.952\\
3.7	9.038\\
3.75	9.116\\
3.8	9.207\\
3.85	9.297\\
3.9	9.383\\
3.95	9.483\\
4	9.552\\
};
\addlegendentry{$\Delta{99}, L = \infty$}

\end{axis}
\end{tikzpicture}%
        \fi        
        \caption{\gls{paoi} percentiles for the $(4,7)$ system with $\varepsilon=0.2$.}
        \label{fig:age_47_02_perc}
    \end{subfigure}
     \caption{Higher percentiles of the \gls{paoi} for different queue sizes and codes with $\mu=1$ as a function of $\tau$.}\vspace{-0.6cm}
 \label{fig:age_perc}
\end{figure}

The difference between the systems with different queue sizes is much more marked if the interval $\tau$ is shorter, i.e., if we increase the inter-block interval $\tau$, as can be noticed from Fig.~\ref{fig:age_CDF_hl}: the $L=\infty$ non-preemptive system is still the worst on average, but it can actually get a better worst-case performance than the preemptive system with $L=1$ for shorter codes, such as the (4,5) system shown in Fig.~\ref{fig:age_45_01_rho06}-\subref{fig:age_45_02_rho06}. As a consequence, optimizing the \gls{paoi} requires optimizing $\tau$, and the optimal value of $\tau$ also depends on $L$. Fig.~\ref{fig:age_perc} shows the 95th and 99th percentiles of the \gls{paoi} for the various considered systems as a function of $\tau$. As the figures show, setting $L=2$ and $L=3$ is functionally equivalent, and often the best choice in terms of \gls{paoi}. The $(4,7)$ system with $L=1$ can almost reach the same performance in the scenario with $\varepsilon=0.1$, but block losses make it significantly worse in all other considered cases. 

On the other hand, the system with $L=\infty$ always performs worse, requiring a far lower load to achieve its optimal \gls{paoi} and showing a gap with the $L=2$ and $L=3$ systems even then. This is to be expected, as it fits the conclusions from~\cite{talak2021age}: while $L=\infty$ and a lower load are the best choices if the system is aimed at having a low latency and reliability, there is an inevitable trade-off with the \gls{paoi}. In the same way, optimizing the \gls{paoi} will lead to a high block loss, even with a longer queues, as frequent preemption remains a very effective method to minimize the age in a system. Interestingly, the \gls{paoi} follows an irregular curve, particularly when $L=1$ in the $(4,5)$ system shown in Fig.~\ref{fig:age_45_01_perc}-\subref{fig:age_45_02_perc}: this might be due to the irregularities we observe in the \gls{paoi} \gls{cdf}.

\subsection{Unbalanced Scenario}

We can also analyze what happens when there are some faster, or slower, paths. In this section, we consider a faster connection, which has 3 paths with $\mu_j=1.25$, while all the others have $\mu_j=1$, e.g., $\bm{\mu}=(1.25,1.25,1.25,1,1)$ if $N=5$. We also look at a slower one, which has 3 paths with $\mu_j=0.75$, while all the others have $\mu_j=1$, e.g., $\bm{\mu}=(0.75,0.75,0.75,1,1)$ if $N=5$. As we discussed above, a shorter queue can represent both an advantage and a disadvantage if the load on the system changes: if there are enough redundant paths, systems with a lower $L$ can deal with a higher load by dropping stragglers, but this can quickly lead to a far lower decoding probability if the redundancy is too low. 

\begin{figure}[t]
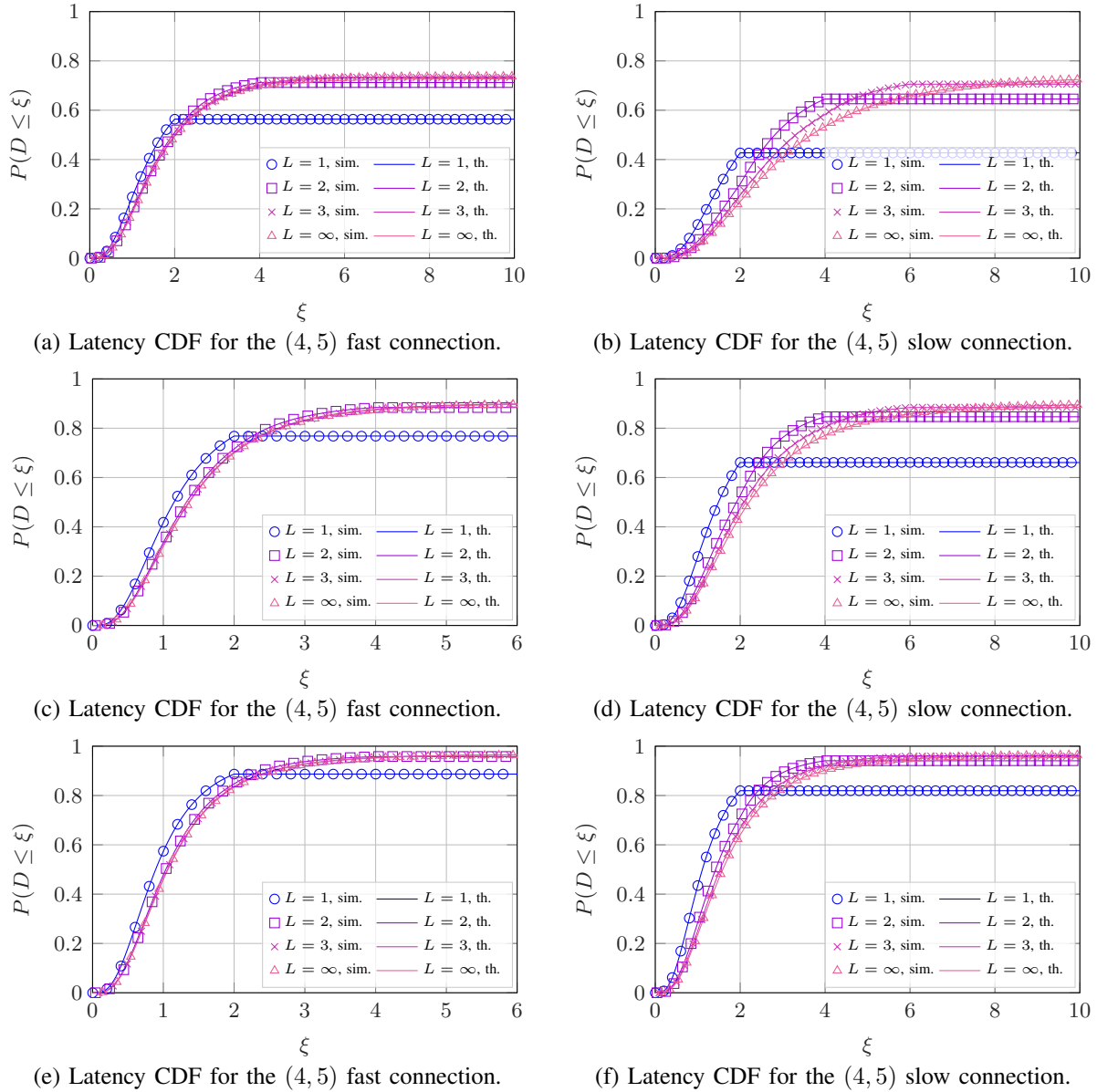

    \centering
	\begin{subfigure}[b]{.49\linewidth}
	    \centering
        \ifdefined\pdffig
            \includegraphics[width=\linewidth]{tikz/paper-figure35}
        \else
            \input{fig/lat_up_45_e_0.2.tex}
        \fi        
        \caption{Latency \gls{cdf} for the $(4,5)$ fast connection.}
        \label{fig:lat_45_up}
    \end{subfigure}	
	\begin{subfigure}[b]{.49\linewidth}
	    \centering
        \ifdefined\pdffig
            \includegraphics[width=\linewidth]{tikz/paper-figure36}
        \else
            \input{fig/lat_down_45_e_0.2.tex}
        \fi        
        \caption{Latency \gls{cdf} for the $(4,5)$ slow connection.}
        \label{fig:lat_45_dn}
    \end{subfigure}	
	\begin{subfigure}[b]{.49\linewidth}
	    \centering
        \ifdefined\pdffig
            \includegraphics[width=\linewidth]{tikz/paper-figure37}
        \else
            \input{fig/lat_up_46_e_0.2.tex}
        \fi        
        \caption{Latency \gls{cdf} for the $(4,5)$ fast connection.}
        \label{fig:lat_46_up}
    \end{subfigure}	
    \begin{subfigure}[b]{.49\linewidth}
	    \centering
        \ifdefined\pdffig
            \includegraphics[width=\linewidth]{tikz/paper-figure38}
        \else
            \input{fig/lat_down_46_e_0.2.tex}
        \fi        
        \caption{Latency \gls{cdf} for the $(4,5)$ slow connection.}
        \label{fig:lat_46_dn}
    \end{subfigure}
    \begin{subfigure}[b]{.49\linewidth}
	    \centering
        \ifdefined\pdffig
            \includegraphics[width=\linewidth]{tikz/paper-figure39}
        \else
            \input{fig/lat_up_47_e_0.2.tex}
        \fi        
        \caption{Latency \gls{cdf} for the $(4,5)$ fast connection.}
        \label{fig:lat_47_up}
    \end{subfigure}
    \begin{subfigure}[b]{.49\linewidth}
	    \centering
        \ifdefined\pdffig
            \includegraphics[width=\linewidth]{tikz/paper-figure40}
        \else
            \input{fig/lat_down_47_e_0.2.tex}
        \fi        
        \caption{Latency \gls{cdf} for the $(4,5)$ slow connection.}
        \label{fig:lat_47_dn}
    \end{subfigure}
     \caption{Latency \gls{cdf} for different queue sizes and codes with $\tau=2$ and $\varepsilon=0.2$. Three out of $N$ paths have a faster (left) or slower (right) service time.\vspace{-0.6cm}}
 \label{fig:lat_CDF_unb}
\end{figure}

This phenomenon is evident in Fig.~\ref{fig:lat_CDF_unb}, which shows the latency \gls{cdf} for various configurations with $\varepsilon=0.2$ and $\tau=2$. The faster connection, shown on the left, reduces the advantage of the $L=1$ system: in this case, and particularly in the $(4,5)$ scenario, the systems with longer queues are able to exploit the faster paths to reduce their latency, while the $L=1$ system still suffers from dropped and erased packets. The opposite happens in the slower connection, shown on the right: in this case, the increased reliability of systems with larger values of $L$ is significantly more expensive in terms of latency, and the blocks that are decoded correctly for the $L=1$ and $L=2$ systems are significantly faster.

\begin{figure}
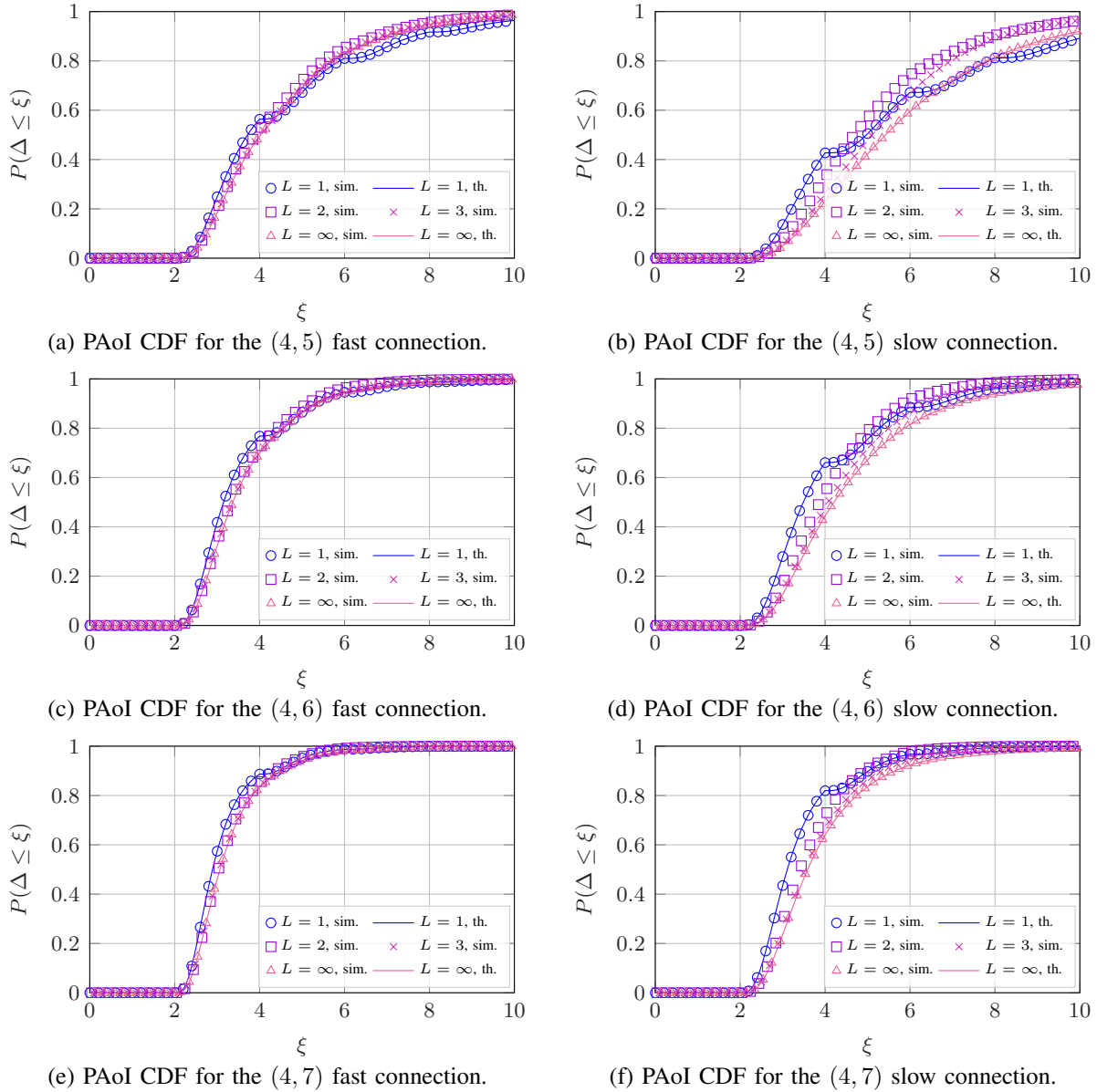

    \centering
	\begin{subfigure}[b]{.49\linewidth}
	    \centering
        \ifdefined\pdffig
            \includegraphics[width=\linewidth]{tikz/paper-figure41}
        \else
            \input{fig/age_up_45_e_0.2.tex}
        \fi        
        \caption{\gls{paoi} \gls{cdf} for the $(4,5)$ fast connection.}
        \label{fig:age_45_up}
    \end{subfigure}	
	\begin{subfigure}[b]{.49\linewidth}
	    \centering
        \ifdefined\pdffig
            \includegraphics[width=\linewidth]{tikz/paper-figure42}
        \else
            \input{fig/age_down_45_e_0.2.tex}
        \fi        
        \caption{\gls{paoi} \gls{cdf} for the $(4,5)$ slow connection.}
        \label{fig:age_45_dn}
    \end{subfigure}	
	\begin{subfigure}[b]{.49\linewidth}
	    \centering
        \ifdefined\pdffig
            \includegraphics[width=\linewidth]{tikz/paper-figure43}
        \else
            \input{fig/age_up_46_e_0.2.tex}
        \fi        
        \caption{\gls{paoi} \gls{cdf} for the $(4,6)$ fast connection.}
        \label{fig:age_46_up}
    \end{subfigure}	
    \begin{subfigure}[b]{.49\linewidth}
	    \centering
        \ifdefined\pdffig
            \includegraphics[width=\linewidth]{tikz/paper-figure44}
        \else
            \input{fig/age_down_46_e_0.2.tex}
        \fi        
        \caption{\gls{paoi} \gls{cdf} for the $(4,6)$ slow connection.}
        \label{fig:age_46_dn}
    \end{subfigure}
    \begin{subfigure}[b]{.49\linewidth}
	    \centering
        \ifdefined\pdffig
            \includegraphics[width=\linewidth]{tikz/paper-figure45}
        \else
            \input{fig/age_up_47_e_0.2.tex}
        \fi        
        \caption{\gls{paoi} \gls{cdf} for the $(4,7)$ fast connection.}
        \label{fig:age_47_up}
    \end{subfigure}
    \begin{subfigure}[b]{.49\linewidth}
	    \centering
        \ifdefined\pdffig
            \includegraphics[width=\linewidth]{tikz/paper-figure46}
        \else
            \input{fig/age_down_47_e_0.2.tex}
        \fi        
        \caption{\gls{paoi} \gls{cdf} for the $(4,7)$ slow connection.}
        \label{fig:age_47_dn}
    \end{subfigure}
     \caption{\gls{paoi} \gls{cdf} for different queue sizes and codes with $\tau=2$ and $\varepsilon=0.2$.}\vspace{-0.6cm}
 \label{fig:age_CDF_unb}
\end{figure}

We can observe a similar pattern with regard to the \gls{paoi} in Fig.~\ref{fig:age_CDF_unb}: in particular, the benefits from having a longer queue in the $(4,5)$ system are more evident in the faster connection, shown on the left side. In this case, the higher percentiles of the distribution are far better for $L>1$, as dropping packets becomes more of a liability than an advantage for the $L=1$ system. However, even adding one path is enough to significantly reduce this disadvantage, as the figure shows. 

On the other hand, there is an interesting phenomenon on the slower connection, shown on the right: setting a too large value of $L$ leads to significantly lower performance in terms of \gls{paoi}, as the queues on the slower paths can become very long, but setting $L=1$ can \emph{also} lead to worse performance, as the dropping probability on those same paths becomes very high. In this case, setting $L=2$ or $L=3$ seems to be the best choice, as those systems avoid dropping packets too soon, but also never build up a large queue. In general, setting the appropriate queue length is a non-trivial optimization, particularly when redundancy is limited: this is a stark difference from the \gls{aoi} optimization on simpler queuing networks with Markovian service, in which preemption is always the best choice.

\subsection{Redundancy Optimization}

In the following, we consider a system with $N=6$, $\varepsilon=0.1$, and $\mu=1$: in this case, $K$ is not fixed, but we consider a data block payload with a set size $M$. We can then choose the value of $K$, and consequently the amount of redundancy in the multipath transmission. However, as the payload size is fixed, adding more redundancy means transmitting bigger packets on each path and increasing the system load. If we consider the average service time to be a linear function of the packet size, we get:
\begin{equation}
 \mu'_j=\frac{M\mu_j}{K}.
\end{equation}
The latency and \gls{paoi} \glspl{pdf} can then be computed simply by substituting $\bm{\mu}'$ for $\bm{\mu}$. We can look at the offered traffic $G$, which corresponds to the system load on the slowest path if the blocks are transmitted without any redundancy, as an independent variable:
\begin{equation}
 G=\frac{M}{N\tau\min_{j\in\{1,\ldots,N\}}\mu_j}.
\end{equation}

\begin{figure}[t]
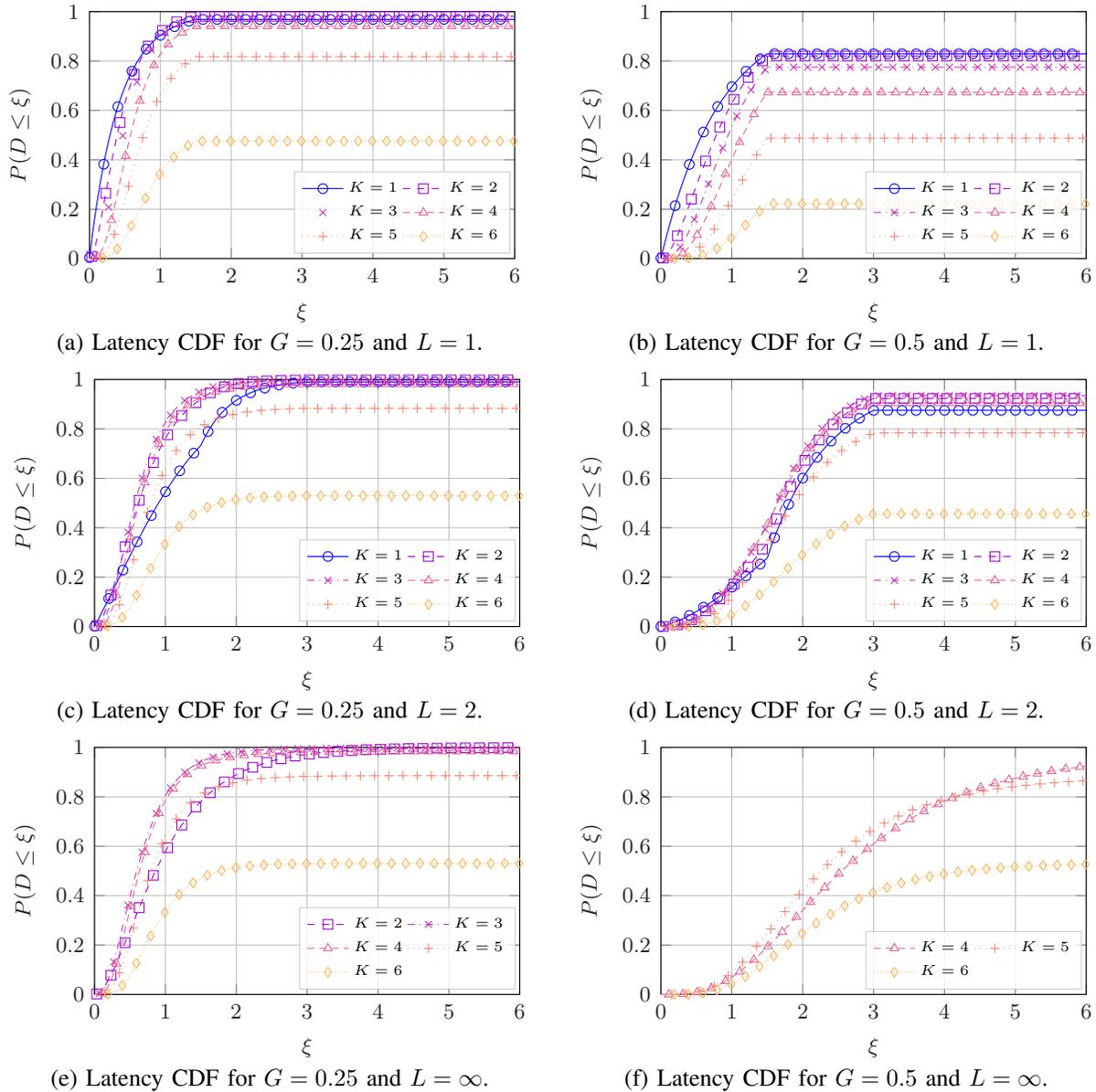

    \centering
	\begin{subfigure}[b]{.49\linewidth}
	    \centering
        \ifdefined\pdffig
            \includegraphics[width=\linewidth]{tikz/paper-figure47}
        \else
            \input{fig/lat_L_1_N_6_M_2.25.tex}
        \fi        
        \caption{Latency \gls{cdf} for $G=0.25$ and $L=1$.}
        \label{fig:lat_red_L1_025}
    \end{subfigure}	
	\begin{subfigure}[b]{.49\linewidth}
	    \centering
        \ifdefined\pdffig
            \includegraphics[width=\linewidth]{tikz/paper-figure48}
        \else
            \input{fig/lat_L_1_N_6_M_4.5.tex}
        \fi        
         \caption{Latency \gls{cdf} for $G=0.5$ and $L=1$.}
        \label{fig:lat_red_L1_075}
    \end{subfigure}	
	\begin{subfigure}[b]{.49\linewidth}
	    \centering
        \ifdefined\pdffig
            \includegraphics[width=\linewidth]{tikz/paper-figure49}
        \else
            \input{fig/lat_L_2_N_6_M_2.25.tex}
        \fi        
         \caption{Latency \gls{cdf} for $G=0.25$ and $L=2$.}
        \label{fig:lat_red_L2_025}
    \end{subfigure}	
    \begin{subfigure}[b]{.49\linewidth}
	    \centering
        \ifdefined\pdffig
            \includegraphics[width=\linewidth]{tikz/paper-figure50}
        \else
            \input{fig/lat_L_2_N_6_M_4.5.tex}
        \fi        
         \caption{Latency \gls{cdf} for $G=0.5$ and $L=2$.}
        \label{fig:lat_red_L2_075}
    \end{subfigure}
    \begin{subfigure}[b]{.49\linewidth}
	    \centering
        \ifdefined\pdffig
            \includegraphics[width=\linewidth]{tikz/paper-figure51}
        \else
            \input{fig/lat_L_inf_N_6_M_2.25.tex}
        \fi        
        \caption{Latency \gls{cdf} for $G=0.25$ and $L=\infty$.}
        \label{fig:lat_red_Linf_025}
    \end{subfigure}
    \begin{subfigure}[b]{.49\linewidth}
	    \centering
        \ifdefined\pdffig
            \includegraphics[width=\linewidth]{tikz/paper-figure52}
        \else
            \input{fig/lat_L_inf_N_6_M_4.5.tex}
        \fi        
        \caption{Latency \gls{cdf} for $G=0.5$ and $L=\infty$.}
        \label{fig:lat_red_Linf_075}
    \end{subfigure}
     \caption{Latency \gls{cdf} for different coding schemes with $N=6$, $\mu=1$ and $\varepsilon=0.1$.}\vspace{-0.6cm}
 \label{fig:lat_red}
\end{figure}

We then examine the latency of the various systems for $G=0.25$ and $G=0.5$, fixing $\tau=1.5$. Fig.~\ref{fig:lat_red} shows the \gls{cdf} of the latency for different queue lengths and codes. The case with low offered traffic, shown on the left, shows a clear advantage for the $L=1$ system: since the offered traffic is low, the system remains stable even using a low coding rate, with the advantage of fully exploiting preemption and protecting blocks from channel erasures: the system that performs best, with $K=1$, has a coding rate $R_c=\frac{K}{N}=\frac{1}{6}$. This is not true for the systems with $L=2$ and $L=\infty$, which seem to do better for $K=3$, but these systems have a significantly higher overall latency. The reader should note that the system with $L=\infty$ and $K=1$ is unstable, as $\frac{1}{\tau}>\mu'$ and there is no preemption, and the latency explodes; this is not a problem in finite queue systems, which can deal with this case by dropping packets when the queue becomes full.

\begin{figure}[t]
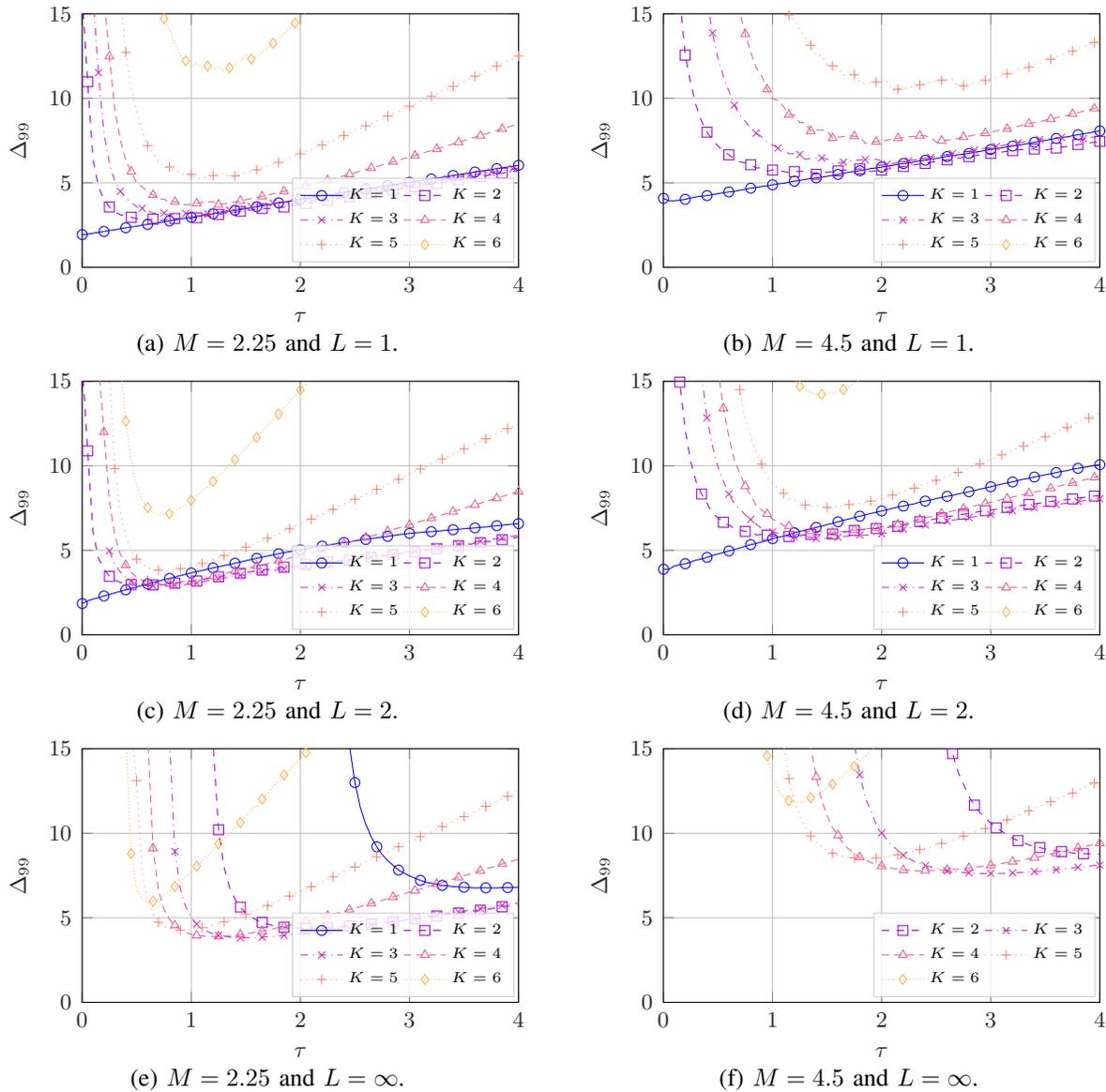

    \centering
	\begin{subfigure}[b]{.49\linewidth}
	    \centering
        \ifdefined\pdffig
            \includegraphics[width=\linewidth]{tikz/paper-figure53}
        \else
            \begin{tikzpicture}

\begin{axis}[%
width=\sfwidth,
height=\sfheight,
xmin=0,
xmax=4,
ymin=0,
ymax=15,
axis background/.style={fill=white},
xlabel style={font=\footnotesize\color{white!15!black}},
xlabel={$\tau$},
ylabel near ticks,
ylabel style={font=\footnotesize\color{white!15!black}},
xticklabel style={font=\footnotesize\color{white!15!black}},
yticklabel style={font=\footnotesize\color{white!15!black}},
ylabel={$\Delta_{99}$},
axis background/.style={fill=white},
xmajorgrids,
ymajorgrids,
legend style={font=\tiny, at={(0.99,0.02)}, anchor=south east, legend columns=2,legend cell align=left, align=left,fill opacity=0.8, draw opacity=1, text opacity=1, draw=white!80!black}
]
\addplot [color=orange_D, mark=o, mark repeat = 4, mark phase=1]
  table[row sep=crcr]{%
0.001   1.934\\
0.05	1.968\\
0.1	2.02\\
0.15	2.066\\
0.2	2.125\\
0.25	2.186\\
0.3	2.22\\
0.35	2.286\\
0.4	2.343\\
0.45	2.389\\
0.5	2.433\\
0.55	2.495\\
0.6	2.54\\
0.65	2.601\\
0.7	2.643\\
0.75	2.7\\
0.8	2.749\\
0.85	2.808\\
0.9	2.861\\
0.95	2.913\\
1	2.963\\
1.05	3.012\\
1.1	3.069\\
1.15	3.114\\
1.2	3.175\\
1.25	3.228\\
1.3	3.274\\
1.35	3.327\\
1.4	3.38\\
1.45	3.425\\
1.5	3.479\\
1.55	3.537\\
1.6	3.588\\
1.65	3.644\\
1.7	3.705\\
1.75	3.751\\
1.8	3.814\\
1.85	3.86\\
1.9	3.917\\
1.95	3.974\\
2	4.03\\
2.05	4.097\\
2.1	4.131\\
2.15	4.187\\
2.2	4.234\\
2.25	4.283\\
2.3	4.337\\
2.35	4.381\\
2.4	4.437\\
2.45	4.485\\
2.5	4.545\\
2.55	4.581\\
2.6	4.639\\
2.65	4.691\\
2.7	4.736\\
2.75	4.785\\
2.8	4.832\\
2.85	4.885\\
2.9	4.936\\
2.95	4.982\\
3	5.035\\
3.05	5.08\\
3.1	5.137\\
3.15	5.184\\
3.2	5.229\\
3.25	5.284\\
3.3	5.329\\
3.35	5.378\\
3.4	5.44\\
3.45	5.485\\
3.5	5.53\\
3.55	5.583\\
3.6	5.64\\
3.65	5.696\\
3.7	5.735\\
3.75	5.785\\
3.8	5.838\\
3.85	5.885\\
3.9	5.935\\
3.95	5.99\\
4	6.038\\
};
\addlegendentry{$K=1$}

\addplot [color=cyan, dashed, mark=square, mark repeat = 4, mark options={solid},mark phase=1]
  table[row sep=crcr]{%
0.001   16\\
0.05	10.981\\
0.1	6.357\\
0.15	4.786\\
0.2	3.995\\
0.25	3.574\\
0.3	3.294\\
0.35	3.111\\
0.4	3.026\\
0.45	2.962\\
0.5	2.881\\
0.55	2.843\\
0.6	2.826\\
0.65	2.852\\
0.7	2.766\\
0.75	2.803\\
0.8	2.852\\
0.85	2.889\\
0.9	2.894\\
0.95	2.832\\
1	2.88\\
1.05	2.933\\
1.1	2.981\\
1.15	3.034\\
1.2	3.088\\
1.25	3.13\\
1.3	3.184\\
1.35	3.228\\
1.4	3.273\\
1.45	3.318\\
1.5	3.372\\
1.55	3.403\\
1.6	3.446\\
1.65	3.485\\
1.7	3.489\\
1.75	3.475\\
1.8	3.528\\
1.85	3.573\\
1.9	3.626\\
1.95	3.668\\
2	3.729\\
2.05	3.773\\
2.1	3.829\\
2.15	3.873\\
2.2	3.929\\
2.25	3.975\\
2.3	4.023\\
2.35	4.079\\
2.4	4.121\\
2.45	4.17\\
2.5	4.221\\
2.55	4.272\\
2.6	4.331\\
2.65	4.374\\
2.7	4.428\\
2.75	4.476\\
2.8	4.535\\
2.85	4.575\\
2.9	4.624\\
2.95	4.675\\
3	4.726\\
3.05	4.771\\
3.1	4.821\\
3.15	4.88\\
3.2	4.925\\
3.25	4.979\\
3.3	5.027\\
3.35	5.077\\
3.4	5.127\\
3.45	5.174\\
3.5	5.226\\
3.55	5.276\\
3.6	5.33\\
3.65	5.384\\
3.7	5.426\\
3.75	5.474\\
3.8	5.533\\
3.85	5.574\\
3.9	5.628\\
3.95	5.673\\
4	5.726\\
};
\addlegendentry{$K=2$}

\addplot [color=green_D, dashdotted, mark=x, mark repeat=4, mark options={solid}]
  table[row sep=crcr]{%
0.05	16\\
0.1	16\\
0.15	11.516\\
0.2	7.916\\
0.25	6.163\\
0.3	5.099\\
0.35	4.503\\
0.4	4.106\\
0.45	3.83\\
0.5	3.583\\
0.55	3.491\\
0.6	3.359\\
0.65	3.194\\
0.7	3.234\\
0.75	3.211\\
0.8	3.09\\
0.85	3.143\\
0.9	3.171\\
0.95	3.194\\
1	3.177\\
1.05	3.072\\
1.1	3.127\\
1.15	3.175\\
1.2	3.233\\
1.25	3.279\\
1.3	3.334\\
1.35	3.395\\
1.4	3.446\\
1.45	3.493\\
1.5	3.544\\
1.55	3.606\\
1.6	3.653\\
1.65	3.701\\
1.7	3.751\\
1.75	3.808\\
1.8	3.843\\
1.85	3.852\\
1.9	3.781\\
1.95	3.832\\
2	3.881\\
2.05	3.931\\
2.1	3.971\\
2.15	4.038\\
2.2	4.083\\
2.25	4.139\\
2.3	4.186\\
2.35	4.244\\
2.4	4.282\\
2.45	4.334\\
2.5	4.381\\
2.55	4.436\\
2.6	4.479\\
2.65	4.537\\
2.7	4.575\\
2.75	4.639\\
2.8	4.679\\
2.85	4.733\\
2.9	4.788\\
2.95	4.839\\
3	4.88\\
3.05	4.935\\
3.1	4.978\\
3.15	5.035\\
3.2	5.083\\
3.25	5.135\\
3.3	5.185\\
3.35	5.234\\
3.4	5.285\\
3.45	5.338\\
3.5	5.379\\
3.55	5.43\\
3.6	5.486\\
3.65	5.526\\
3.7	5.582\\
3.75	5.634\\
3.8	5.682\\
3.85	5.73\\
3.9	5.783\\
3.95	5.834\\
4	5.873\\
};
\addlegendentry{$K=3$}

\addplot [color=violet, densely dashed, mark=triangle, mark repeat=4, mark options={solid}]
table[row sep=crcr]{%
0.05	16\\
0.1	16\\
0.15	16\\
0.2	16\\
0.25	12.487\\
0.3	9.293\\
0.35	7.573\\
0.4	6.337\\
0.45	5.668\\
0.5	4.996\\
0.55	4.76\\
0.6	4.505\\
0.65	4.283\\
0.7	4.06\\
0.75	4.042\\
0.8	3.853\\
0.85	3.882\\
0.9	3.857\\
0.95	3.693\\
1	3.742\\
1.05	3.784\\
1.1	3.847\\
1.15	3.878\\
1.2	3.877\\
1.25	3.713\\
1.3	3.759\\
1.35	3.819\\
1.4	3.884\\
1.45	3.94\\
1.5	4.003\\
1.55	4.073\\
1.6	4.14\\
1.65	4.199\\
1.7	4.278\\
1.75	4.353\\
1.8	4.428\\
1.85	4.501\\
1.9	4.575\\
1.95	4.659\\
2	4.736\\
2.05	4.815\\
2.1	4.908\\
2.15	4.99\\
2.2	5.073\\
2.25	5.153\\
2.3	5.236\\
2.35	5.338\\
2.4	5.415\\
2.45	5.508\\
2.5	5.599\\
2.55	5.698\\
2.6	5.781\\
2.65	5.878\\
2.7	5.96\\
2.75	6.058\\
2.8	6.154\\
2.85	6.247\\
2.9	6.339\\
2.95	6.439\\
3	6.517\\
3.05	6.619\\
3.1	6.724\\
3.15	6.814\\
3.2	6.914\\
3.25	7.007\\
3.3	7.108\\
3.35	7.203\\
3.4	7.302\\
3.45	7.399\\
3.5	7.5\\
3.55	7.602\\
3.6	7.7\\
3.65	7.79\\
3.7	7.893\\
3.75	7.99\\
3.8	8.085\\
3.85	8.191\\
3.9	8.291\\
3.95	8.383\\
4	8.489\\
};
\addlegendentry{$K=4$}

\addplot [color=blue, dotted, mark=+, mark repeat=4, mark options={solid}]
  table[row sep=crcr]{%
0.05	16\\
0.1	16\\
0.15	16\\
0.2	16\\
0.25	16\\
0.3	16\\
0.35	16\\
0.4	12.728\\
0.45	10.611\\
0.5	9.208\\
0.55	8.035\\
0.6	7.198\\
0.65	6.828\\
0.7	6.268\\
0.75	5.955\\
0.8	5.938\\
0.85	5.71\\
0.9	5.608\\
0.95	5.456\\
1	5.501\\
1.05	5.222\\
1.1	5.283\\
1.15	5.363\\
1.2	5.434\\
1.25	5.508\\
1.3	5.555\\
1.35	5.342\\
1.4	5.411\\
1.45	5.494\\
1.5	5.587\\
1.55	5.666\\
1.6	5.771\\
1.65	5.873\\
1.7	5.986\\
1.75	6.104\\
1.8	6.21\\
1.85	6.334\\
1.9	6.457\\
1.95	6.587\\
2	6.713\\
2.05	6.849\\
2.1	6.979\\
2.15	7.112\\
2.2	7.237\\
2.25	7.381\\
2.3	7.508\\
2.35	7.666\\
2.4	7.798\\
2.45	7.945\\
2.5	8.078\\
2.55	8.21\\
2.6	8.37\\
2.65	8.49\\
2.7	8.647\\
2.75	8.779\\
2.8	8.947\\
2.85	9.096\\
2.9	9.231\\
2.95	9.385\\
3	9.53\\
3.05	9.663\\
3.1	9.826\\
3.15	9.966\\
3.2	10.117\\
3.25	10.265\\
3.3	10.412\\
3.35	10.551\\
3.4	10.704\\
3.45	10.85\\
3.5	11.005\\
3.55	11.148\\
3.6	11.304\\
3.65	11.438\\
3.7	11.601\\
3.75	11.745\\
3.8	11.906\\
3.85	12.051\\
3.9	12.199\\
3.95	12.344\\
4	12.507\\
};
\addlegendentry{$K=5$}

\addplot [color=red, densely dotted, mark=diamond, mark repeat=4, mark options={solid}]
  table[row sep=crcr]{%
0.5	16\\
0.55	16\\
0.6	16\\
0.65	16\\
0.7	15.908\\
0.75	14.705\\
0.8	14.066\\
0.85	13.26\\
0.9	12.532\\
0.95	12.221\\
1	11.887\\
1.05	12.025\\
1.1	11.795\\
1.15	11.91\\
1.2	11.696\\
1.25	11.86\\
1.3	11.608\\
1.35	11.813\\
1.4	12.012\\
1.45	12.238\\
1.5	12.504\\
1.55	12.327\\
1.6	12.493\\
1.65	12.739\\
1.7	12.949\\
1.75	13.265\\
1.8	13.529\\
1.85	13.833\\
1.9	14.139\\
1.95	14.45\\
2	14.774\\
2.05	15.048\\
2.1	15.417\\
2.15	15.735\\
2.2	16\\
2.25	16\\
2.3	16\\
2.35	16\\
2.4	16\\
2.45	16\\
2.5	16\\
2.55	16\\
2.6	16\\
2.65	16\\
2.7	16\\
2.75	16\\
2.8	16\\
2.85	16\\
2.9	16\\
2.95	16\\
3	16\\
3.05	16\\
3.1	16\\
3.15	16\\
3.2	16\\
3.25	16\\
3.3	16\\
3.35	16\\
3.4	16\\
3.45	16\\
3.5	16\\
3.55	16\\
3.6	16\\
3.65	16\\
3.7	16\\
3.75	16\\
3.8	16\\
3.85	16\\
3.9	16\\
3.95	16\\
4	16\\
};
\addlegendentry{$K=6$}

\end{axis}
\end{tikzpicture}%
        \fi        
        \caption{$M=2.25$ and $L=1$.}
        \label{fig:age_red_L1_025}
    \end{subfigure}	
	\begin{subfigure}[b]{.49\linewidth}
	    \centering
        \ifdefined\pdffig
            \includegraphics[width=\linewidth]{tikz/paper-figure54}
        \else
            \begin{tikzpicture}

\begin{axis}[%
width=\sfwidth,
height=\sfheight,
xmin=0,
xmax=4,
ymin=0,
ymax=15,
axis background/.style={fill=white},
xlabel style={font=\footnotesize\color{white!15!black}},
xlabel={$\tau$},
ylabel near ticks,
ylabel style={font=\footnotesize\color{white!15!black}},
xticklabel style={font=\footnotesize\color{white!15!black}},
yticklabel style={font=\footnotesize\color{white!15!black}},
ylabel={$\Delta_{99}$},
axis background/.style={fill=white},
xmajorgrids,
ymajorgrids,
legend style={font=\tiny, at={(0.99,0.02)}, anchor=south east, legend columns=2,legend cell align=left, align=left,fill opacity=0.8, draw opacity=1, text opacity=1, draw=white!80!black}
]
\addplot [color=orange_D, mark=o, mark repeat = 4, mark phase=1]
  table[row sep=crcr]{%
0.001	4.081\\
0.05	3.942\\
0.1	3.939\\
0.15	3.976\\
0.2	4.025\\
0.25	4.102\\
0.3	4.173\\
0.35	4.21\\
0.4	4.247\\
0.45	4.306\\
0.5	4.364\\
0.55	4.414\\
0.6	4.465\\
0.65	4.493\\
0.7	4.565\\
0.75	4.619\\
0.8	4.685\\
0.85	4.727\\
0.9	4.788\\
0.95	4.815\\
1	4.886\\
1.05	4.939\\
1.1	4.97\\
1.15	5.033\\
1.2	5.1\\
1.25	5.148\\
1.3	5.201\\
1.35	5.255\\
1.4	5.293\\
1.45	5.339\\
1.5	5.386\\
1.55	5.449\\
1.6	5.499\\
1.65	5.576\\
1.7	5.626\\
1.75	5.669\\
1.8	5.718\\
1.85	5.769\\
1.9	5.851\\
1.95	5.88\\
2	5.936\\
2.05	5.989\\
2.1	6.035\\
2.15	6.099\\
2.2	6.145\\
2.25	6.174\\
2.3	6.242\\
2.35	6.299\\
2.4	6.338\\
2.45	6.396\\
2.5	6.437\\
2.55	6.504\\
2.6	6.554\\
2.65	6.602\\
2.7	6.658\\
2.75	6.692\\
2.8	6.74\\
2.85	6.812\\
2.9	6.86\\
2.95	6.919\\
3	6.971\\
3.05	7.02\\
3.1	7.069\\
3.15	7.129\\
3.2	7.175\\
3.25	7.234\\
3.3	7.275\\
3.35	7.334\\
3.4	7.399\\
3.45	7.456\\
3.5	7.503\\
3.55	7.571\\
3.6	7.622\\
3.65	7.688\\
3.7	7.716\\
3.75	7.783\\
3.8	7.848\\
3.85	7.9\\
3.9	7.946\\
3.95	8.021\\
4	8.069\\
};
\addlegendentry{$K=1$}

\addplot [color=cyan, dashed, mark=square, mark repeat = 4, mark options={solid}]
  table[row sep=crcr]{%
0.05	16\\
0.1	16\\
0.15	15.978\\
0.2	12.557\\
0.25	10.91\\
0.3	9.577\\
0.35	8.644\\
0.4	7.994\\
0.45	7.55\\
0.5	7.174\\
0.55	6.9\\
0.6	6.657\\
0.65	6.402\\
0.7	6.24\\
0.75	6.186\\
0.8	6.071\\
0.85	5.883\\
0.9	5.886\\
0.95	5.7\\
1	5.755\\
1.05	5.755\\
1.1	5.676\\
1.15	5.609\\
1.2	5.66\\
1.25	5.681\\
1.3	5.683\\
1.35	5.649\\
1.4	5.51\\
1.45	5.586\\
1.5	5.623\\
1.55	5.665\\
1.6	5.706\\
1.65	5.754\\
1.7	5.771\\
1.75	5.799\\
1.8	5.824\\
1.85	5.798\\
1.9	5.668\\
1.95	5.714\\
2	5.758\\
2.05	5.816\\
2.1	5.867\\
2.15	5.91\\
2.2	5.968\\
2.25	6.019\\
2.3	6.063\\
2.35	6.112\\
2.4	6.152\\
2.45	6.216\\
2.5	6.259\\
2.55	6.321\\
2.6	6.36\\
2.65	6.409\\
2.7	6.452\\
2.75	6.509\\
2.8	6.552\\
2.85	6.603\\
2.9	6.642\\
2.95	6.697\\
3	6.743\\
3.05	6.782\\
3.1	6.816\\
3.15	6.863\\
3.2	6.901\\
3.25	6.944\\
3.3	6.969\\
3.35	6.992\\
3.4	6.999\\
3.45	6.899\\
3.5	6.95\\
3.55	7.001\\
3.6	7.06\\
3.65	7.118\\
3.7	7.151\\
3.75	7.198\\
3.8	7.267\\
3.85	7.298\\
3.9	7.356\\
3.95	7.397\\
4	7.452\\
};
\addlegendentry{$K=2$}

\addplot [color=green_D, dashdotted, mark=x, mark repeat=4, mark options={solid}]
  table[row sep=crcr]{%
0.4	15.905\\
0.45	13.866\\
0.5	12.366\\
0.55	11.31\\
0.6	10.425\\
0.65	9.594\\
0.7	9.001\\
0.75	8.642\\
0.8	8.21\\
0.85	7.939\\
0.9	7.719\\
0.95	7.384\\
1	7.239\\
1.05	7.054\\
1.1	6.981\\
1.15	6.719\\
1.2	6.727\\
1.25	6.678\\
1.3	6.39\\
1.35	6.444\\
1.4	6.454\\
1.45	6.468\\
1.5	6.413\\
1.55	6.146\\
1.6	6.187\\
1.65	6.224\\
1.7	6.277\\
1.75	6.305\\
1.8	6.362\\
1.85	6.374\\
1.9	6.39\\
1.95	6.393\\
2	6.298\\
2.05	6.106\\
2.1	6.141\\
2.15	6.203\\
2.2	6.248\\
2.25	6.308\\
2.3	6.358\\
2.35	6.407\\
2.4	6.458\\
2.45	6.506\\
2.5	6.571\\
2.55	6.616\\
2.6	6.66\\
2.65	6.737\\
2.7	6.777\\
2.75	6.83\\
2.8	6.876\\
2.85	6.937\\
2.9	6.982\\
2.95	7.053\\
3	7.097\\
3.05	7.157\\
3.1	7.198\\
3.15	7.263\\
3.2	7.31\\
3.25	7.379\\
3.3	7.416\\
3.35	7.454\\
3.4	7.522\\
3.45	7.568\\
3.5	7.604\\
3.55	7.643\\
3.6	7.674\\
3.65	7.698\\
3.7	7.709\\
3.75	7.69\\
3.8	7.563\\
3.85	7.611\\
3.9	7.666\\
3.95	7.718\\
4	7.745\\
};
\addlegendentry{$K=3$}

\addplot [color=violet, densely dashed, mark=triangle, mark repeat=4, mark options={solid}]
table[row sep=crcr]{%
0.5	16\\
0.55	16\\
0.6	16\\
0.65	16\\
0.7	15.137\\
0.75	13.806\\
0.8	12.715\\
0.85	11.815\\
0.9	11.328\\
0.95	10.538\\
1	9.982\\
1.05	9.88\\
1.1	9.509\\
1.15	9.027\\
1.2	9.014\\
1.25	8.556\\
1.3	8.56\\
1.35	8.063\\
1.4	8.116\\
1.45	8.131\\
1.5	8.079\\
1.55	7.648\\
1.6	7.699\\
1.65	7.739\\
1.7	7.776\\
1.75	7.754\\
1.8	7.729\\
1.85	7.34\\
1.9	7.381\\
1.95	7.428\\
2	7.458\\
2.05	7.527\\
2.1	7.59\\
2.15	7.638\\
2.2	7.678\\
2.25	7.72\\
2.3	7.744\\
2.35	7.788\\
2.4	7.778\\
2.45	7.677\\
2.5	7.41\\
2.55	7.489\\
2.6	7.518\\
2.65	7.58\\
2.7	7.625\\
2.75	7.71\\
2.8	7.763\\
2.85	7.822\\
2.9	7.868\\
2.95	7.937\\
3	7.982\\
3.05	8.067\\
3.1	8.137\\
3.15	8.195\\
3.2	8.273\\
3.25	8.349\\
3.3	8.421\\
3.35	8.497\\
3.4	8.562\\
3.45	8.619\\
3.5	8.702\\
3.55	8.778\\
3.6	8.844\\
3.65	8.922\\
3.7	8.991\\
3.75	9.079\\
3.8	9.153\\
3.85	9.245\\
3.9	9.313\\
3.95	9.399\\
4	9.5\\
};
\addlegendentry{$K=4$}

\addplot [color=blue, dotted, mark=+, mark repeat=4, mark options={solid}]
  table[row sep=crcr]{%
0.5	16\\
0.55	16\\
0.6	16\\
0.65	16\\
0.7	16\\
0.75	16\\
0.8	16\\
0.85	16\\
0.9	16\\
0.95	16\\
1	16\\
1.05	16\\
1.1	16\\
1.15	14.93\\
1.2	14.648\\
1.25	14.05\\
1.3	13.641\\
1.35	13.148\\
1.4	12.522\\
1.45	12.541\\
1.5	11.928\\
1.55	11.927\\
1.6	11.87\\
1.65	11.361\\
1.7	11.407\\
1.75	11.398\\
1.8	10.951\\
1.85	10.845\\
1.9	10.904\\
1.95	10.966\\
2	10.994\\
2.05	11.006\\
2.1	10.451\\
2.15	10.531\\
2.2	10.573\\
2.25	10.632\\
2.3	10.712\\
2.35	10.799\\
2.4	10.902\\
2.45	10.951\\
2.5	11.04\\
2.55	11.068\\
2.6	11.112\\
2.65	11.093\\
2.7	10.71\\
2.75	10.738\\
2.8	10.86\\
2.85	10.909\\
2.9	10.958\\
2.95	11.066\\
3	11.161\\
3.05	11.247\\
3.1	11.338\\
3.15	11.453\\
3.2	11.564\\
3.25	11.669\\
3.3	11.746\\
3.35	11.842\\
3.4	11.994\\
3.45	12.088\\
3.5	12.203\\
3.55	12.316\\
3.6	12.461\\
3.65	12.567\\
3.7	12.685\\
3.75	12.795\\
3.8	12.934\\
3.85	13.066\\
3.9	13.172\\
3.95	13.311\\
4	13.456\\
};
\addlegendentry{$K=5$}

\addplot [color=red, densely dotted, mark=diamond, mark repeat=4, mark options={solid}]
  table[row sep=crcr]{%
0.5	16\\
0.55	16\\
0.6	16\\
0.65	16\\
0.7	16\\
0.75	16\\
0.8	16\\
0.85	16\\
0.9	16\\
0.95	16\\
1	16\\
1.05	16\\
1.1	16\\
1.15	16\\
1.2	16\\
1.25	16\\
1.3	16\\
1.35	16\\
1.4	16\\
1.45	16\\
1.5	16\\
1.55	16\\
1.6	16\\
1.65	16\\
1.7	16\\
1.75	16\\
1.8	16\\
1.85	16\\
1.9	16\\
1.95	16\\
2	16\\
2.05	16\\
2.1	16\\
2.15	16\\
2.2	16\\
2.25	16\\
2.3	16\\
2.35	16\\
2.4	16\\
2.45	16\\
2.5	16\\
2.55	16\\
2.6	16\\
2.65	16\\
2.7	16\\
2.75	16\\
2.8	16\\
2.85	16\\
2.9	16\\
2.95	16\\
3	16\\
3.05	16\\
3.1	16\\
3.15	16\\
3.2	16\\
3.25	16\\
3.3	16\\
3.35	16\\
3.4	16\\
3.45	16\\
3.5	16\\
3.55	16\\
3.6	16\\
3.65	16\\
3.7	16\\
3.75	16\\
3.8	16\\
3.85	16\\
3.9	16\\
3.95	16\\
4	16\\
};
\addlegendentry{$K=6$}

\end{axis}
\end{tikzpicture}%
        \fi        
        \caption{$M=4.5$ and $L=1$.}
        \label{fig:age_red_L1_075}
    \end{subfigure}	
	\begin{subfigure}[b]{.49\linewidth}
	    \centering
        \ifdefined\pdffig
            \includegraphics[width=\linewidth]{tikz/paper-figure55}
        \else
            \begin{tikzpicture}

\begin{axis}[%
width=\sfwidth,
height=\sfheight,
xmin=0,
xmax=4,
ymin=0,
ymax=15,
axis background/.style={fill=white},
xlabel style={font=\footnotesize\color{white!15!black}},
xlabel={$\tau$},
ylabel near ticks,
ylabel style={font=\footnotesize\color{white!15!black}},
xticklabel style={font=\footnotesize\color{white!15!black}},
yticklabel style={font=\footnotesize\color{white!15!black}},
ylabel={$\Delta_{99}$},
axis background/.style={fill=white},
xmajorgrids,
ymajorgrids,
legend style={font=\tiny, at={(0.99,0.02)}, anchor=south east, legend columns=2,legend cell align=left, align=left,fill opacity=0.8, draw opacity=1, text opacity=1, draw=white!80!black}
]
\addplot [color=orange_D, mark=o, mark repeat = 4, mark phase=1]
  table[row sep=crcr]{%
0.001   1.851\\
0.05	2.01\\
0.1	2.115\\
0.15	2.197\\
0.2	2.297\\
0.25	2.391\\
0.3	2.484\\
0.35	2.569\\
0.4	2.662\\
0.45	2.748\\
0.5	2.831\\
0.55	2.93\\
0.6	3.01\\
0.65	3.103\\
0.7	3.177\\
0.75	3.265\\
0.8	3.338\\
0.85	3.423\\
0.9	3.507\\
0.95	3.584\\
1	3.661\\
1.05	3.741\\
1.1	3.809\\
1.15	3.889\\
1.2	3.961\\
1.25	4.027\\
1.3	4.099\\
1.35	4.162\\
1.4	4.234\\
1.45	4.308\\
1.5	4.383\\
1.55	4.455\\
1.6	4.526\\
1.65	4.603\\
1.7	4.667\\
1.75	4.73\\
1.8	4.803\\
1.85	4.863\\
1.9	4.925\\
1.95	4.973\\
2	5.03\\
2.05	5.091\\
2.1	5.144\\
2.15	5.207\\
2.2	5.254\\
2.25	5.301\\
2.3	5.357\\
2.35	5.425\\
2.4	5.465\\
2.45	5.498\\
2.5	5.563\\
2.55	5.609\\
2.6	5.643\\
2.65	5.7\\
2.7	5.741\\
2.75	5.785\\
2.8	5.83\\
2.85	5.876\\
2.9	5.926\\
2.95	5.959\\
3	6.004\\
3.05	6.034\\
3.1	6.054\\
3.15	6.094\\
3.2	6.104\\
3.25	6.127\\
3.3	6.159\\
3.35	6.19\\
3.4	6.219\\
3.45	6.249\\
3.5	6.28\\
3.55	6.303\\
3.6	6.328\\
3.65	6.364\\
3.7	6.378\\
3.75	6.428\\
3.8	6.457\\
3.85	6.49\\
3.9	6.523\\
3.95	6.554\\
4	6.583\\
};
\addlegendentry{$K=1$}

\addplot [color=cyan, dashed, mark=square, mark repeat = 4, mark options={solid}]
  table[row sep=crcr]{%
0.001   16\\
0.05	10.882\\
0.1	6.071\\
0.15	4.537\\
0.2	3.889\\
0.25	3.461\\
0.3	3.233\\
0.35	3.101\\
0.4	3.028\\
0.45	2.971\\
0.5	2.929\\
0.55	2.945\\
0.6	2.919\\
0.65	2.953\\
0.7	2.975\\
0.75	2.956\\
0.8	3.013\\
0.85	3.059\\
0.9	3.097\\
0.95	3.14\\
1	3.163\\
1.05	3.185\\
1.1	3.244\\
1.15	3.309\\
1.2	3.38\\
1.25	3.434\\
1.3	3.49\\
1.35	3.547\\
1.4	3.604\\
1.45	3.646\\
1.5	3.696\\
1.55	3.74\\
1.6	3.783\\
1.65	3.831\\
1.7	3.869\\
1.75	3.917\\
1.8	3.964\\
1.85	4\\
1.9	4.046\\
1.95	4.078\\
2	4.134\\
2.05	4.167\\
2.1	4.209\\
2.15	4.242\\
2.2	4.273\\
2.25	4.306\\
2.3	4.34\\
2.35	4.372\\
2.4	4.406\\
2.45	4.427\\
2.5	4.476\\
2.55	4.515\\
2.6	4.561\\
2.65	4.588\\
2.7	4.627\\
2.75	4.664\\
2.8	4.715\\
2.85	4.75\\
2.9	4.793\\
2.95	4.837\\
3	4.873\\
3.05	4.914\\
3.1	4.957\\
3.15	5.008\\
3.2	5.049\\
3.25	5.097\\
3.3	5.134\\
3.35	5.18\\
3.4	5.224\\
3.45	5.27\\
3.5	5.313\\
3.55	5.362\\
3.6	5.413\\
3.65	5.457\\
3.7	5.499\\
3.75	5.543\\
3.8	5.598\\
3.85	5.635\\
3.9	5.69\\
3.95	5.735\\
4	5.784\\
};
\addlegendentry{$K=2$}

\addplot [color=green_D, dashdotted, mark=x, mark repeat=4, mark options={solid}]
  table[row sep=crcr]{%
0.25	4.968\\
0.3	4.162\\
0.35	3.715\\
0.4	3.412\\
0.45	3.144\\
0.5	3.036\\
0.55	3.002\\
0.6	2.899\\
0.65	2.9\\
0.7	2.849\\
0.75	2.862\\
0.8	2.894\\
0.85	2.923\\
0.9	2.939\\
0.95	2.961\\
1	2.985\\
1.05	3.058\\
1.1	3.126\\
1.15	3.188\\
1.2	3.255\\
1.25	3.307\\
1.3	3.363\\
1.35	3.421\\
1.4	3.472\\
1.45	3.515\\
1.5	3.56\\
1.55	3.619\\
1.6	3.659\\
1.65	3.712\\
1.7	3.76\\
1.75	3.809\\
1.8	3.857\\
1.85	3.897\\
1.9	3.951\\
1.95	3.986\\
2	4.033\\
2.05	4.078\\
2.1	4.11\\
2.15	4.162\\
2.2	4.2\\
2.25	4.251\\
2.3	4.287\\
2.35	4.335\\
2.4	4.369\\
2.45	4.416\\
2.5	4.455\\
2.55	4.505\\
2.6	4.546\\
2.65	4.597\\
2.7	4.632\\
2.75	4.692\\
2.8	4.729\\
2.85	4.78\\
2.9	4.832\\
2.95	4.877\\
3	4.918\\
3.05	4.969\\
3.1	5.01\\
3.15	5.065\\
3.2	5.11\\
3.25	5.163\\
3.3	5.209\\
3.35	5.256\\
3.4	5.308\\
3.45	5.356\\
3.5	5.396\\
3.55	5.446\\
3.6	5.503\\
3.65	5.541\\
3.7	5.595\\
3.75	5.645\\
3.8	5.692\\
3.85	5.742\\
3.9	5.794\\
3.95	5.844\\
4	5.881\\
};
\addlegendentry{$K=3$}

\addplot [color=violet, densely dashed, mark=triangle, mark repeat=4, mark options={solid}]
table[row sep=crcr]{%
0.05	16\\
0.1	16\\
0.15	16\\
0.2	11.996\\
0.25	7.905\\
0.3	5.881\\
0.35	4.781\\
0.4	4.126\\
0.45	3.686\\
0.5	3.411\\
0.55	3.233\\
0.6	3.157\\
0.65	3.082\\
0.7	3.06\\
0.75	3.005\\
0.8	3.05\\
0.85	3.084\\
0.9	3.122\\
0.95	3.156\\
1	3.19\\
1.05	3.237\\
1.1	3.292\\
1.15	3.383\\
1.2	3.458\\
1.25	3.542\\
1.3	3.62\\
1.35	3.699\\
1.4	3.779\\
1.45	3.855\\
1.5	3.921\\
1.55	4.004\\
1.6	4.073\\
1.65	4.143\\
1.7	4.226\\
1.75	4.301\\
1.8	4.382\\
1.85	4.456\\
1.9	4.54\\
1.95	4.613\\
2	4.702\\
2.05	4.775\\
2.1	4.869\\
2.15	4.951\\
2.2	5.035\\
2.25	5.111\\
2.3	5.206\\
2.35	5.3\\
2.4	5.383\\
2.45	5.479\\
2.5	5.57\\
2.55	5.676\\
2.6	5.758\\
2.65	5.855\\
2.7	5.937\\
2.75	6.037\\
2.8	6.135\\
2.85	6.234\\
2.9	6.323\\
2.95	6.423\\
3	6.505\\
3.05	6.606\\
3.1	6.714\\
3.15	6.805\\
3.2	6.903\\
3.25	7\\
3.3	7.101\\
3.35	7.195\\
3.4	7.294\\
3.45	7.392\\
3.5	7.493\\
3.55	7.597\\
3.6	7.695\\
3.65	7.786\\
3.7	7.889\\
3.75	7.985\\
3.8	8.081\\
3.85	8.187\\
3.9	8.289\\
3.95	8.38\\
4	8.487\\
};
\addlegendentry{$K=4$}

\addplot [color=blue, dotted, mark=+, mark repeat=4, mark options={solid}]
  table[row sep=crcr]{%
0.05    16\\
0.1 16\\
0.15    16\\
0.2 16\\
0.25	15.422\\
0.3	9.834\\
0.35	7.253\\
0.4	5.847\\
0.45	4.946\\
0.5	4.475\\
0.55	4.201\\
0.6	3.987\\
0.65	3.843\\
0.7	3.825\\
0.75	3.771\\
0.8	3.812\\
0.85	3.852\\
0.9	3.909\\
0.95	3.958\\
1	4.039\\
1.05	4.125\\
1.1	4.23\\
1.15	4.337\\
1.2	4.45\\
1.25	4.558\\
1.3	4.68\\
1.35	4.81\\
1.4	4.929\\
1.45	5.05\\
1.5	5.182\\
1.55	5.316\\
1.6	5.454\\
1.65	5.572\\
1.7	5.718\\
1.75	5.844\\
1.8	5.988\\
1.85	6.124\\
1.9	6.274\\
1.95	6.418\\
2	6.557\\
2.05	6.704\\
2.1	6.85\\
2.15	6.99\\
2.2	7.122\\
2.25	7.285\\
2.3	7.419\\
2.35	7.592\\
2.4	7.729\\
2.45	7.88\\
2.5	8.021\\
2.55	8.154\\
2.6	8.318\\
2.65	8.451\\
2.7	8.606\\
2.75	8.743\\
2.8	8.916\\
2.85	9.065\\
2.9	9.205\\
2.95	9.356\\
3	9.508\\
3.05	9.643\\
3.1	9.808\\
3.15	9.95\\
3.2	10.101\\
3.25	10.256\\
3.3	10.405\\
3.35	10.54\\
3.4	10.696\\
3.45	10.841\\
3.5	11\\
3.55	11.143\\
3.6	11.298\\
3.65	11.432\\
3.7	11.598\\
3.75	11.741\\
3.8	11.902\\
3.85	12.047\\
3.9	12.197\\
3.95	12.341\\
4	12.506\\
};
\addlegendentry{$K=5$}

\addplot [color=red, densely dotted, mark=diamond, mark repeat=4, mark options={solid}]
  table[row sep=crcr]{%
0.05    16\\
0.1 16\\
0.15    16\\
0.2 16\\
0.25	16\\
0.3	16\\
0.35	16\\
0.4	12.643\\
0.45	10.218\\
0.5	8.879\\
0.55	7.964\\
0.6	7.537\\
0.65	7.259\\
0.7	7.125\\
0.75	7.131\\
0.8	7.169\\
0.85	7.345\\
0.9	7.512\\
0.95	7.69\\
1	7.956\\
1.05	8.25\\
1.1	8.505\\
1.15	8.76\\
1.2	9.079\\
1.25	9.407\\
1.3	9.738\\
1.35	10.049\\
1.4	10.371\\
1.45	10.702\\
1.5	11.052\\
1.55	11.363\\
1.6	11.68\\
1.65	12.047\\
1.7	12.381\\
1.75	12.705\\
1.8	13.064\\
1.85	13.452\\
1.9	13.767\\
1.95	14.124\\
2	14.487\\
2.05	14.791\\
2.1	15.178\\
2.15	15.534\\
2.2	15.85\\
2.25	16\\
2.3	16\\
2.35	16\\
2.4	16\\
2.45	16\\
2.5	16\\
2.55	16\\
2.6	16\\
2.65	16\\
2.7	16\\
2.75	16\\
2.8	16\\
2.85	16\\
2.9	16\\
2.95	16\\
3	16\\
3.05	16\\
3.1	16\\
3.15	16\\
3.2	16\\
3.25	16\\
3.3	16\\
3.35	16\\
3.4	16\\
3.45	16\\
3.5	16\\
3.55	16\\
3.6	16\\
3.65	16\\
3.7	16\\
3.75	16\\
3.8	16\\
3.85	16\\
3.9	16\\
3.95	16\\
4	16\\
};
\addlegendentry{$K=6$}

\end{axis}
\end{tikzpicture}%
        \fi        
        \caption{$M=2.25$ and $L=2$.}
        \label{fig:age_red_L2_025}
    \end{subfigure}	
    \begin{subfigure}[b]{.49\linewidth}
	    \centering
        \ifdefined\pdffig
            \includegraphics[width=\linewidth]{tikz/paper-figure56}
        \else
            \begin{tikzpicture}

\begin{axis}[%
width=\sfwidth,
height=\sfheight,
xmin=0,
xmax=4,
ymin=0,
ymax=15,
axis background/.style={fill=white},
xlabel style={font=\footnotesize\color{white!15!black}},
xlabel={$\tau$},
ylabel near ticks,
ylabel style={font=\footnotesize\color{white!15!black}},
xticklabel style={font=\footnotesize\color{white!15!black}},
yticklabel style={font=\footnotesize\color{white!15!black}},
ylabel={$\Delta_{99}$},
axis background/.style={fill=white},
xmajorgrids,
ymajorgrids,
legend style={font=\tiny, at={(0.99,0.02)}, anchor=south east, legend columns=2,legend cell align=left, align=left,fill opacity=0.8, draw opacity=1, text opacity=1, draw=white!80!black}
]
\addplot [color=orange_D, mark=o, mark repeat = 4, mark phase=1]
  table[row sep=crcr]{%
0.001   3.884\\
0.05	3.876\\
0.1	4.073\\
0.15	4.148\\
0.2	4.205\\
0.25	4.307\\
0.3	4.422\\
0.35	4.489\\
0.4	4.585\\
0.45	4.698\\
0.5	4.782\\
0.55	4.877\\
0.6	4.964\\
0.65	5.042\\
0.7	5.14\\
0.75	5.238\\
0.8	5.339\\
0.85	5.407\\
0.9	5.503\\
0.95	5.574\\
1	5.687\\
1.05	5.781\\
1.1	5.836\\
1.15	5.926\\
1.2	6.041\\
1.25	6.126\\
1.3	6.198\\
1.35	6.28\\
1.4	6.358\\
1.45	6.431\\
1.5	6.524\\
1.55	6.6\\
1.6	6.69\\
1.65	6.783\\
1.7	6.857\\
1.75	6.941\\
1.8	7.02\\
1.85	7.097\\
1.9	7.179\\
1.95	7.253\\
2	7.331\\
2.05	7.402\\
2.1	7.489\\
2.15	7.561\\
2.2	7.62\\
2.25	7.694\\
2.3	7.754\\
2.35	7.856\\
2.4	7.915\\
2.45	7.978\\
2.5	8.059\\
2.55	8.129\\
2.6	8.194\\
2.65	8.256\\
2.7	8.329\\
2.75	8.399\\
2.8	8.465\\
2.85	8.532\\
2.9	8.618\\
2.95	8.684\\
3	8.761\\
3.05	8.842\\
3.1	8.919\\
3.15	8.984\\
3.2	9.047\\
3.25	9.133\\
3.3	9.209\\
3.35	9.261\\
3.4	9.336\\
3.45	9.404\\
3.5	9.467\\
3.55	9.531\\
3.6	9.599\\
3.65	9.67\\
3.7	9.708\\
3.75	9.801\\
3.8	9.85\\
3.85	9.904\\
3.9	9.958\\
3.95	10.019\\
4	10.07\\
};
\addlegendentry{$K=1$}

\addplot [color=cyan, dashed, mark=square, mark repeat = 4, mark options={solid}]
  table[row sep=crcr]{%
0.05	16\\
0.1	16\\
0.15	14.951\\
0.2	12.145\\
0.25	10.249\\
0.3	9.066\\
0.35	8.332\\
0.4	7.784\\
0.45	7.187\\
0.5	6.922\\
0.55	6.662\\
0.6	6.495\\
0.65	6.305\\
0.7	6.191\\
0.75	6.12\\
0.8	6.058\\
0.85	5.918\\
0.9	5.941\\
0.95	5.898\\
1	5.841\\
1.05	5.872\\
1.1	5.876\\
1.15	5.823\\
1.2	5.825\\
1.25	5.883\\
1.3	5.919\\
1.35	5.946\\
1.4	5.939\\
1.45	5.934\\
1.5	5.904\\
1.55	5.968\\
1.6	6.024\\
1.65	6.082\\
1.7	6.109\\
1.75	6.181\\
1.8	6.214\\
1.85	6.243\\
1.9	6.275\\
1.95	6.293\\
2	6.343\\
2.05	6.365\\
2.1	6.389\\
2.15	6.413\\
2.2	6.485\\
2.25	6.558\\
2.3	6.62\\
2.35	6.698\\
2.4	6.742\\
2.45	6.808\\
2.5	6.865\\
2.55	6.92\\
2.6	6.99\\
2.65	7.037\\
2.7	7.097\\
2.75	7.134\\
2.8	7.21\\
2.85	7.242\\
2.9	7.299\\
2.95	7.345\\
3	7.387\\
3.05	7.431\\
3.1	7.484\\
3.15	7.541\\
3.2	7.583\\
3.25	7.621\\
3.3	7.668\\
3.35	7.716\\
3.4	7.747\\
3.45	7.797\\
3.5	7.832\\
3.55	7.891\\
3.6	7.941\\
3.65	7.968\\
3.7	7.996\\
3.75	8.042\\
3.8	8.1\\
3.85	8.127\\
3.9	8.176\\
3.95	8.212\\
4	8.253\\
};
\addlegendentry{$K=2$}

\addplot [color=green_D, dashdotted, mark=x, mark repeat=4, mark options={solid}]
  table[row sep=crcr]{%
0.05    16\\
0.1 16\\
0.15    16\\
0.2	16\\
0.25	16\\
0.3	16\\
0.35	15.397\\
0.4	12.839\\
0.45	11.175\\
0.5	9.937\\
0.55	9.086\\
0.6	8.329\\
0.65	7.763\\
0.7	7.438\\
0.75	7.099\\
0.8	6.818\\
0.85	6.564\\
0.9	6.292\\
0.95	6.262\\
1	6.089\\
1.05	6.011\\
1.1	6.009\\
1.15	5.813\\
1.2	5.808\\
1.25	5.812\\
1.3	5.806\\
1.35	5.776\\
1.4	5.693\\
1.45	5.669\\
1.5	5.716\\
1.55	5.756\\
1.6	5.775\\
1.65	5.819\\
1.7	5.847\\
1.75	5.864\\
1.8	5.902\\
1.85	5.903\\
1.9	5.922\\
1.95	5.936\\
2	5.969\\
2.05	6.042\\
2.1	6.113\\
2.15	6.186\\
2.2	6.245\\
2.25	6.325\\
2.3	6.383\\
2.35	6.446\\
2.4	6.506\\
2.45	6.561\\
2.5	6.62\\
2.55	6.679\\
2.6	6.728\\
2.65	6.795\\
2.7	6.826\\
2.75	6.896\\
2.8	6.935\\
2.85	6.986\\
2.9	7.044\\
2.95	7.101\\
3	7.13\\
3.05	7.185\\
3.1	7.227\\
3.15	7.287\\
3.2	7.332\\
3.25	7.383\\
3.3	7.43\\
3.35	7.472\\
3.4	7.529\\
3.45	7.573\\
3.5	7.611\\
3.55	7.655\\
3.6	7.725\\
3.65	7.75\\
3.7	7.802\\
3.75	7.842\\
3.8	7.889\\
3.85	7.931\\
3.9	7.979\\
3.95	8.036\\
4	8.051\\
};
\addlegendentry{$K=3$}

\addplot [color=violet, densely dashed, mark=triangle, mark repeat=4, mark options={solid}]
table[row sep=crcr]{%
0.5	15.77\\
0.55	13.404\\
0.6	11.718\\
0.65	10.362\\
0.7	9.568\\
0.75	8.793\\
0.8	8.259\\
0.85	7.649\\
0.9	7.377\\
0.95	7.153\\
1	6.842\\
1.05	6.712\\
1.1	6.461\\
1.15	6.429\\
1.2	6.328\\
1.25	6.145\\
1.3	6.144\\
1.35	6.148\\
1.4	6.138\\
1.45	6.1\\
1.5	6.008\\
1.55	6.053\\
1.6	6.088\\
1.65	6.116\\
1.7	6.155\\
1.75	6.199\\
1.8	6.238\\
1.85	6.274\\
1.9	6.308\\
1.95	6.349\\
2	6.386\\
2.05	6.425\\
2.1	6.475\\
2.15	6.528\\
2.2	6.578\\
2.25	6.661\\
2.3	6.743\\
2.35	6.845\\
2.4	6.923\\
2.45	7.002\\
2.5	7.073\\
2.55	7.178\\
2.6	7.24\\
2.65	7.331\\
2.7	7.39\\
2.75	7.47\\
2.8	7.554\\
2.85	7.633\\
2.9	7.699\\
2.95	7.769\\
3	7.834\\
3.05	7.919\\
3.1	8.012\\
3.15	8.064\\
3.2	8.153\\
3.25	8.225\\
3.3	8.291\\
3.35	8.386\\
3.4	8.447\\
3.45	8.514\\
3.5	8.598\\
3.55	8.674\\
3.6	8.755\\
3.65	8.836\\
3.7	8.915\\
3.75	8.992\\
3.8	9.062\\
3.85	9.154\\
3.9	9.232\\
3.95	9.31\\
4	9.408\\
};
\addlegendentry{$K=4$}

\addplot [color=blue, dotted, mark=+, mark repeat=4, mark options={solid}]
  table[row sep=crcr]{%
0.05    16\\
0.1 16\\
0.15    16\\
0.2 16\\
0.25	16\\
0.3	16\\
0.35	16\\
0.4	16\\
0.45	16\\
0.5	16\\
0.55	16\\
0.6	16\\
0.65	16\\
0.7	14.5\\
0.75	12.796\\
0.8	11.703\\
0.85	10.755\\
0.9	10.096\\
0.95	9.37\\
1	8.928\\
1.05	8.634\\
1.1	8.374\\
1.15	8.041\\
1.2	7.97\\
1.25	7.833\\
1.3	7.684\\
1.35	7.652\\
1.4	7.632\\
1.45	7.608\\
1.5	7.537\\
1.55	7.554\\
1.6	7.629\\
1.65	7.647\\
1.7	7.709\\
1.75	7.762\\
1.8	7.789\\
1.85	7.856\\
1.9	7.911\\
1.95	7.976\\
2	8.055\\
2.05	8.15\\
2.1	8.269\\
2.15	8.38\\
2.2	8.459\\
2.25	8.573\\
2.3	8.678\\
2.35	8.816\\
2.4	8.908\\
2.45	9.027\\
2.5	9.142\\
2.55	9.238\\
2.6	9.384\\
2.65	9.474\\
2.7	9.608\\
2.75	9.716\\
2.8	9.865\\
2.85	10.01\\
2.9	10.11\\
2.95	10.262\\
3	10.371\\
3.05	10.479\\
3.1	10.644\\
3.15	10.76\\
3.2	10.898\\
3.25	11.044\\
3.3	11.165\\
3.35	11.279\\
3.4	11.438\\
3.45	11.562\\
3.5	11.718\\
3.55	11.848\\
3.6	11.988\\
3.65	12.118\\
3.7	12.271\\
3.75	12.391\\
3.8	12.564\\
3.85	12.702\\
3.9	12.824\\
3.95	12.975\\
4	13.138\\
};
\addlegendentry{$K=5$}

\addplot [color=red, densely dotted, mark=diamond, mark repeat=4, mark options={solid}]
  table[row sep=crcr]{%
0.05    16\\
0.1 16\\
0.15    16\\
0.2 16\\
0.25	16\\
0.3	16\\
0.35	16\\
0.4	16\\
0.45	16\\
0.5	16\\
0.55	16\\
0.6	16\\
0.65	16\\
0.7	16\\
0.75	16\\
0.8	16\\
0.85	16\\
0.9	16\\
0.95	16\\
1	16\\
1.05	16\\
1.1	16\\
1.15	15.46\\
1.2	15.147\\
1.25	14.712\\
1.3	14.447\\
1.35	14.348\\
1.4	14.233\\
1.45	14.22\\
1.5	14.297\\
1.55	14.282\\
1.6	14.353\\
1.65	14.507\\
1.7	14.678\\
1.75	14.915\\
1.8	15.063\\
1.85	15.231\\
1.9	15.439\\
1.95	15.678\\
2	15.923\\
2.05	16\\
2.1	16\\
2.15	16\\
2.2	16\\
2.25	16\\
2.3	16\\
2.35	16\\
2.4	16\\
2.45	16\\
2.5	16\\
2.55	16\\
2.6	16\\
2.65	16\\
2.7	16\\
2.75	16\\
2.8	16\\
2.85	16\\
2.9	16\\
2.95	16\\
3	16\\
3.05	16\\
3.1	16\\
3.15	16\\
3.2	16\\
3.25	16\\
3.3	16\\
3.35	16\\
3.4	16\\
3.45	16\\
3.5	16\\
3.55	16\\
3.6	16\\
3.65	16\\
3.7	16\\
3.75	16\\
3.8	16\\
3.85	16\\
3.9	16\\
3.95	16\\
4	16\\
};
\addlegendentry{$K=6$}

\end{axis}
\end{tikzpicture}%
        \fi        
        \caption{$M=4.5$ and $L=2$.}
        \label{fig:age_red_L2_075}
    \end{subfigure}
    \begin{subfigure}[b]{.49\linewidth}
	    \centering
        \ifdefined\pdffig
            \includegraphics[width=\linewidth]{tikz/paper-figure57}
        \else
            \begin{tikzpicture}

\begin{axis}[%
width=\sfwidth,
height=\sfheight,
xmin=0,
xmax=4,
ymin=0,
ymax=15,
axis background/.style={fill=white},
xlabel style={font=\footnotesize\color{white!15!black}},
xlabel={$\tau$},
ylabel near ticks,
ylabel style={font=\footnotesize\color{white!15!black}},
xticklabel style={font=\footnotesize\color{white!15!black}},
yticklabel style={font=\footnotesize\color{white!15!black}},
ylabel={$\Delta_{99}$},
axis background/.style={fill=white},
xmajorgrids,
ymajorgrids,
legend style={font=\tiny, at={(0.99,0.02)}, anchor=south east, legend columns=2,legend cell align=left, align=left,fill opacity=0.8, draw opacity=1, text opacity=1, draw=white!80!black}
]
\addplot [color=orange_D, mark=o, mark repeat = 4]
  table[row sep=crcr]{%
0.5	16\\
0.55	16\\
0.6	16\\
0.65	16\\
0.7	16\\
0.75	16\\
0.8	16\\
0.85	16\\
0.9	16\\
0.95	16\\
1	16\\
1.05	16\\
1.1	16\\
1.15	16\\
1.2	16\\
1.25	16\\
1.3	16\\
1.35	16\\
1.4	16\\
1.45	16\\
1.5	16\\
1.55	16\\
1.6	16\\
1.65	16\\
1.7	16\\
1.75	16\\
1.8	16\\
1.85	16\\
1.9	16\\
1.95	16\\
2	16\\
2.05	16\\
2.1	16\\
2.15	16\\
2.2	16\\
2.25	16\\
2.3	16\\
2.35	16\\
2.4	16\\
2.45	15.21\\
2.5	13\\
2.55	11.514\\
2.6	10.525\\
2.65	9.783\\
2.7	9.201\\
2.75	8.734\\
2.8	8.386\\
2.85	8.102\\
2.9	7.823\\
2.95	7.649\\
3	7.478\\
3.05	7.347\\
3.1	7.217\\
3.15	7.106\\
3.2	7.05\\
3.25	6.989\\
3.3	6.92\\
3.35	6.895\\
3.4	6.844\\
3.45	6.827\\
3.5	6.81\\
3.55	6.788\\
3.6	6.766\\
3.65	6.778\\
3.7	6.773\\
3.75	6.769\\
3.8	6.766\\
3.85	6.787\\
3.9	6.788\\
3.95	6.798\\
4	6.814\\
};
\addlegendentry{$K=1$}

\addplot [color=cyan, dashed, mark=square, mark repeat = 4, mark options={solid}]
  table[row sep=crcr]{%
0.5	16\\
0.55	16\\
0.6	16\\
0.65	16\\
0.7	16\\
0.75	16\\
0.8	16\\
0.85	16\\
0.9	16\\
0.95	16\\
1	16\\
1.05	16\\
1.1	16\\
1.15	16\\
1.2	15.301\\
1.25	10.207\\
1.3	7.994\\
1.35	6.812\\
1.4	6.104\\
1.45	5.631\\
1.5	5.269\\
1.55	5.058\\
1.6	4.858\\
1.65	4.729\\
1.7	4.632\\
1.75	4.535\\
1.8	4.493\\
1.85	4.441\\
1.9	4.418\\
1.95	4.399\\
2	4.386\\
2.05	4.373\\
2.1	4.381\\
2.15	4.387\\
2.2	4.396\\
2.25	4.42\\
2.3	4.434\\
2.35	4.458\\
2.4	4.479\\
2.45	4.513\\
2.5	4.524\\
2.55	4.565\\
2.6	4.589\\
2.65	4.623\\
2.7	4.667\\
2.75	4.693\\
2.8	4.732\\
2.85	4.77\\
2.9	4.807\\
2.95	4.855\\
3	4.89\\
3.05	4.931\\
3.1	4.972\\
3.15	5.011\\
3.2	5.056\\
3.25	5.102\\
3.3	5.148\\
3.35	5.184\\
3.4	5.23\\
3.45	5.282\\
3.5	5.323\\
3.55	5.365\\
3.6	5.405\\
3.65	5.459\\
3.7	5.499\\
3.75	5.538\\
3.8	5.597\\
3.85	5.655\\
3.9	5.694\\
3.95	5.733\\
4	5.782\\
};
\addlegendentry{$K=2$}

\addplot [color=green_D, dashdotted, mark=x, mark repeat=4, mark options={solid}]
  table[row sep=crcr]{%
0.5	16\\
0.55	16\\
0.6	16\\
0.65	16\\
0.7	16\\
0.75	16\\
0.8	15.693\\
0.85	8.923\\
0.9	6.715\\
0.95	5.639\\
1	5.031\\
1.05	4.62\\
1.1	4.362\\
1.15	4.181\\
1.2	4.057\\
1.25	3.963\\
1.3	3.908\\
1.35	3.862\\
1.4	3.825\\
1.45	3.813\\
1.5	3.82\\
1.55	3.82\\
1.6	3.829\\
1.65	3.855\\
1.7	3.876\\
1.75	3.89\\
1.8	3.928\\
1.85	3.963\\
1.9	3.987\\
1.95	4.035\\
2	4.064\\
2.05	4.095\\
2.1	4.138\\
2.15	4.176\\
2.2	4.213\\
2.25	4.248\\
2.3	4.295\\
2.35	4.343\\
2.4	4.376\\
2.45	4.416\\
2.5	4.472\\
2.55	4.505\\
2.6	4.552\\
2.65	4.597\\
2.7	4.65\\
2.75	4.688\\
2.8	4.728\\
2.85	4.781\\
2.9	4.814\\
2.95	4.883\\
3	4.924\\
3.05	4.973\\
3.1	5.023\\
3.15	5.07\\
3.2	5.109\\
3.25	5.159\\
3.3	5.203\\
3.35	5.263\\
3.4	5.308\\
3.45	5.348\\
3.5	5.399\\
3.55	5.452\\
3.6	5.497\\
3.65	5.548\\
3.7	5.592\\
3.75	5.651\\
3.8	5.698\\
3.85	5.737\\
3.9	5.789\\
3.95	5.839\\
4	5.892\\
};
\addlegendentry{$K=3$}

\addplot [color=violet, densely dashed, mark=triangle, mark repeat=4, mark options={solid}]
table[row sep=crcr]{%
0.5	16\\
0.55	16\\
0.6	16\\
0.65	9.116\\
0.7	6.633\\
0.75	5.516\\
0.8	4.899\\
0.85	4.557\\
0.9	4.299\\
0.95	4.134\\
1	4.025\\
1.05	3.95\\
1.1	3.92\\
1.15	3.899\\
1.2	3.887\\
1.25	3.905\\
1.3	3.927\\
1.35	3.936\\
1.4	3.967\\
1.45	4.015\\
1.5	4.06\\
1.55	4.103\\
1.6	4.163\\
1.65	4.213\\
1.7	4.288\\
1.75	4.344\\
1.8	4.419\\
1.85	4.484\\
1.9	4.56\\
1.95	4.631\\
2	4.708\\
2.05	4.794\\
2.1	4.874\\
2.15	4.954\\
2.2	5.036\\
2.25	5.126\\
2.3	5.223\\
2.35	5.304\\
2.4	5.399\\
2.45	5.486\\
2.5	5.58\\
2.55	5.662\\
2.6	5.76\\
2.65	5.856\\
2.7	5.939\\
2.75	6.035\\
2.8	6.132\\
2.85	6.233\\
2.9	6.324\\
2.95	6.424\\
3	6.523\\
3.05	6.612\\
3.1	6.705\\
3.15	6.809\\
3.2	6.908\\
3.25	7.005\\
3.3	7.101\\
3.35	7.19\\
3.4	7.296\\
3.45	7.397\\
3.5	7.49\\
3.55	7.586\\
3.6	7.692\\
3.65	7.789\\
3.7	7.892\\
3.75	7.989\\
3.8	8.07\\
3.85	8.193\\
3.9	8.28\\
3.95	8.373\\
4	8.476\\
};
\addlegendentry{$K=4$}

\addplot [color=blue, dotted, mark=+, mark repeat=4, mark options={solid}]
  table[row sep=crcr]{%
0.45    16\\
0.5	13.113\\
0.55	7.709\\
0.6	5.934\\
0.65	5.198\\
0.7	4.735\\
0.75	4.492\\
0.8	4.379\\
0.85	4.307\\
0.9	4.259\\
0.95	4.258\\
1	4.298\\
1.05	4.35\\
1.1	4.407\\
1.15	4.475\\
1.2	4.558\\
1.25	4.649\\
1.3	4.763\\
1.35	4.853\\
1.4	4.986\\
1.45	5.085\\
1.5	5.21\\
1.55	5.333\\
1.6	5.466\\
1.65	5.59\\
1.7	5.72\\
1.75	5.88\\
1.8	6.004\\
1.85	6.147\\
1.9	6.28\\
1.95	6.438\\
2	6.56\\
2.05	6.697\\
2.1	6.842\\
2.15	6.984\\
2.2	7.134\\
2.25	7.279\\
2.3	7.433\\
2.35	7.572\\
2.4	7.728\\
2.45	7.864\\
2.5	8.009\\
2.55	8.168\\
2.6	8.32\\
2.65	8.451\\
2.7	8.602\\
2.75	8.753\\
2.8	8.905\\
2.85	9.064\\
2.9	9.205\\
2.95	9.346\\
3	9.502\\
3.05	9.644\\
3.1	9.802\\
3.15	9.942\\
3.2	10.101\\
3.25	10.25\\
3.3	10.403\\
3.35	10.541\\
3.4	10.684\\
3.45	10.842\\
3.5	10.987\\
3.55	11.166\\
3.6	11.278\\
3.65	11.443\\
3.7	11.6\\
3.75	11.744\\
3.8	11.892\\
3.85	12.048\\
3.9	12.19\\
3.95	12.347\\
4	12.497\\
};
\addlegendentry{$K=5$}

\addplot [color=red, densely dotted, mark=diamond, mark repeat=4, mark options={solid}]
  table[row sep=crcr]{%
0.4	16\\
0.45	8.798\\
0.5	6.683\\
0.55	6.063\\
0.6	5.924\\
0.65	5.983\\
0.7	6.1\\
0.75	6.34\\
0.8	6.581\\
0.85	6.853\\
0.9	7.1\\
0.95	7.404\\
1	7.726\\
1.05	8.051\\
1.1	8.341\\
1.15	8.691\\
1.2	9.024\\
1.25	9.372\\
1.3	9.671\\
1.35	9.987\\
1.4	10.336\\
1.45	10.642\\
1.5	11.014\\
1.55	11.373\\
1.6	11.647\\
1.65	12.033\\
1.7	12.371\\
1.75	12.766\\
1.8	13.077\\
1.85	13.454\\
1.9	13.729\\
1.95	14.123\\
2	14.448\\
2.05	14.765\\
2.1	15.216\\
2.15	15.498\\
2.2	15.821\\
2.25	16\\
2.3	16\\
2.35	16\\
2.4	16\\
2.45	16\\
2.5	16\\
2.55	16\\
2.6	16\\
2.65	16\\
2.7	16\\
2.75	16\\
2.8	16\\
2.85	16\\
2.9	16\\
2.95	16\\
3	16\\
3.05	16\\
3.1	16\\
3.15	16\\
3.2	16\\
3.25	16\\
3.3	16\\
3.35	16\\
3.4	16\\
3.45	16\\
3.5	16\\
3.55	16\\
3.6	16\\
3.65	16\\
3.7	16\\
3.75	16\\
3.8	16\\
3.85	16\\
3.9	16\\
3.95	16\\
4	16\\
};
\addlegendentry{$K=6$}

\end{axis}
\end{tikzpicture}%
        \fi        
        \caption{$M=2.25$ and $L=\infty$.}
        \label{fig:age_red_Linf_025}
    \end{subfigure}
    \begin{subfigure}[b]{.49\linewidth}
	    \centering
        \ifdefined\pdffig
            \includegraphics[width=\linewidth]{tikz/paper-figure58}
        \else
            \begin{tikzpicture}

\begin{axis}[%
width=\sfwidth,
height=\sfheight,
xmin=0,
xmax=4,
ymin=0,
ymax=15,
axis background/.style={fill=white},
xlabel style={font=\footnotesize\color{white!15!black}},
xlabel={$\tau$},
ylabel near ticks,
ylabel style={font=\footnotesize\color{white!15!black}},
xticklabel style={font=\footnotesize\color{white!15!black}},
yticklabel style={font=\footnotesize\color{white!15!black}},
ylabel={$\Delta_{99}$},
axis background/.style={fill=white},
xmajorgrids,
ymajorgrids,
legend style={font=\tiny, at={(0.99,0.02)}, anchor=south east, legend columns=2,legend cell align=left, align=left,fill opacity=0.8, draw opacity=1, text opacity=1, draw=white!80!black}
]

\addplot [color=cyan, dashed, mark=square, mark repeat = 4, mark options={solid}]
  table[row sep=crcr]{%
0.5	16\\
0.55	16\\
0.6	16\\
0.65	16\\
0.7	16\\
0.75	16\\
0.8	16\\
0.85	16\\
0.9	16\\
0.95	16\\
1	16\\
1.05	16\\
1.1	16\\
1.15	16\\
1.2	16\\
1.25	16\\
1.3	16\\
1.35	16\\
1.4	16\\
1.45	16\\
1.5	16\\
1.55	16\\
1.6	16\\
1.65	16\\
1.7	16\\
1.75	16\\
1.8	16\\
1.85	16\\
1.9	16\\
1.95	16\\
2	16\\
2.05	16\\
2.1	16\\
2.15	16\\
2.2	16\\
2.25	16\\
2.3	16\\
2.35	16\\
2.4	16\\
2.45	16\\
2.5	16\\
2.55	16\\
2.6	16\\
2.65	14.718\\
2.7	13.631\\
2.75	12.878\\
2.8	12.172\\
2.85	11.663\\
2.9	11.228\\
2.95	10.912\\
3	10.606\\
3.05	10.323\\
3.1	10.119\\
3.15	9.894\\
3.2	9.706\\
3.25	9.568\\
3.3	9.443\\
3.35	9.32\\
3.4	9.228\\
3.45	9.16\\
3.5	9.065\\
3.55	9.025\\
3.6	8.96\\
3.65	8.916\\
3.7	8.897\\
3.75	8.841\\
3.8	8.801\\
3.85	8.797\\
3.9	8.788\\
3.95	8.778\\
4	8.757\\
};
\addlegendentry{$K=2$}

\addplot [color=green_D, dashdotted, mark=x, mark repeat=4, mark options={solid}]
  table[row sep=crcr]{%
0.5	16\\
0.55	16\\
0.6	16\\
0.65	16\\
0.7	16\\
0.75	16\\
0.8	16\\
0.85	16\\
0.9	16\\
0.95	16\\
1	16\\
1.05	16\\
1.1	16\\
1.15	16\\
1.2	16\\
1.25	16\\
1.3	16\\
1.35	16\\
1.4	16\\
1.45	16\\
1.5	16\\
1.55	16\\
1.6	16\\
1.65	16\\
1.7	16\\
1.75	15.276\\
1.8	13.462\\
1.85	12.139\\
1.9	11.322\\
1.95	10.528\\
2	10.018\\
2.05	9.593\\
2.1	9.27\\
2.15	8.975\\
2.2	8.706\\
2.25	8.512\\
2.3	8.344\\
2.35	8.23\\
2.4	8.093\\
2.45	8.017\\
2.5	7.922\\
2.55	7.863\\
2.6	7.779\\
2.65	7.753\\
2.7	7.73\\
2.75	7.674\\
2.8	7.664\\
2.85	7.644\\
2.9	7.625\\
2.95	7.617\\
3	7.624\\
3.05	7.624\\
3.1	7.636\\
3.15	7.632\\
3.2	7.657\\
3.25	7.673\\
3.3	7.691\\
3.35	7.718\\
3.4	7.732\\
3.45	7.76\\
3.5	7.79\\
3.55	7.831\\
3.6	7.839\\
3.65	7.892\\
3.7	7.913\\
3.75	7.96\\
3.8	7.993\\
3.85	8.032\\
3.9	8.059\\
3.95	8.102\\
4	8.117\\
};
\addlegendentry{$K=3$}

\addplot [color=violet, densely dashed, mark=triangle, mark repeat=4, mark options={solid}]
table[row sep=crcr]{%
0.5	16\\
0.55	16\\
0.6	16\\
0.65	16\\
0.7	16\\
0.75	16\\
0.8	16\\
0.85	16\\
0.9	16\\
0.95	16\\
1	16\\
1.05	16\\
1.1	16\\
1.15	16\\
1.2	16\\
1.25	16\\
1.3	16\\
1.35	15.115\\
1.4	13.332\\
1.45	11.945\\
1.5	11.051\\
1.55	10.338\\
1.6	9.882\\
1.65	9.353\\
1.7	9.061\\
1.75	8.81\\
1.8	8.59\\
1.85	8.442\\
1.9	8.264\\
1.95	8.187\\
2	8.061\\
2.05	7.992\\
2.1	7.927\\
2.15	7.878\\
2.2	7.823\\
2.25	7.817\\
2.3	7.8\\
2.35	7.77\\
2.4	7.796\\
2.45	7.796\\
2.5	7.823\\
2.55	7.819\\
2.6	7.831\\
2.65	7.865\\
2.7	7.878\\
2.75	7.913\\
2.8	7.95\\
2.85	8.01\\
2.9	8.016\\
2.95	8.076\\
3	8.116\\
3.05	8.134\\
3.1	8.212\\
3.15	8.276\\
3.2	8.327\\
3.25	8.371\\
3.3	8.42\\
3.35	8.502\\
3.4	8.558\\
3.45	8.617\\
3.5	8.697\\
3.55	8.771\\
3.6	8.818\\
3.65	8.907\\
3.7	8.972\\
3.75	9.042\\
3.8	9.127\\
3.85	9.2\\
3.9	9.275\\
3.95	9.348\\
4	9.416\\
};
\addlegendentry{$K=4$}

\addplot [color=blue, dotted, mark=+, mark repeat=4, mark options={solid}]
  table[row sep=crcr]{%
0.5	16\\
0.55	16\\
0.6	16\\
0.65	16\\
0.7	16\\
0.75	16\\
0.8	16\\
0.85	16\\
0.9	16\\
0.95	16\\
1	16\\
1.05	16\\
1.1	15.318\\
1.15	13.231\\
1.2	11.855\\
1.25	11.026\\
1.3	10.348\\
1.35	9.842\\
1.4	9.511\\
1.45	9.206\\
1.5	9.035\\
1.55	8.815\\
1.6	8.704\\
1.65	8.644\\
1.7	8.552\\
1.75	8.555\\
1.8	8.541\\
1.85	8.523\\
1.9	8.554\\
1.95	8.542\\
2	8.612\\
2.05	8.647\\
2.1	8.667\\
2.15	8.731\\
2.2	8.777\\
2.25	8.876\\
2.3	8.966\\
2.35	9.036\\
2.4	9.125\\
2.45	9.217\\
2.5	9.337\\
2.55	9.401\\
2.6	9.498\\
2.65	9.603\\
2.7	9.689\\
2.75	9.839\\
2.8	9.959\\
2.85	10.07\\
2.9	10.183\\
2.95	10.284\\
3	10.418\\
3.05	10.558\\
3.1	10.679\\
3.15	10.795\\
3.2	10.937\\
3.25	11.064\\
3.3	11.198\\
3.35	11.326\\
3.4	11.474\\
3.45	11.59\\
3.5	11.722\\
3.55	11.863\\
3.6	11.998\\
3.65	12.131\\
3.7	12.279\\
3.75	12.379\\
3.8	12.555\\
3.85	12.696\\
3.9	12.849\\
3.95	12.969\\
4	13.116\\
};
\addlegendentry{$K=5$}

\addplot [color=red, densely dotted, mark=diamond, mark repeat=4, mark options={solid}]
  table[row sep=crcr]{%
0.5	16\\
0.55	16\\
0.6	16\\
0.65	16\\
0.7	16\\
0.75	16\\
0.8	16\\
0.85	16\\
0.9	16\\
0.95	14.58\\
1	13.305\\
1.05	12.58\\
1.1	12.106\\
1.15	11.937\\
1.2	11.822\\
1.25	11.855\\
1.3	11.925\\
1.35	12.099\\
1.4	12.271\\
1.45	12.455\\
1.5	12.709\\
1.55	12.905\\
1.6	13.185\\
1.65	13.42\\
1.7	13.731\\
1.75	13.968\\
1.8	14.275\\
1.85	14.532\\
1.9	14.878\\
1.95	15.113\\
2	15.508\\
2.05	15.747\\
2.1	16\\
2.15	16\\
2.2	16\\
2.25	16\\
2.3	16\\
2.35	16\\
2.4	16\\
2.45	16\\
2.5	16\\
2.55	16\\
2.6	16\\
2.65	16\\
2.7	16\\
2.75	16\\
2.8	16\\
2.85	16\\
2.9	16\\
2.95	16\\
3	16\\
3.05	16\\
3.1	16\\
3.15	16\\
3.2	16\\
3.25	16\\
3.3	16\\
3.35	16\\
3.4	16\\
3.45	16\\
3.5	16\\
3.55	16\\
3.6	16\\
3.65	16\\
3.7	16\\
3.75	16\\
3.8	16\\
3.85	16\\
3.9	16\\
3.95	16\\
4	16\\
};
\addlegendentry{$K=6$}

\end{axis}
\end{tikzpicture}%
        \fi        
        \caption{$M=4.5$ and $L=\infty$.}
        \label{fig:age_red_Linf_075}
    \end{subfigure}
     \caption{99th percentile $\Delta_{99}$ of the \gls{paoi} as a function of $\tau$ for different coding schemes with $N=6$, $\mu=1$ and $\varepsilon=0.1$.}\vspace{-0.6cm}
 \label{fig:age_red}
\end{figure}

On the other hand, a higher offered traffic leads to different trade-offs, as shown on the right side of Fig.~\ref{fig:lat_red} for $G=0.5$. In this case, setting $L=1$ runs into the reliability problems we discussed above due to excessive dropping. However, it is interesting to note that setting the maximum possible amount of redundancy performs rather well in terms of latency, even though the load of the system is much higher. For small values of $L$, the system can remain stable thanks to preemption, and only one packet is needed to recover the whole block, which is a good trade-off in such a system. On the other hand, the system with $L=\infty$ is not protected by preemption, and systems with $K<4$ become unstable. It is insteresting to note that, as for the case with less traffic, the system with $L=2$ performs best when there is enough redundancy to protect the transmission, but not so much that packets are dropped too frequently: in this case, the best setting is $K=3$.

In general, the optimal choice when considering latency seems to be exploiting redundancy as much as possible without overloading the system: the fewer packets need to get through, the higher the probability that they will be delivered before they are dropped. In cases where the offered traffic is already high, as for $G=0.5$, strict preemption becomes too aggressive, and allowing a short queue such as $L=2$ or $L=3$ is the best option. In almost every case, setting $L=\infty$ is not a good choice, as queuing delay becomes an important factor in increasing the latency. However, it does provide the best reliability when the offered traffic is low, as no packets are dropped, and as such, it might be the optimal choice in connections with very high erasure probabilities.

Increasing redundancy seems to be the optimal choice to minimize the \gls{paoi} as well, as Fig.~\ref{fig:age_red} shows: we considered a fixed block size of $M=2.25$ and $M=4.5$ (corresponding to $G=0.25$ and $G=0.5$ with $\tau=1.5$), and plotted the 99th percentile $\Delta_{99}$ of the \gls{paoi} as a function of $\tau$. Setting $K=1$ and a very low $\tau$ is clearly the optimal choice to minimize the \gls{paoi}, even though it leads to a very low reliability. This result holds for both $M=2.25$ and $M=4.5$, as well as for $L=1$ and $L=2$. In the case with $L=\infty$, the stability of the queue becomes a concern, and the optimal setting is actually $K=3$, but this system has a significant disadvantage in terms of \gls{paoi} with respect to the ones with finite, short queues. Interestingly, the results from single-link transmission hold, as sending data as fast as possible with preemption seems to be the best choice. Furthermore, the trade-off between age and latency-reliability described by Talak \emph{et al.}~\cite{talak2021age} is clearly still crucial, as the optimal settings to minimize \gls{paoi} lead to a very low reliability (having a \gls{paoi} close to 2 with $\tau=0.05$, as for the $L=1$ system with $M=2.25$, means that almost 40 consecutive blocks are lost).

\section{Conclusion}\label{sec:conc}

In this paper, we have analyzed the $D/M/(K,N)$ fork-join queue with packet-level coding, deriving the latency \gls{pdf} for an arbitrary queue length $L$ under a \gls{fcfs} policy with preemption. We also derived the \gls{pdf} of the \gls{paoi} for $L\in\{1,2,\infty\}$. These analytical derivations may be useful in the future theoretical development of the field, as well as to inspire system design in distributed computing and multipath communication.

Our results show that maintaining a short queue can be beneficial to the system in terms of both latency and \gls{paoi}, as strict preemption can make the decoding of data blocks more difficult by dropping packets too frequently. Having a longer queue is beneficial when $K$ is close to $N$, i.e., the system has little redundancy, and the packet erasure probability $\varepsilon$ is high: in those cases, packets that would have been critical to decode the block end up being dropped. The optimization of the queue length with respect to the expected load is less trivial, as it involves a trade-off between not dropping packets too often and maintaining few packets in the queue at all times.

Future work on the subject involves the investigation of the \gls{lcfs} queuing policy, which can exploit preemption while maintaining a high reliability, for the cases with $L>1$, as well as more practical models oriented at multipath communication. In particular, high-throughput real-time applications such as \gls{ar} and \gls{vr} are an interesting application of these models, and integrating traffic models for them into the framework will be an interesting development for 5G and beyond.

\section*{Acknowledgment}
This work is part of the IntellIoT project that received funding from the European Union's Horizon 2020 research and innovation program under grant agreement No. 957218.

\bibliographystyle{IEEEtran}
\bibliography{parallel.bib}

\end{document}